%% file: paper_main.tex
\documentclass[aps,twocolumn,showpacs,preprintnumbers,amsmath,amssymb,pra,nofootinbib,10pt]{revtex4-1}
\pdfoutput=1
\usepackage{graphicx}
\usepackage{dcolumn}
\usepackage[tight]{subfigure}
\usepackage{amsmath}
\usepackage{verbatim}
\usepackage{color}
\usepackage{bm} 
\usepackage{bbm}
\usepackage{mathbbol}
\usepackage{natbib}
\usepackage{xspace}
\usepackage{marginnote}
\usepackage{mathtools}
\usepackage{dsfont}
\usepackage{hyperref} \hypersetup{colorlinks=true,linktoc=all,linkcolor=blue,breaklinks=true,citecolor=blue,urlcolor=blue}
\usepackage{etoolbox}
\usepackage[normalem]{ulem}
\usepackage[bottom]{footmisc}

\input{paper_latex}

\begin{document}
%
% Title
%
\input{paper_title}
%
% Abstract
%
\input{paper_abstract}
\maketitle
\input{paper_introduction}
\input{paper_single_mode}
\input{paper_model}
\input{paper_duality_qme}
\input{paper_duality_rate}
\input{paper_quantum_dot}
\input{paper_quantum_dot2}
\input{paper_summary}
\input{paper_ack}
\appendix
\input{paper_app_linearization}
\input{paper_app_duality}
\input{paper_app_unexpected_L}
\input{paper_app_crossrelation}
% \bibliography{cite}
%
\end{document}

%% file: paper_latex.tex
\robustify{\uparrow}
\robustify{\downarrow}
\robustify{\sum}
\robustify{\int}
\robustify{\nonumber}
\robustify{\cite}
\robustify{\footnote}
\newrobustcmd{\Figure}[2]{
  \begin{figure}[ht]
    \includegraphics[width=1.0\linewidth]{#1}
    \caption{#2}
  \end{figure}
}
\expandafter\newrobustcmd\csname Figure2\endcsname[3]{
\begin{figure}[ht]
  \includegraphics[width=0.9\linewidth]{#1}
  \\
  \includegraphics[width=0.9\linewidth]{#2}
  \caption{#3}
\end{figure}
}
\renewrobustcmd{\Re}{{\text{Re}}}
\renewrobustcmd{\Im}{{\text{Im}}}
\newrobustcmd{\eff}{\text{eff}} 
\newrobustcmd{\dagtot}{{\dag_\tot}}
\newrobustcmd{\dagres}{{\dag_\res}}
\newrobustcmd{\Ttot}{{\text{T}_\tot}}
\newrobustcmd{\Tres}{{\text{T}_\res}}
\newrobustcmd{\T}{{\text{T}}}
\newrobustcmd{\tot}{\text{tot}}
\newrobustcmd{\tun}{\text{T}}
\newrobustcmd{\res}{\text{R}}
\newrobustcmd{\lead}{\text{res}}
\newrobustcmd{\un}{\text{i}}   
\newrobustcmd{\In}{\text{0}}   
\newrobustcmd{\state}{inverted stationary state\xspace}   
\newrobustcmd{\K}{\mathcal{K}}
\newrobustcmd{\D}{\mathcal{I}}
\renewrobustcmd{\P}{\mathcal{P}}   
\newrobustcmd{\W}{\mathcal{W}}
\newrobustcmd{\Temp}{T}
\newrobustcmd{\dual}[1]{\bar{#1}}
\newrobustcmd{\one}{\mathds{1}}
\newrobustcmd{\ket}[1]{|#1\rangle}
\newrobustcmd{\bra}[1]{\langle#1|}
\newrobustcmd{\brkt}[1]{\langle #1 \rangle}
\newrobustcmd{\braket}[2]{\langle #1 | #2 \rangle}
\newrobustcmd{\Ket}[1]{\bm{|}#1\bm{)}}
\newrobustcmd{\Bra}[1]{\bm{(}#1\bm{|}}
\newrobustcmd{\Braket}[2]{\bm{(}#1\bm{|}#2\bm{)}}
\newrobustcmd{\Brkt}[1]{\bm{(} #1 \bm{)}}
\newrobustcmd{\op}[1]{\hat{#1}}
\newrobustcmd{\psiIn}{\psi_{\mathrm{I}}}
\newrobustcmd{\kvecIn}{\boldsymbol{k}_{\mathrm{I}}}
\newrobustcmd{\kvecInParr}{\boldsymbol{k}^{\|}_{\mathrm{I}}}
\DeclareMathOperator{\Tr}{Tr}
\newrobustcmd{\tr}{\underset{\res}{\Tr}}
\newrobustcmd{\Trlead}{\underset{\lead}{\Tr}}
\newrobustcmd{\Tralp}{\underset{\alpha}{\Tr}}
\newrobustcmd{\Trbet}{\underset{\beta}{\Tr}}
\newrobustcmd{\tri}{\Tr_\res}
\newrobustcmd{\col}[4]{{\begin{bmatrix}#1 \\ #2 \\ #3 \\ #4 \end{bmatrix}}}
\newrobustcmd{\row}[4]{{\begin{bmatrix}#1 &  #2  & #3  & #4 \end{bmatrix}}}
\newrobustcmd{\suppmat}{\cite{Schulenborg15Suppmat}}
\newrobustcmd{\Eq}[1]{Eq.~(\ref{#1})}
\newrobustcmd{\Eqs}[1]{Eqs.~(\ref{#1})}
\newrobustcmd{\BrackEq}[1]{[Eq.~(\ref{#1})]}
\newrobustcmd{\BrackApp}[1]{[App.~(\ref{#1})]}
\newrobustcmd{\BrackSec}[1]{[Sec.~(\ref{#1})]}
\newrobustcmd{\eq}[1]{(\ref{#1})}
\newrobustcmd{\Fig}[1]{Fig.~\ref{#1}}
\newrobustcmd{\BrackFig}[1]{[Fig.~\ref{#1}]}
\newrobustcmd{\fig}[1]{\ref{#1}}
\newrobustcmd{\Figs}[1]{Figs.~\ref{#1}}
\newrobustcmd{\Sec}[1]{Sec.~\ref{#1}}
\newrobustcmd{\Ref}[1]{Ref.~\cite{#1}}
\newrobustcmd{\Refs}[1]{Refs.~\cite{#1}}
\newrobustcmd{\App}[1]{App.~\ref{#1}}
\newrobustcmd{\fpop}{(-\one)^N}
\newrobustcmd{\fpOp}{(-\one)^N}
\newrobustcmd{\fpOpHat}{(-\one)^{\hat{N}}}
\newrobustcmd{\hatfpop}{(-\one)^{\hat{N}}}
\newrobustcmd{\hatfpOp}{(-\one)^{\hat{N}}}
\newrobustcmd{\zin}{z_{\text{i}}}
\newrobustcmd{\hatzin}{\hat{z}_{\text{i}}}
\newrobustcmd{\Phif}{\Phi_\mathrm{f}}
\newrobustcmd{\Phil}{\Phi_\mathrm{l}}
\newrobustcmd{\kvec}{\boldsymbol{k}}
\newrobustcmd{\kvecperp}{\boldsymbol{k}^{\perp}}
\newrobustcmd{\kvecparr}{\boldsymbol{k}^{\|}}
\newrobustcmd{\xvec}{\boldsymbol{x}}
\newrobustcmd{\pvec}{\boldsymbol{p}}
\newrobustcmd{\Vvec}{\boldsymbol{V}}
\newrobustcmd{\qZvec}{\boldsymbol{q}_0}
\newrobustcmd{\qZvecParr}{\boldsymbol{q}^{\|}_0}
\newrobustcmd{\kZvec}{\boldsymbol{k}_0}
\newrobustcmd{\uvec}{\boldsymbol{u}}
\newrobustcmd{\vvec}{\boldsymbol{v}}
\newrobustcmd{\kZvecperp}{\boldsymbol{k}^{\perp}_0}
\newrobustcmd{\kZvecalpbet}{\kvec_{\alpha\beta}}
\newrobustcmd{\kZvecalpbetParr}{\kvec^{\|}_{\alpha\beta}}
\newrobustcmd{\kZvecalpbetperp}{\kvec^{\perp}_{\alpha\beta}}
\newrobustcmd{\hOp}{\hat{h}}
\newrobustcmd{\hSC}{\hat{h}_{S}}
\newrobustcmd{\HSC}{H_{S}}
\newrobustcmd{\hW}{h_{\mathrm{W}}}
\newrobustcmd{\hWalpbet}{h_{\mathrm{W},\alpha\beta}}
\newrobustcmd{\PsiBCS}{\Psi_{BCS}}
\newrobustcmd{\HBCS}{H_{BCS}}
\newrobustcmd{\Htot}{H_{\text{tot}}}
\newrobustcmd{\hatHtot}{\hat{H}_{\text{tot}}}
\newrobustcmd{\Htun}{H_{\text{tun}}}
\newrobustcmd{\hatHtun}{\hat{H}_{\text{tun}}}
\newrobustcmd{\Hpair}{H_{\mathrm{pair}}}
\newrobustcmd{\Jens}[1]{ {\color{red}(JENS: #1)\color{black}} }
\newrobustcmd{\EBCS}{E_{BCS}}
\newrobustcmd{\EWSM}{E_{W}}
\newrobustcmd{\Csb}{C_{\mathrm{sb}}}
\newrobustcmd{\sgnfn}[1]{\mathrm{sgn}\left(#1\right)}
\newrobustcmd{\markinred}[1]{\color{red}{#1}\color{black}}
\newrobustcmd{\rsepa}{\,,\quad}
\newrobustcmd{\lsepa}{\quad ,\,}
\newrobustcmd{\lrsepa}{\quad , \quad}
\newrobustcmd{\colvec}[1]{\begin{pmatrix}#1\end{pmatrix}}
\newrobustcmd{\nbrack}[1]{\left(#1\right)}
\newrobustcmd{\sqbrack}[1]{\left[#1\right]}
\newrobustcmd{\cbrack}[1]{\left\{#1\right\}}
\newrobustcmd{\nlbrack}[1]{\left(#1\right.}
\newrobustcmd{\sqlbrack}[1]{\left[#1\right.}
\newrobustcmd{\clbrack}[1]{\left\{#1\right.}
\newrobustcmd{\nrbrack}[1]{\left.#1\right)}
\newrobustcmd{\sqrbrack}[1]{\left.#1\right]}
\newrobustcmd{\crbrack}[1]{\left.#1\right\}}
\newrobustcmd{\mean}[1]{\left\langle#1\right\rangle}
\newrobustcmd{\MeanO}[1]{\mean{#1}_0}
\newrobustcmd{\Mset}[1]{\left\{#1\right\}}
\newrobustcmd{\Mserset}[2]{\left\{#1,\dotsc,#2\right\}}
\newrobustcmd{\Forall}{\quad\forall}
\newrobustcmd{\Exists}{\quad\exists}
\newrobustcmd{\ExistsOne}{\quad\exists_1}
\newrobustcmd{\existsOne}{\exists_1}
\newrobustcmd{\MTupel}[1]{\left(#1\right)}
\newrobustcmd{\MserTup}[2]{\left(#1,\cdots,#2\right)}
\newrobustcmd{\Thetafn}[1]{\varTheta(#1)}
\newrobustcmd{\Deltafn}[1]{\delta(#1)}
\newrobustcmd{\expfn}[1]{\exp\left(#1\right)}
\newrobustcmd{\lnfn}[1]{\ln\left(#1\right)}
\newrobustcmd{\lnP}[1]{\ln_P\left(#1\right)}
\newrobustcmd{\Fermfn}[2]{\Over{\expfn{\beta\left(#1-#2\right)}+1}}
\newrobustcmd{\fermfn}[2]{f_{#2}(#1)}
\newrobustcmd{\bosefn}[2]{b_{#2}(#1)}
\newrobustcmd{\abs}[1]{\left|#1\right|}
\renewrobustcmd{\Re}{\mathrm{Re}}
\renewrobustcmd{\Im}{\mathrm{Im}}
\newrobustcmd{\rhofn}[2]{\rho_{#1}(#2)}
\newrobustcmd{\cosfn}[1]{\cos{\left(#1\right)}}
\newrobustcmd{\arcsinfn}[1]{\mathrm{arcsin}\left(#1\right)}
\newrobustcmd{\sinfn}[1]{\sin{\left(#1\right)}}
\newrobustcmd{\tanfn}[1]{\tan{\left(#1\right)}}
\newrobustcmd{\cotfn}[1]{\cot{\left(#1\right)}}
\newrobustcmd{\cothfn}[1]{\coth{\left(#1\right)}}
\newrobustcmd{\tanhfn}[1]{\tanh\!\left(#1\right)}
\newrobustcmd{\sumsub}[2]{\sum_{\substack{{#1}\\{#2}}}}
\newrobustcmd{\sumsubsub}[3]{\sum_{\substack{{#1}\\{#2}\\{#3}}}}
\newrobustcmd{\prodsub}[2]{\prod_{\substack{{#1}\\{#2}}}}
\newrobustcmd{\cc}[1]{\overline{#1}}
\newrobustcmd{\equalby}[1]{\overset{#1}{=}}
\newrobustcmd{\equalbys}[2]{\underset{#2}{\overset{#1}{=}}}
\newrobustcmd{\equalbyeqn}[1]{\overset{\eqref{#1}}{=}}
\newrobustcmd{\equalbyeqns}[2]{\underset{\eqref{#2}}{\overset{\eqref{#1}}{=}}}
\newrobustcmd{\evalAtEqui}[1]{#1\big|_{\text{eq}}}
\newrobustcmd{\evalAtBal}[1]{\left.#1\right|_{\text{bal}}}
\newrobustcmd{\zeq}{z_{\text{eq}}}
\newrobustcmd{\hatzeq}{\hat{z}_{\text{eq}}}
\newrobustcmd{\zieq}{z_{\text{i,eq}}}
\newrobustcmd{\hatzieq}{\hat{z}_{\text{i,eq}}}
\newrobustcmd{\zia}{z_{\text{i}\alpha}}
\newrobustcmd{\za}{z_{\alpha}}
\newrobustcmd{\hatzia}{\hat{z}_{\text{i}\alpha}}
\newrobustcmd{\nzia}{n_{\text{i}\alpha}}
\newrobustcmd{\nziap}{n_{\text{i}\alpha'}}
\newrobustcmd{\pzia}{p_{\text{i}\alpha}}
\newrobustcmd{\Eeq}{E_{\text{eq}}}
\newrobustcmd{\nza}{n_{z\alpha}}
\newrobustcmd{\pza}{p_{z\alpha}}
\newrobustcmd{\nzap}{n_{z\alpha'}}
\newrobustcmd{\nz}{n_{z}}
\newrobustcmd{\nzeq}{n_{z,\text{eq}}}
\newrobustcmd{\delnsqzeq}{\delta n^2_{z,\text{eq}}}
\newrobustcmd{\pzeq}{p_{z,\text{eq}}}
\newrobustcmd{\pzieq}{p_{i,\text{eq}}}
\newrobustcmd{\ceq}{c_{\text{eq}}}
\newrobustcmd{\cpeq}{c'_{\text{eq}}}
\newrobustcmd{\peq}{p_{\text{eq}}}
\newrobustcmd{\ppeq}{p'_{\text{eq}}}
\newrobustcmd{\gamc}{\gamma_c}
\newrobustcmd{\gamp}{\gamma_p}
\newrobustcmd{\gamca}{\gamma_{c\alpha}}
\newrobustcmd{\gampa}{\gamma_{p\alpha}}
\newrobustcmd{\gamceq}{\gamma_{c,\text{eq}}}
\newrobustcmd{\gamceff}{\gamma^{\text{eff}}_{c}}
\newrobustcmd{\gamceffa}{\gamma^{\text{eff}}_{c\alpha}}
\newrobustcmd{\gamceffeq}{\gamma^{\text{eff}}_{c,\text{eq}}}
\newrobustcmd{\gampeff}{\gamma^{\text{eff}}_{p}}
\newrobustcmd{\gampeffa}{\gamma^{\text{eff}}_{p\alpha}}
\newrobustcmd{\gampeffeq}{\gamma^{\text{eff}}_{p,\text{eq}}}
\newrobustcmd{\gamceps}{\gamma_{c\epsilon}}
\newrobustcmd{\gamcepsa}{\gamma_{c\epsilon,\alpha}}
\newrobustcmd{\gamcepseq}{\gamma_{c\epsilon,\text{eq}}}
\newrobustcmd{\gamcU}{\gamma_{cU}}
\newrobustcmd{\gamcUa}{\gamma_{cU,\alpha}}
\newrobustcmd{\gamcUeq}{\gamma_{cU,\text{eq}}}
\newrobustcmd{\gameq}{\gamma_{\text{eq}}}
\newrobustcmd{\gampeq}{\gamma_{p,\text{eq}}}
\newrobustcmd{\gamci}{\dual{\gamma}_c}
\newrobustcmd{\gampi}{\dual{\gamma}_p}
\newrobustcmd{\gamcia}{\dual{\gamma}_{c\alpha}}
\newrobustcmd{\gampia}{\dual{\gamma}_{p\alpha}}
\newrobustcmd{\gamcieq}{\dual{\gamma}_{c,\text{eq}}}
\newrobustcmd{\gamceffi}{\dual{\gamma}^{\text{eff}}_{c}}
\newrobustcmd{\gamceffia}{\dual{\gamma}^{\text{eff}}_{c\alpha}}
\newrobustcmd{\gamceffieq}{\dual{\gamma}^{\text{eff}}_{c,\text{eq}}}
\newrobustcmd{\gampeffi}{\dual{\gamma}^{\text{eff}}_{p}}
\newrobustcmd{\gampeffia}{\dual{\gamma}^{\text{eff}}_{p\alpha}}
\newrobustcmd{\gampeffieq}{\dual{\gamma}^{\text{eff}}_{p,\text{eq}}}
\newrobustcmd{\gamcepsi}{\dual{\gamma}_{c\epsilon}}
\newrobustcmd{\gamcepsia}{\dual{\gamma}_{c\epsilon,\alpha}}
\newrobustcmd{\gamcepsieq}{\dual{\gamma}_{c\epsilon,\text{eq}}}
\newrobustcmd{\gamcUi}{\dual{\gamma}_{cU}}
\newrobustcmd{\gamcUia}{\dual{\gamma}_{cU,\alpha}}
\newrobustcmd{\gamcUieq}{\dual{\gamma}_{cU,\text{eq}}}
\newrobustcmd{\gamieq}{\dual{\gamma}_{\text{eq}}}
\newrobustcmd{\gampieq}{\dual{\gamma}_{p,\text{eq}}}
\newrobustcmd{\nzieq}{n_{\text{i,eq}}}
\newrobustcmd{\delnsqzieq}{\delta n^2_{\text{i,eq}}}
\newrobustcmd{\delnsqzia}{\delta n^2_{\text{i}\alpha}}
\newrobustcmd{\delnsqzi}{\delta n^2_{\text{i}}}
\newrobustcmd{\delnsqz}{\delta n^2_{z}}
\newrobustcmd{\delnsqza}{\delta n^2_{z\alpha}}
\newrobustcmd{\nzi}{n_{\text{i}}}
\newrobustcmd{\pzi}{p_{\text{i}}}
\newrobustcmd{\kBT}{k_{\text{B}}T}
\newrobustcmd{\kB}{k_{\text{B}}}
\newrobustcmd{\Hdot}{H}
\newrobustcmd{\Hleads}{H_{\text{res}}}
\newrobustcmd{\Hlead}{H_{\text{res}}}
\newrobustcmd{\hatHleads}{\hat{H}_{\text{R}}}
\newrobustcmd{\INa}{I_{\text{N}}^{\alpha}}
\newrobustcmd{\INap}{I_{\text{N}}^{\alpha'}}
\newrobustcmd{\IEa}{I_{\text{E}}^{\alpha}}
\newrobustcmd{\GamR}{\Gamma_{\text{R}}}
\newrobustcmd{\GamL}{\Gamma_{\text{L}}}
\newrobustcmd{\muR}{\mu_{\text{R}}}
\newrobustcmd{\muL}{\mu_{\text{L}}}
\newrobustcmd{\mua}{\mu_{\alpha}}
\newrobustcmd{\TR}{T_{\text{R}}}
\newrobustcmd{\TL}{T_{\text{L}}}
\newrobustcmd{\Ta}{T_{\alpha}}
\newrobustcmd{\IR}{I^{\text{R}}}
\newrobustcmd{\IL}{I^{\text{L}}}
\newrobustcmd{\INR}{I^{\text{R}}_N}
\newrobustcmd{\INL}{I^{\text{L}}_N}
\newrobustcmd{\IER}{I^{\text{R}}_E}
\newrobustcmd{\IEL}{I^{\text{L}}_E}
\newrobustcmd{\JR}{J^{\text{R}}}
\newrobustcmd{\JL}{J^{\text{L}}}
\newrobustcmd{\mueq}{\mu}
\newrobustcmd{\sudag}{{\boldsymbol{\dagger}}}
\newrobustcmd{\N}{\mathds{N}}
\newrobustcmd{\Gamop}{\mathbb{\Gamma}}
\newrobustcmd{\Gamopa}{\mathbb{\Gamma}_{\alpha}}
\newrobustcmd{\cP}{\mathcal{P}}
\newrobustcmd{\Gamepsa}{\Gamma_{\epsilon\alpha}}
\newrobustcmd{\GamepsL}{\Gamma_{\epsilon\text{L}}}
\newrobustcmd{\GamepsR}{\Gamma_{\epsilon\text{R}}}
\newrobustcmd{\Gamepsap}{\Gamma_{\epsilon\alpha'}}
\newrobustcmd{\GamUa}{\Gamma_{U\alpha}}
\newrobustcmd{\Gama}{\Gamma_{\alpha}}
\newrobustcmd{\GamUL}{\Gamma_{U\text{L}}}
\newrobustcmd{\GamUR}{\Gamma_{U\text{R}}}
\newrobustcmd{\GamUap}{\Gamma_{U\alpha'}}
\newrobustcmd{\Gameps}{\Gamma_{\epsilon}}
\newrobustcmd{\Repsaap}{R^{\alpha\alpha'}_{\epsilon}}
\newrobustcmd{\RepsLL}{R^{\text{LL}}_{\epsilon}}
\newrobustcmd{\RepsLR}{R^{\text{LR}}_{\epsilon}}
\newrobustcmd{\RepsRL}{R^{\text{RL}}_{\epsilon}}
\newrobustcmd{\RepsRR}{R^{\text{RR}}_{\epsilon}}
\newrobustcmd{\RUaap}{R^{\alpha\alpha'}_{U}}
\newrobustcmd{\RULL}{R^{\text{LL}}_{U}}
\newrobustcmd{\RULR}{R^{\text{LR}}_{U}}
\newrobustcmd{\RURL}{R^{\text{RL}}_{U}}
\newrobustcmd{\RURR}{R^{\text{RR}}_{U}}
\newrobustcmd{\GamU}{\Gamma_{U}}
\newrobustcmd{\nph}{n_{\text{ph}}}
\newrobustcmd{\nphz}{n_{\text{ph},z}}

%% file: paper_title.tex
\newrobustcmd{\suppmattitle}{}
\title{\suppmattitle Duality for open fermion systems:\\
energy-dependent weak coupling and quantum master equations
}
\author{J. Schulenborg$^{(1)}$}
\author{J. Splettstoesser$^{(1)}$}
\author{M. R. Wegewijs$^{(2,3,4)}$}
\affiliation{
  (1) Department of Microtechnology and Nanoscience (MC2), Chalmers University of Technology, SE-41298 G{\"o}teborg, Sweden
  \\
  (2) Institute for Theory of Statistical Physics,
      RWTH Aachen, 52056 Aachen,  Germany
  \\
  (3) JARA- Fundamentals of Future Information Technology
  \\
  (4) Peter Gr{\"u}nberg Institut,
      Forschungszentrum J{\"u}lich, 52425 J{\"u}lich,  Germany
}

%% file: paper_abstract.tex
\begin{abstract}
Open \emph{fermion} systems with energy-independent bilinear coupling to a \emph{fermionic} environment have been shown to obey a general \emph{duality} relation [Phys. Rev. B 93, 81411 (2016)]
which allows for a drastic simplification of time-evolution calculations.
In the weak-coupling limit, such a system can be associated with a unique dual physical system in which all energies are inverted, in particular the internal interaction.
This paper generalizes this fermionic duality in two ways:
we allow for weak coupling with arbitrary \emph{energy dependence}
and describe both occupations and \emph{coherences} coupled by a quantum master equation for the density operator.
We also show that whenever generalized detailed balance holds (Kolmogorov criterion),
the stationary probabilities for the dual system can be expressed explicitly in terms of the \emph{stationary recurrence times} of the original system, even at large bias.\par

We illustrate the generalized duality by a detailed analysis of the rate equation for a quantum dot with strong onsite Coulomb repulsion,
going beyond the commonly assumed wideband limit.
We present predictions for 
(i) the decay rates for transient charge and heat currents after a gate-voltage quench
and
(ii) the thermoelectric linear response coefficients in the stationary limit.
We show that even for pronouncedly energy-dependent coupling, all nontrivial parameter dependence in these problems
is entirely captured by just two well-understood stationary variables,
the average charge of the system and of the dual system.
Remarkably, it is the latter that often dictates the most striking features of the measurable quantities (e.g., positions of resonances),
underscoring the importance of the dual system for understanding the actual one.
\end{abstract}

%% file: paper_introduction.tex
\section{Introduction}
\label{sec_introduction}

Recent progress in few-electron devices aimed at applications such as energy-harvesting and single-electron sources has relied critically either on quantum coherent effects, or on an explicit, often specifically \emph{engineered} energy-dependence in the coupling of the small quantum system of interest and its environment. For the latter systems, the challenge of analytically understanding their complex behavior is not so much limited by strong coupling corrections, but instead by the energy-structure introduced by the tunnel barrier and the density of states in the electrodes. 
While these complications already arise for stationary phenomena~\cite{Sanchez2011Feb,Hartmann2015Apr,Thierschmann2015Oct,Roche2015Apr,Sanchez2017Nov,Golovach2011Feb,Rossello2017Jun}, they tend to proliferate when turning to more complicated situations involving the time-dependent response or coherent quantum effects or a combination thereof.

In order to simplify the description of such situations as much as possible, one typically exploits symmetries, conservation laws, fluctuation relations and dualities.
A well-known example of the latter is detailed balance, which for classical Markovian processes expresses the absence of any probability flow between the different states of the system. By relating transition rates to stationary-state occupations, it dictates both the absence of entropy production in the stationary limit
---distinguishing local equilibrium states, possibly at large bias, from 'true' non-equilibrium stationary states (NESS)~\cite{Seifert2012Nov}---
as well as purely exponential, non-oscillatory decay in the transient regime.
Dualities extending into the quantum and/or far non-equilibrium regime exist ---such as quantum detailed balance~\cite{Kossakowski1977Jun,Fagnola2015Sep,Alhambra2017Aug,Carlen2017Sep,Duvenhage2018Mar} and non-equilibrium (quantum) fluctuation relations~\cite{Campisi2011Jul,Seifert2012Nov,Jarzynski2017Jan}--- but are more complicated to apply and still under continued active research.\par

One particular useful duality relation which we have recently discovered~\cite{Schulenborg2016Feb} is restricted to open \emph{fermionic} systems, but otherwise extends straightforwardly into the quantum regime and also remains valid even if local detailed balance is broken. 
This \emph{fermionic duality}~\footnote{Previously referred to as ``fermion-parity duality'' by us~\cite{Schulenborg2016Feb,Vanherck2017Mar,Schulenborg2017Dec}.} has been shown to hold for discrete but otherwise arbitrarily complex fermionic systems; this includes both strong coupling to several reservoirs at low temperatures,
as well as strong local interactions, such as Coulomb repulsion between electrons in quantum dots.\par

This general applicability of the fermionic duality reflects that its derivation in \cite{Schulenborg2016Feb} relies mostly on two independent fundamental physical principles for fermionic systems:
Pauli's exclusion principle (antisymmetrization of states) and fermion-parity superselection (universal conservation of parity observable).
Despite this generality, its applications so far have been limited to rate equations for fermionic systems~\cite{Schulenborg2016Feb,Vanherck2017Mar,Schulenborg2017Dec}, excluding coherences.
Also, the coupling was assumed to be energy-independent, referred to as \emph{wideband limit} in the following.
The central aim of this paper is to overcome these two restrictions.

The first main result reported here is that for weak coupling, energy-dependence does not in any way inhibit the existence and practical usefulness of a fermionic duality. In fact, we show that the known wideband result can easily be modified to incorporate weak but otherwise arbitrarily \emph{energy-dependent} couplings.
One class of experimentally relevant setups for which this allows striking simplifications and insights is time-dependent decay after a parameter switch,
a basic building block for complex device operations that aim at controlled charge and energy exchange via individual electrons. Examples include coherent single- or few-electron sources~\cite{Feve2007May,Bocquillon2013Mar,Ubbelohde2014Dec,Bauerle2018Apr}, charge pumps~\cite{Blumenthal2007Apr,Giblin2012Jul,Pekola2013Oct,Kaestner2015Sep}, or heat engines~\cite{Esposito2010Feb,Esposito2010Apr,Lim2013Nov,Juergens2013Jun,Ludovico2016Jul,Bruch2016Mar,Dare2016Jan,Whitney2018May} and driven thermoelectrics~\cite{Crepieux2011Apr,Zhou2015Oct,Ludovico2016Feb}.

The second key result is that the above generalization for energy-dependent, weak coupling also extends to the broader class of systems that require a \emph{quantum master equation} description for the density matrix, in which the probabilities are coupled to quantum coherences.
Examples for applications in quantum transport where nontrivial coherences play a role~\cite{Konig2001Apr,Wunsch2005Nov,Hartle2014Dec,Wenderoth2016Sep}
include 
quantum-dot spin-valves~\cite{Hauptmann2008Mar,Gaass2011Oct}
for noncollinear spintronics~\cite{Konig2003Apr,Braun2004Nov,Braig2005May,Weymann2007Apr,Splettstoesser2008May,Sothmann2010Dec,Baumgartel2011Aug,Misiorny2013Oct,Sothmann2014Oct,Hell2015May}
multi-orbital~\cite{Begemann2008May,Darau2009Jun,Schultz2009Jul,Donarini2009Aug,Niklas2017Mar} and vibration-assisted~\cite{Schultz2010Oct} transport in molecular electronics,
but also quantum-dot sensing~\cite{Hell2014May,Hell2016Jan} of qubits.
The extended approach that we report underlines the usefulness of the fermionic duality for ongoing research on such few-electron systems.
Moreover, the presented derivation is simpler than the one used to address strong, yet energy-independent coupling~\cite{Schulenborg2016Feb} and provides a clearer picture.

The paper is organized as follows. Since the fermionic duality requires a non-standard way of solving quantum master equations, we first outline in \Sec{sec_idea} the main idea for the most simple example, and highlight several features that will be generalized in the remainder of the paper.
In \Sec{sec_model}, we formulate a very general model with energy-dependent bilinear coupling,
review the weak-coupling quantum master equation,
and derive a useful linear response formula that is valid beyond the wideband limit.
Then, in \Sec{sec_duality_qme}, we discuss how the fermionic duality for the weak-coupling quantum master equation for the density operator
can be extended beyond the wideband limit
and outline how it can be exploited.
This clarifies that fermionic duality is a novel concept \emph{not} equivalent to detailed balance, 
and also \emph{not} equivalent to other known symmetries under electron-hole transformations~\cite{Vanherck2017Mar}.
In \Sec{sec_duality_rate}, we highlight this by specializing to simpler rate equations (probabilities only)
and \emph{combining} the duality with the implications of classical detailed balance.
This leads to an explicit relation between dual stationary-state probability vectors
and nontrivial orthogonality relations to other vectors that simplify calculations beyond the wideband limit.
In \Sec{sec_quantum_dot} we illustrate some of the practical advantages and interesting insights that the fermionic duality offers which are otherwise impossible to infer.
For a strongly interacting quantum dot, we discuss the transient decay of time-dependent charge and heat currents after a quench, and, secondly, the stationary-state thermoelectric properties in the linear response to a voltage and thermal bias.
We show that the complex dependence of these effects on many parameters can be fully understood in a simple way by exploiting the extended fermionic duality.
We thereby demonstrate that the conclusions of \Refs{Schulenborg2016Feb,Vanherck2017Mar,Schulenborg2017Dec} extend to pronouncedly energy-dependent couplings,
and, in addition, identify several measurable effects specific to strongly energy-dependent couplings.

Throughout the paper we use units $|e|=\hbar=k_\text{B}=1$.

%% file: paper_single_mode.tex
\section{Single-mode fermionic duality}
\label{sec_idea}
To illustrate how the fermionic duality in principle works,
we consider the simplest possible, self-contained example:
a single fermion mode described by a rate equation.
Although one might think that the relations discussed in this section derive from the simplicity of this example, the remainder of the paper will demonstrate that they in fact hold for any number of fermion modes interacting arbitrarily in an open system weakly coupled to fermionic reservoirs.\par

%%%%%%%%%%%%%%%%%%%%%%%%%%%%%%%%%%%%%%%%%%%%%%%%%%%%%%%%%%%%%
\subsection{Rate equation and mode-amplitude decomposition}
%%%%%%%%%%%%%%%%%%%%%%%%%%%%%%%%%%%%%%%%%%%%%%%%%%%%%%%%%%%%%
Consider a single fermion mode at energy $\epsilon$ that is weakly coupled to a noninteracting fermion bath with chemical potential $\mu$ and temperature $T$. The probabilities $P_i(t)$ of finding the state $i=\text{u,o}$ of the fermion mode being unoccupied or occupied at time $t$ are governed by the rate equation
\begin{equation}
 \partial_t \begin{bmatrix}P_\text{u}(t) \\ P_\text{o}(t)\end{bmatrix} = \Gamma(\epsilon)
 \begin{bmatrix} -f^+(\frac{\epsilon - \mu}{T}) & f^-(\frac{\epsilon - \mu}{T}) \\ f^+(\frac{\epsilon - \mu}{T}) & -f^-(\frac{\epsilon - \mu}{T})\end{bmatrix}
 \begin{bmatrix}P_\text{u}(t) \\ P_\text{o}(t)\end{bmatrix}.
 \label{eq_master_equation_simple}
\end{equation}
Here, $f^\pm(x) = \sqbrack{\expfn{\pm x} + 1}^{-1}$ is the probability of finding a particle $(+)$ / hole $(-)$
at the corresponding energy
in the fermionic reservoir.
The coupling $\Gamma(E)$ is evaluated at the energy $E = \epsilon$ of the mode.

The time-evolution problem \eq{eq_master_equation_simple} has the form
$\partial_t  \Ket{\rho(t)}=W \Ket{\rho(t)}$, where $\Ket{\rho(t)}$ stands for the column vector of probabilities.
To compute the time evolution of $\Ket{\rho(t)}$, one diagonalizes the kernel $W=\sum_{k} \lambda_k \Ket{m_k} \Bra{a_k}$
to evaluate the formal exponential solution:
\begin{align}
	\Ket{\rho(t)} = e^{Wt} \Ket{\rho(0)}
	= \sum_k e^{\lambda_k t} \Ket{m_k} \Braket{a_k}{\rho(0)} 
	.
	\label{eq_rho_expansion}
\end{align}
Here, the right eigenvectors $\Ket{m_k}$ are the evolution \emph{modes},
i.e., components of $\Ket{\rho(t)}$ with a definite time-evolution factor $e^{\lambda_k t}$ with eigenvalue $\lambda_k \leq 0$ equal to minus the decay rate. The corresponding left eigenvectors $\Bra{a_k}$, written as row vectors in our current example, determine the amplitude $\Braket{a_k}{\rho(0)}$ of this mode for a given initial state $\Ket{\rho(0)}$.
The left and right eigenvectors need to be distinguished not just because of their different roles:
since $W$ is not a normal matrix, they
are not simply connected by taking the adjoint.
In fact, the relation between these vectors is a theme of the paper.\par

For the matrix $W$ in \Eq{eq_master_equation_simple}, we obtain the decomposition in Table \ref{tab}.
%%%%%%%%%%%%%%%%%%%%%%%%%%%%%%%%%%%%%%%%%%%%%%%%%%%%%%%%%%%%%
\begingroup \squeezetable
\begin{table}[t]
	\caption{\label{tab}
		Spectral decomposition of the rate matrix $W$.
	}
	\begin{ruledtabular}
		\begin{tabular}{l|l|l||l}
			$k$ & Decay rate & Amplitude & Mode
			\\
			    & $-\lambda_k$ & $\Bra{a_k}$ & $\Ket{m_k}$
			\\
			\hline
			$0$
			&
			$0$
			&
			$
			\begin{bmatrix}
				1 , 1
			\end{bmatrix}
			$
			&
			$
			\frac{1}{W_\text{uo} + W_\text{ou}}
			\begin{bmatrix}
				W_\text{uo} \\ W_\text{ou}
			\end{bmatrix}
			$
			\\
			$1$
			&
			$\Gamma(\epsilon)$
			&
			$\frac{1}{W_\text{uo} + W_\text{ou}}
			\begin{bmatrix}
			W_\text{ou} , -W_\text{uo}
			\end{bmatrix}
			$
			&
			$
			\begin{bmatrix}
			1 \\ -1
			\end{bmatrix}
			$
		\end{tabular}
	\end{ruledtabular}
\end{table}\endgroup
%%%%%%%%%%%%%%%%%%%%%%%%%%%%%%%%%%%%%%%%%%%%%%%%%%%%%%%%%%%%%
Inserting these into the mode expansion \eq{eq_rho_expansion}, one obtains the solution
$P_i(t) = P_i + [P_i(0)-P_i] e^{-\Gamma(\epsilon)t}$
where $P_i$ without time-argument denotes the stationary value $\lim_{t\to \infty} P_i(t)=P_i$ as given by the components of $\Ket{m_0}$.

%%%%%%%%%%%%%%%%%%%%%%%%%%%%%%%%%%%%%%%%%%%%%%%%%%%%%%%%%%%%%
\subsection{Mode-amplitude cross relations}
%%%%%%%%%%%%%%%%%%%%%%%%%%%%%%%%%%%%%%%%%%%%%%%%%%%%%%%%%%%%%
Of course, the above solution could have been derived and understood more easily by directly decoupling the equations \Eq{eq_master_equation_simple} using $P_\text{u}(t)+P_\text{o}(t)=1$, thereby avoiding diagonalization of $W$.
However, this strategy does not significantly simplify more general fermionic problems.
Here, instead we wish to highlight a general structure of the kernel $W$ which \emph{does} simplify more general problems:
although the matrix $W$ is neither symmetric nor antisymmetric,
Table \ref{tab} indicates that its eigenvectors show striking \emph{cross} relations.

Consider first the \emph{smaller} of the two decay rate, $-\lambda_0=0$, and its left eigenvector $\Bra{a_0}$.
Their expressions are well-known to be dictated by probability normalization, and are therefore independent of the rates $W_\text{uo}$ and $W_\text{ou}$.
As the table shows, the right eigenvector $\Ket{m_1}$ for the \emph{larger} of the two decay rates, $-\lambda_1=\Gamma(\epsilon)$, is also independent of the rates and differs only by a relative sign between the components.
The remaining left eigenvector $\Bra{a_1}$ and right eigenvector $\Ket{m_0}$ display a similar cross relation.
They are already similar in that they both depend explicitly on the rates $W_\text{uo}$ and $W_\text{ou}$. Their components are also related by introducing a relative sign and, additionally, exchanging the initial and final states of the rates, $W_\text{uo} \leftrightarrow W_\text{ou}$.

We thus find that the left eigenvectors \emph{cross}-determine the right ones in a way that is much simpler than the general relation between left and right eigenvectors.\footnote
	{The left and right eigenvectors of a matrix $M$ which is diagonalizable by similarity $S$ to a diagonal matrix $\Lambda$ are always related in general:
	if $M=S\Lambda S^{-1}$ then $l_k= S e_k$ and $r_k=S^{-1} e_k$. However, this general relation involves a complicated \emph{algebraic} inversion of the similarity matrix $S$. Instead, the fermionic duality, which is responsible for the cross relations between eigenvectors presented here, yields a relation between  \emph{matrix functions} $S$ and $S^{-1}$ basically consisting in simple replacements of function arguments  by parameters of a dual model.}
The general upshot for practical analytical calculations is that one can avoid half of the work.
Moreover, as we clarify later on in this paper, one can understand the often quite unexpected parameter dependence of some of the eigenvectors for more general systems~\cite{Schulenborg2016Feb}.

We next observe that in the above cross relation, the fermionic nature of the system and the reservoirs enters in two ways.
First, due to Pauli's exclusion principle, each reservoir mode can be filled with only one particle with probability $f^+$, or one hole with probability $f^-$. This restricts the larger of the two decay rates, which is found to be the sum of rates for adding and removing one particle, to
\begin{align}
W_\text{ou}+W_\text{uo} = \Gamma(\epsilon)(f^+ + f^-) = \Gamma(\epsilon).
\label{eq_exclusion}
\end{align}
This decay rate is independent of the reservoir's energies ($\mu$, $T$) since the Fermi-Dirac environment statistics sum to $1$.
Second, the special symmetry $f^\pm(-x) = f^\mp(x)$ of the Fermi-Dirac statistics enables to swap initial and final states of the rates $W_{ij}$, as required in the mode-amplitude cross relation, by inverting all energies on the system ($\epsilon \to -\epsilon$) and on the reservoir ($\mu \to -\mu$),
\begin{subequations}
	\begin{align}
	W_\text{ou}  = \Gamma(\epsilon) f( (\epsilon-\mu)/T )
	\quad
	\to
	\\
	\dual{W}_\text{ou}  := W_\text{uo} = \Gamma(\epsilon) f(-(\epsilon-\mu)/T),
	\end{align}\label{eq_inversion}\end{subequations}
given that we \emph{maintain the original energy dependence of the coupling} $\Gamma(\epsilon)$. We indicate this parameter transformation by an overbar.\footnote{
In previous formulations of fermionic duality, the different treatment of the couplings compared to the statistical factors did not play a role since the wideband limit $\Gamma(\epsilon)=\Gamma$ was considered.}

We summarize all above relations with the help of a sign matrix
that keeps track of the \emph{fermion-parity} $(-1)^N$ of the fermion number $N=0,1$ corresponding to the states $i=\text{u,o}$:
\begin{equation}
 W + {\scriptstyle\begin{bmatrix}1 & 0 \\ 0 & -1\end{bmatrix}}\dual{W}^\sudag{\scriptstyle\begin{bmatrix}1 & 0 \\ 0 & -1\end{bmatrix}} = -\Gamma(\epsilon) \,  {\scriptstyle\begin{bmatrix}1 & 0 \\ 0 & 1\end{bmatrix}}.
 \label{eq_matrix_duality}
\end{equation}
This equation in its generalized form is the fermionic duality, Eq.~\eqref{eq_duality_generators}, which is going to be derived and exploited in the remainder of this paper. The right hand side comes from the exclusion principle, also resulting in \Eq{eq_exclusion}.
The fermion-parity signs on the left hand side are required since ---in contrast to $W^\sudag$--- the matrix $\dual{W}$ is a physical rate matrix.

Indeed, the rate matrix $\dual{W}$ describes a \emph{dual model} of a fermion at energy $-\epsilon$ and a reservoir at chemical potential $-\mu$, with coupling value $\Gamma(\epsilon)$ taken at the energy $\epsilon$.
In this simple example, the dual system is hence qualitatively the same as the original system,
explaining why the calculations are so simple.
However, in interacting multi-mode systems the dual system shows nontrivial, \emph{qualitatively different physics}, yielding unexpected insights into the physics of the original model.

Relation \eq{eq_matrix_duality} ties the kernel $W$ to its adjoint $W^\sudag$ modulo signs and energy inversion, and is therefore~\cite{Schulenborg2016Feb} responsible for the above observed cross relation between the left and right eigenvectors of $W$.
Thereby, it also links the decay rate $\Gamma(\epsilon)$ in the transient
response to the zero eigenvalue belonging to the stationary state.
Interestingly, we later generalize this to a link
between the smallest and largest decay rate for complex fermionic
models. The right hand side of \Eq{eq_matrix_duality} in fact
imposes ---in its more general form Eq.~\eqref{eq_duality_generators}--- a
fundamental restriction on the largest decay rate of $W$, providing an
upper bound that is independent of the reservoirs ($\mu, T$) and, in the
wideband limit $\Gamma(\epsilon) \rightarrow \Gamma$, even independent of
the open-system parameters.

Finally, we combine the above with a generally independent observation, namely,
that for our simple master equation, the stationary probabilities $P_i:=\lim_{t\to \infty} P_i(t)$,
given by the components of $\Ket{m_0}$ in Table 
\ref{tab}, obey detailed balance:
${P_\text{o}}/{P_\text{u}} = {W_\text{ou}}/{W_\text{uo}}$.
The stationary occupations $\dual{P}_i$ for $\dual{W}$, describing the stationary \emph{dual system}, then also satisfy detailed balance but with the inverse ratio:
\begin{equation}
\frac{\dual{P}_\text{o}}{\dual{P}_\text{u}}
=
\frac{\dual{W}_\text{ou}}{\dual{W}_\text{uo}}
=
\frac{W_\text{uo}}{W_\text{ou}}
=
\frac{P_\text{u}}{P_\text{o}}
\label{eq_detailed_balance_intro}.
\end{equation}
We see that the product of stationary occupation probability for the original system and for its dual,
$P_i \dual{P}_i=C$, is equal to some function $C$ of the rate matrix elements but  independent of the state $i$.
Probability normalization $\sum_i P_i =1=\sum_i \dual{P}_i$ implies a simple relation for $C$, $1=\sum_i C/P_i$. This allows to directly compute the stationary state of a system with inverted energies from the original system via
\begin{equation}
 \dual{P}_i = \frac{1}{P_i}\Big( \sum_{j}\frac{1}{P_{j}} \Big)^{-1}
 \label{eq_pbar}
.
\end{equation}
In the present simple example, this expresses that starting from 
$P_\text{o}=f((\epsilon-\mu)/T)$, one finds as expected
$\dual{P}_\text{o}=f(-(\epsilon-\mu)/T)$, which is equal to $P_\text{u}$  and related to $P_\text{o}$ by replacing parameters by the ones of the dual model.
However, it will turn out that formula \eq{eq_pbar} also applies to stationary states of arbitrary interacting many-fermion systems, providing nontrivial insights [\Eq{eq_orthogonalities_detailed_balance}].
This is also a canonical example of how a complicated combination of probabilities
[right hand side of \Eq{eq_pbar}] for one problem
may be expressed in a single expression ($\dual{P}_i$) whose parameter dependence is easily understood by considering another physical problem, that of the energy-inverted, dual model.

%%%%%%%%%%%%%%%%%%%%%%%%%%%%%%%%%%%%%%%%%%%%%%%%%%%%%%%%%%%%%
\subsection{Outlook on general fermionic duality}
%%%%%%%%%%%%%%%%%%%%%%%%%%%%%%%%%%%%%%%%%%%%%%%%%%%%%%%%%%%%%
The duality \eq{eq_matrix_duality}
is responsible for the mode-amplitude cross relations in Table \ref{tab},
the bound $\Gamma(\epsilon)$ on decay eigenvalues,
and the stationary-state duality \Eq{eq_pbar}.
It is surprising that these are not mere artifacts of the simplicity of the example.
At the origin of this are two independent postulates of many-body quantum physics that we noted earlier:
the antisymmetrization of fermionic states (exclusion principle limiting $N=0,1$ in our example) and the fermion-parity superselection (related to the signs $(-1)^N$).

As a result, the fermionic duality also extends to the quantum master equation for the density matrix,
including off-diagonal elements associated with coherent superpositions,
and not just to rate equations for the diagonal elements, the above considered probabilities.
Although this was in principle already clear in \Ref{Schulenborg2016Feb}
for bilinear \emph{energy-independent} coupling,
the procedure for exploiting the duality for quantum master equations was not yet worked out.
This introduces new aspects ---even in the wideband limit--- which will be addressed here.

For the same reason, the duality even extends to open fermionic systems with energy-independent, yet arbitrarily strong coupling to noninteracting reservoirs at arbitrarily low temperature. Notably, such open systems are no longer described by the rate equation, as in the our example, or even by a quantum master equation as discussed later on,
but require a time-nonlocal (Nakajima-Zwanzig) approach.
The crucial advance made in the present paper is that we lift the limiting assumption of energy-independent coupling made in \Ref{Schulenborg2016Feb} while keeping the coherences in the description;
the only restriction is the assumption of weak coupling.
Indeed, in the above example, we have already observed that the $\epsilon$-dependence of $\Gamma$ does not really obstruct any of the considerations. This holds more generally, and all its consequences are worked out in the following.

%% file: paper_model.tex
\section{Model and method}
\label{sec_model}

\begin{figure}
	\includegraphics[width=0.9\linewidth]{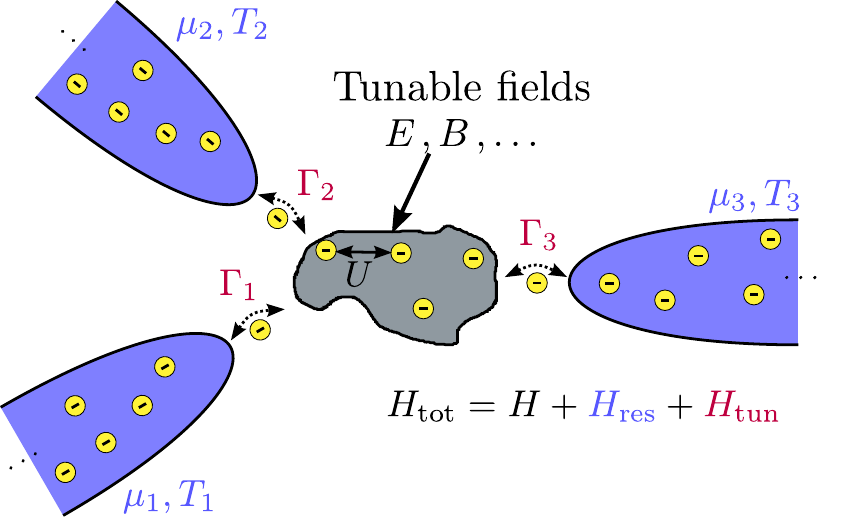}
	\caption{
	Example of the type of open systems to which the fermionic duality applies:
	a central device hosting a number of electrons tunnel-coupled with rates $\Gamma_\alpha$ \BrackEq{eq_coupling_rates} several non-interacting electronic reservoirs $\alpha = 1,2,3$ at different electrochemical potentials $\mua$ and temperatures $\Ta$. Arbitrary many-body interactions such as Coulomb repulsion  are allowed, as are externally tunable electric $(E)$ and magnetic fields $(B)$, etc. .}
	\label{fig_open_system}
\end{figure}

%%%%%%%%%%%%%%%%%%%%%%%%%%%%%%%%%%%%%%%%%%%%%%%%%%%%%%%%
\subsection{Generic open fermionic system}
%%%%%%%%%%%%%%%%%%%%%%%%%%%%%%%%%%%%%%%%%%%%%%%%%%%%%%%%

We are interested in \emph{fermionic} open systems, consisting of a discrete set of fermion modes labeled $\ell$
coupled to a continuum of fermionic modes in reservoirs labeled $\alpha$, as sketched in \Fig{fig_open_system}.
The system is not restricted in any way
except that it is fermionic:
the Hamiltonian $H$ and any observable of the system
act on antisymmetric states with well-defined fermion-parity
and must\footnote{
 	As a brief reminder, the parity can be expressed as $(-\one)^N=e^{-i 2\pi S_z}$
 	with the total spin of the $N$ fermions $S_z$, corresponding to the rotation by $2\pi$.
 	Without imposing superselection, superpositions of even (integer spin) and odd parity (half-integer spin) will therefore develop a \emph{relative} phase factor when rotating the system by a full angle $2\pi$. By standard rules of quantum mechanics this would be observable in interference, in contradiction with the requirement that a full rotation should not modify an observation.
 	To exclude this, superselection must be imposed \emph{together} with antisymmetrization:
 	physical state vectors have definite parity,
 	$(-\one)^N\ket{\psi}=\pm \ket{\psi}$, or, equivalently, state operators commute with parity,
 	$[\rho,(-\one)^N]=0$.
 	The same must then hold for all observables, including the Hamiltonian.
 	}
conserve this parity: $[H,\fpOp] = 0$
(superselection principle~\cite{Wick1952Oct,Aharonov1967Mar,Streater2000Dec}).
Here, the fermion-parity operator and the particle number operator of the open system,
\begin{equation}
\fpOp := \expfn{i\pi N} \lrsepa N := \sum_{\ell}d^\dagger_\ell d_\ell
,
\end{equation}
are defined in terms of creation $(d^\dagger_\ell)$ and annihilation $(d_\ell)$ fields of fermions.
The multi-index $\ell = \sigma, \ldots $
labels the fermion modes by
a spin projection $\sigma\in\{\uparrow,\downarrow\}$
and further possible degrees of freedom (e.g. orbital) which are not indicated.
We allow for \emph{arbitrary} types of parity-conserving interactions
of \emph{any} strength between fermions in the open system.
Furthermore, we do not require that the fermion number is separately conserved on the system, $[H,N]=0$,
which would be a stronger condition than parity conservation. Hence, 
$H$ is allowed to mix charge states differing by an \emph{even} number of fermions, as for example due to the superconducting proximity effect.
We mention this here to stress the general nature of the duality; for the applications presented below in \Sec{sec_quantum_dot}, $[H,N]=0$ does however hold.

%%%%%%%%%%%%%%%%%%%%%%%%%%%%%%%%%%%%%%%%%%%%%%%%%%%%%%%%
\subsection{Reservoirs and energy-dependent tunneling}
%%%%%%%%%%%%%%%%%%%%%%%%%%%%%%%%%%%%%%%%%%%%%%%%%%%%%%%%

The reservoirs consist of noninteracting fermion modes with energies $\epsilon_{\kappa\alpha\nu}$, described by
\begin{gather}
 \Hlead := \sum_\alpha H_\alpha \lrsepa H_\alpha := \sum_{\kappa\nu}\epsilon_{\kappa\alpha\nu} c^\dagger_{\kappa\alpha\nu}c_{\kappa\alpha\nu}\label{eq_hamiltonian_lead},
\end{gather}
with $c^\dagger_{\kappa\alpha\nu} (c_{\kappa\alpha\nu})$ creating (annihilating) a fermion in reservoir $\alpha$ with
orbital $\kappa$ and discrete degrees of freedom $\nu = \tau,\dotsc$ such as the spin projection $\tau$, and possible further ones.

The reservoir orbitals $\kappa$ form a continuum and the bilinear coupling
\begin{equation}
 \Htun := \sum_{\kappa\alpha\nu}\sqbrack{\tau_{\kappa\alpha\nu;\ell}
 	\,  d^\dagger_{\ell}c_{\kappa\alpha\nu} + \text{H.c.}}\label{eq_hamiltonian_tun},
\end{equation}
describes tunneling to / from the open system.
The tunnel junctions are characterized by the Hermitian \emph{coupling matrix} $(\hbar = 1)$
\begin{subequations}
\begin{align}
\Gamma_{\alpha\nu;\ell\ell'}(E)
&:= 2\pi\sum_\kappa \delta\nbrack{E - \epsilon_{\kappa\alpha\nu}}
\tau_{\kappa\alpha\nu;\ell}^{*} \, \tau_{\kappa\alpha\nu;\ell'}
\\
&=
\Gamma_{\alpha\nu;\ell'\ell}(E)^{*}
,
\end{align}\label{eq_coupling_rates}\end{subequations}
which accounts for the joint effect of the tunneling amplitudes $\tau_{\kappa\alpha\nu;\ell}$ and  the density of states in reservoir $\alpha$.
The diagonal couplings ($\ell = \ell'$) are real, positive semidefinite,
and enter into the decay of occupation probabilities
as the typical inverse tunneling times. We do not assume that spin ($\tau \neq \sigma$) or other quantum numbers are conserved by the tunneling, allowing for applications where nontrivial coherences play a role. The
off-diagonal couplings $(\ell \neq \ell')$ are therefore in general complex valued
and enter into the evolution of quantum superpositions on the system induced by coupling to the environment.

In \Eq{eq_coupling_rates}, $E$ is the energy of the \emph{reservoir} electron involved in the tunneling process.
The crucial step made in this work is that we allow for arbitrary energy-dependence of the coupling matrices $\Gamma_{\alpha \nu}(E)$ characterizing the tunnel junctions.
The only simplifying assumption that we make is that this $E$-dependence is analytic.
We stress that it is a quite common assumption to neglect the energy structure of the tunnel barriers (wideband limit), also in prior work on fermion-parity duality~\cite{Schulenborg2016Feb,Vanherck2017Mar,Schulenborg2017Dec}.
However, especially for applications to energy-harvesting~\cite{Sanchez2011Feb,Hartmann2015Apr,Thierschmann2015Oct,Sanchez2017Nov} and charge pumps~\cite{Giblin2012Jul,Fletcher2013Nov,Kataoka2017Mar}, it is crucial to account for this energy dependence.\footnote{Open-system single-particle states $\ket{\ell}$ entering the couplings $\Gamma_{\alpha\nu;\ell\ell'}(E)$ \BrackEq{eq_coupling_rates} can be states obtained by hybridizing localized states, for example, on dot 1 and 2 of a double quantum dot system, $c_1 \ket{1} + c_2 \ket{2}$.
The coefficients $c_1$, $c_2$ may depend on \emph{differences} between the \emph{open-system} energies $\epsilon_1$, $\epsilon_2$ of these localized states in the \emph{absence of hybridization}, and enter $\Gamma_{\alpha\nu;\ell\ell'}(E)$ via the tunneling amplitudes $\tau_{\alpha\kappa,\ell}$ to the reservoirs.
This type of 'energy dependence' is included in the wideband limit, and must be clearly distinguished from the \emph{reservoir} energy dependence ($E$) that we  and, e.g., \Refs{Sanchez2011Feb,Thierschmann2015Oct,Sanchez2017Nov} are concerned about.
}

%%%%%%%%%%%%%%%%%%%%%%%%%%%%%%%%%%%%%%%%%%%%%%%%%%%%%%%%
\subsection{Non-equilibrium Born-Markov dynamics}
\label{sec_born_markov}
%%%%%%%%%%%%%%%%%%%%%%%%%%%%%%%%%%%%%%%%%%%%%%%%%%%%%%%%
The fermionic duality arises naturally within the reduced density-operator formulation of open-system dynamics
in which we trace out the reservoirs in the density operator, 
$\rho(t) := \text{Tr}_{\lead} \rho^\tot(t)$. 
To obtain $\rho^\tot(t)$, we evolve  an initially factorized state $\rho^\tot_0 = \rho_0\cdot\rho^\lead_0$ with $\Htot$,
where each reservoir $\alpha$ is initially a grand-canonical equilibrium with
electrochemical potential $\mu_\alpha$ and temperature $T_\alpha$
\begin{equation}
 \rho^\lead_0 := \prod_\alpha
 \frac{
 	e^{-\nbrack{H_\alpha - \mu_\alpha N_\alpha}/T_\alpha}
 	}{\Trbet\sqbrack{
 	e^{-\nbrack{H_\beta - \mu_\beta N_\beta}/T_\beta}
 	}
 	}
 	.
 \label{eq_lead_state}
\end{equation}
Here, $N_\alpha = \sum_{\kappa\nu}c^\dagger_{\kappa\alpha\nu}c_{\kappa\alpha\nu}$ is the fermion number operator for reservoir $\alpha$.
With the above assumptions,
tracing out the reservoirs~\cite{Schoeller1994Dec,Konig1996Dec,Timm2008May} gives a time-nonlocal quantum master equation for $\rho(t)$:
\begin{equation}
 \partial_t\rho(t) = -iL\rho(t) + \int_0^t d t'\mathcal{W}(t-t')\rho(t').\label{eq_generalized_master_equation}
\end{equation}
Whereas the Liouvillian $L\bullet = [H,\bullet]$ is the system part, the non-Hermitian time-nonlocal kernel $\W$ is due to the particle exchange with the reservoirs.
The fermionic duality already applies to this quite general situation, possibly describing a strongly coupled system,
if one only makes the wideband approximation~\cite{Schulenborg2016Feb}.
However, in order to go beyond the limitation of this wideband approximation,
we focus on the weak-coupling, high-temperature limit~\footnote{$\Gamma$ denotes the typical scale of the diagonal couplings $\Gamma_{\alpha\nu;\ell\ell}$.}
$\Gamma \ll T_\alpha$.
Then, for times $t \gtrsim 1/\Gamma$, the state $\rho(t)$ is governed
by the simpler, time-local Born-Markov master equation
\begin{equation}
 \partial_t\rho(t) = (-i L + W ) \rho(t)
 \label{eq_master_equation},
\end{equation}
with the kernel $W$ given by the zero-frequency limit of the Laplace transformation of $\mathcal{W}$ evaluated to leading order in the system-reservoir couplings,
\begin{equation}
W := \lim_{\eta\downarrow 0}\int_0^\infty d t'\W_1 (t')e^{-\eta t'}.
\label{eq_instantaneous_kernel}
\end{equation}
The special structure of \emph{fermionic} kernels $W$ will be studied in \Sec{sec_duality_qme},
affording general insights into the transient evolution and stationary limit of $\rho(t)$.

To distinguish superoperators (such as $L$ and $W$) from the ordinary operators on which they act ($\rho$),
we adopt the Liouville-space notation of \Refs{Breuer2002,Saptsov2012Dec}.
An ordinary operator $y$ is denoted by a rounded ket $\Ket{y}$
to indicate that it is considered as a vector in Liouville space, called supervector. This is consistent with \Sec{sec_idea}, in which rounded kets are represented by column vectors containing only the diagonal elements of the operators $y$ in the many-body energy eigenbasis of the open system.
The covectors which correspond to these supervectors with respect to the Hilbert-Schmidt scalar product
are linear functions $\Bra{x}\bullet = \Tr\sqbrack{x^\dagger \bullet}$ of an operator argument $\bullet$, each function being uniquely parametrized by an operator $x$.
The kernel $W$ of interest is a linear superoperator that acts on the Liouville-space of supervectors.
Its superhermitian adjoint -- of central importance here --
is then defined relative to the Hilbert-Schmidt scalar product, 
$\Bra{x}W \Ket{y}
:= \Braket{ W^\sudag x}{y}
= \sqbrack{ \Bra{y} W^\sudag \Ket{x} }^{*}$.

%%%%%%%%%%%%%%%%%%%%%%%%%%%%%%%%%%%%%%%%%%%%%%%%%%%%%%%%
\subsection{Linear response beyond the wideband limit}
\label{sec_linear_response}
%%%%%%%%%%%%%%%%%%%%%%%%%%%%%%%%%%%%%%%%%%%%%%%%%%%%%%%%

One of the applications of the fermionic duality
is the simplification of linear-response calculations based on the stationary state, yielding the long-time limit solution to the
quantum master equation \eq{eq_master_equation}.
For this, one needs to linearize this stationary state in $\mu_\alpha$ and $T_\alpha$.
We here focus on the rate-equation limit, which we consider also in our later illustration in \Sec{sec_stationary}.

The main problem is that by naively computing $\mu_\alpha$- or $T_\alpha$-derivatives, the expressions
quickly become analytically unwieldy and uninformative,
even for small systems and in particular for energy-dependent couplings $\Gamma(E)$.
To not spoil the advantages offered by the fermionic duality, this linearization should be
compatible with the mode-amplitude expansion of the kernel [\Eq{eq_rho_expansion}]
and thus formulated in terms of eigen supervectors of the superoperator $W$.
Also, it should be applicable for energy-dependent couplings $\Gamma(E)$, which the well-known linearization~\cite{Beenakker1991Jul,Beenakker1992Oct} and its recent extension~\cite{Erdman2017Jun} do not account for.

A more general approach starts by noting that in the weak-coupling limit we can always\footnote
	{This is possible because coherent processes simultaneously involving more than one reservoir are neglected in this approximation.}
decompose the full kernel $W$ into a sum of partial kernels $W_\alpha$ describing the effect of each reservoir $\alpha$ separately:
\begin{equation}
W = \sum_\alpha W_\alpha.\label{eq_kernel_lead_sum}
\end{equation}
We will assume throughout that the sum $W$ has a non-degenerate zero eigenvalue and a corresponding, unique stationary state $\Ket{m_0}$.
If the joint environment is in equilibrium,
this stationary state should become the grand-canonical ensemble
\begin{equation}
	\evalAtEqui{\Ket{m_0}} =
	\frac{e^{-(H - \mu N)/T}}{\Tr\sqbrack{e^{-(H - \mu N)/T}}}
	\label{eq_boltzmann_factor},
\end{equation}
where $\cdot |_\text{eq}$ denotes evaluation at $T_\alpha=T$ and $\mu_\alpha=\mu$ for all reservoirs $\alpha$.
Since each $W_\alpha$ describes the coupling to reservoir $\alpha$ only, one possible\footnote{
	The stationary occupation of those orbitals which are completely disconnected from the specific reservoir $\alpha$ is not \emph{uniquely} determined by the stationary-state equation $W_\alpha\Ket{z_\alpha} = 0$. This is a physically natural consequence of the partial kernel decomposition and not an artifact of the model.}
stationary state of $W_\alpha$ should always be 
the state of \emph{equilibrium} with the single reservoir $\alpha$
\begin{equation}
	\Ket{m_{0,\alpha}} = \frac{e^{-(H - \mu_\alpha N)/T_\alpha}}{\Tr\sqbrack{e^{-(H - \mu_\alpha N)/T_\alpha}}}
.
\label{eq_boltzmann_factor_lead}
\end{equation}
The linearization of the non-equilibrium stationary state $\Ket{m_0}$ with respect to any variable $x$ can then be expressed as a weighted sum of \emph{equilibrium} quantities
evaluated at $\mu_\alpha=\mu$, $T_\alpha=T$ (see Appendix~\ref{sec_app_state_linearization}):
\begin{equation}
\evalAtEqui{\partial_x\Ket{m_0}} =
\sum_\alpha
\frac{1}{W|_{\text{eq}}}
\evalAtEqui{W_\alpha}
\evalAtEqui{\partial_x \Ket{m_{0,\alpha}}}
\label{eq_state_linearization}.
\end{equation}
Here, the reflexive generalized inverse $W^{-1}$ is defined naturally on the support of $W$ using the spectral decomposition~\cite{Ben-Israel2003,Hunter2014Apr}: due to the single, non-degenerate zero eigenvalue we have
\begin{equation}
W\frac{1}{W} = \frac{1}{W}W = \mathcal{I} - \Ket{m_0}\Bra{\one}\label{eq_pseudo_inverse},
\end{equation}
where $\mathcal{I}$ is the unit superoperator.
For the quantum dot studied in \Sec{sec_quantum_dot}, it is straightforward to compute $W^{-1}$ using the fermionic duality.

The linearization \eq{eq_state_linearization} does not rely on the wideband limit,
nor does it require local detailed balance [\Eq{eq_detailed_balance}] to hold for the open subsystem when system plus reservoirs deviate from equilibrium.
Its practical advantage is that it prevents the unnecessary proliferation of derivatives of rate expressions as they occur in a brute-force calculation.
Using \Eq{eq_state_linearization}, we can instead exploit the special structure of the explicitly appearing fermionic kernels and thereby avoid the calculation of the stationary nonequilibrium state as we will illustrate in \Sec{sec_quantum_dot}.

%% file: paper_duality_qme.tex
\section{Fermionic duality for quantum master equations}
\label{sec_duality_qme}

The fermionic duality in the wideband limit $\Gamma_{\alpha\nu;\ell\ell'}(E) = \Gamma_{\alpha\nu;\ell\ell'}$ as derived in \Ref{Schulenborg2016Feb}
is fully nonperturbative in the coupling strength at arbitrary temperature.
The limit $\Gamma \ll T_\alpha$ of this result was shown to lead to a weak-coupling, wideband fermionic duality
\begin{equation}
W +
\cP \dual{W}^\sudag \cP
= -\Gamma \, \mathcal{I}
,
\label{eq_duality_original}
\end{equation}
where $\mathcal{I}$ denotes the Liouville-space identity. The superoperator
\begin{align}
\cP\bullet := \fpOp\bullet 
= \cP^{-1} = \cP^\sudag
\label{eq_parity}
\end{align}
is associated with the \emph{left} multiplication by the fermion parity operator, $\Gamma$ denotes the \emph{lump sum} of constant, non-negative diagonal couplings:
\begin{equation}
\Gamma := \sum_{\alpha\nu \ell}\Gamma_{\alpha\nu;\ell\ell} \geq 0
\label{eq_lump_sum}
.
\end{equation}
The duality \eq{eq_duality_original} relates the superadjoint of the kernel $W^\sudag$ not to itself, but to a \emph{dual} kernel in which the sign of all local energies $L=[H,\bullet]$ and the electrochemical potentials $\mu_\alpha$ of all reservoirs have been inverted. Since the kernel is a function $F$ of the local Liouvillian and all reservoir electrochemical potentials, $W=F(L, \{\mu_\alpha \})$, this means
\begin{align}
\dual{W} := F(-L, \{-\mu_\alpha \})
\label{eq_Wbar_def}.
\end{align}
Focusing on a nonperturbative formulation in the tunnel coupling,
the original derivation of the fermionic duality \eq{eq_duality_original} in \Ref{Schulenborg2016Feb} relies on a renormalized perturbation theory~\cite{Schoeller2009Feb,Saptsov2012Dec} in which an initial resummation leads to a concise self-consistent treatment of dissipative effects. However, this explicitly requires the wideband limit from the beginning, 
thereby prohibiting the analysis of energy-dependent couplings of interest here.
In the weak-coupling limit, a related but more direct approach is possible, which altogether avoids the wideband limit and leads to a more general form of the weak-coupling fermionic duality than the one given in Eq.~\eqref{eq_duality_original}.

%%%%%%%%%%%%%%%%%%%%%%%%%%%%%%%%%%%%%%%%%%%%%%%%%%%%%%%%%%%%%%%
\subsection{Superfermions}
%%%%%%%%%%%%%%%%%%%%%%%%%%%%%%%%%%%%%%%%%%%%%%%%%%%%%%%%%%%%%%%
This extended, non-wideband fermionic duality emerges most naturally when expressing the time-evolution kernels in terms of
fermionic superoperators~\cite{Saptsov2012Dec,Saptsov2014Jul} first introduced in real-time renormalization group studies~\cite{Schoeller2009Feb}.
More precisely, one introduces fermionic field \emph{super}operators acting on the system
\begin{equation}
G^q_{\eta \ell}\bullet := \frac{1}{\sqrt{2}}\sqbrack{d_{\eta \ell}\bullet +q\fpOp\bullet\fpOp d_{\eta \ell}},\label{eq_field_superoperator} 
\end{equation}
where $q=\pm$ labels superadjoint expressions [see \Eq{eq_field_superoperator_properties} below].
The further index $\eta=\pm$ distinguishes
ordinary field operators for creating a particle,
$d_{+\ell}:=d^\dagger_\ell$ or a hole $d_{-\ell}:= d_\ell$ on the system.
The field superoperators \eq{eq_field_superoperator} act on the system Liouville-Fock space containing all many-body density operators. As such, they can be used to set up a systematic second quantization formalism for \emph{mixed} states \cite{Saptsov2012Dec,Saptsov2014Jul}.
Most importantly, they fulfill fermionic anti-commutation relations
\begin{align}
\{G^q_{\eta \ell},G^{q'}_{\eta' \ell'}\} = \delta_{\eta,{-\eta'}} \, \delta_{\ell,\ell'} \, \delta_{q,{-q'}}
\label{eq_anticomm}
\end{align}
by virtue of the explicit inclusion of the fermion-parity in their definition~\cite{Saptsov2012Dec}. Furthermore, using the definition given in Eq.~\eqref{eq_field_superoperator} they can be shown to obey
\begin{equation}
\nbrack{G^q_{\eta \ell}}^\sudag\bullet = G^{{-q}}_{{-\eta} \ell}\bullet \lrsepa \cP G^q_{\eta \ell}\cP\bullet = -G^{{-q}}_{\eta \ell}\bullet.
\label{eq_field_superoperator_properties}
\end{equation}
The properties \eq{eq_anticomm} and \eq{eq_field_superoperator_properties} are the main reason why the superoperators $G^q_{\eta \ell}$ are particularly well-adapted to analyze the fermionic duality, as we shall see now.

%%%%%%%%%%%%%%%%%%%%%%%%%%%%%%%%%%%%%%%%%%%%%%%%%%%%%%%%%%%%%%%
\subsection{Fermionic duality}\label{sec_duality_extended}
%%%%%%%%%%%%%%%%%%%%%%%%%%%%%%%%%%%%%%%%%%%%%%%%%%%%%%%%%%%%%%%

We can now give the explicit functional form of the kernel $W$ and its superadjoint $W^\sudag$, as required in Eq.~\eqref{eq_Wbar_def}.
Using the above notation, the leading order in the tunnel coupling can be expressed concisely as~\cite{Leijnse2008Dec,Schoeller2009Feb,Saptsov2012Dec,Saptsov2014Jul}
\begin{align}
&W=F(L,\{\mu_\alpha\}) =
\frac{1}{2\pi i}\int_{-\infty}^\infty d E
\sumsub{\ell\ell'}{\eta q}
\sum_{\alpha\nu}
\Gamma^{\eta}_{\alpha\nu;\ell\ell'}(E)\notag\\
&\phantom{=}\times \sqbrack{f^{\eta}_\alpha(E) - qf^{{-\eta}}_\alpha(E)}G^+_{{-\eta}\ell}\frac{1}{\eta E - L + i0_+}G^q_{\eta \ell'},
\label{eq_born_markov_kernel}
\end{align}
where we explicitly indicate the dependence on the system Liouvillian $L=[H,\bullet]$
and the chemical potentials $\mu_\alpha$, with the latter entering via the Fermi functions $f^\eta_\alpha(x) = \sqbrack{\exp\nbrack{\eta\frac{x - \mu_\alpha}{T_\alpha}} + 1}^{-1}$ containing the reservoir temperatures $T_\alpha$.
We denote complex-conjugates of the coupling matrix by $\Gamma^{\eta}_{\alpha\nu}$ with
\begin{align}
 \Gamma^{+}_{\alpha\nu}= \Gamma_{\alpha\nu} \text{ and } \Gamma^{-}_{\alpha\nu}=\Gamma_{\alpha\nu}^{*}
\quad
\text{for } \eta=\pm.\label{eq_coupling_conjugate}
\end{align}
We only require $H=H^\dagger$ and parity superselection $[\fpOp,H]=0$, or equivalently~\footnote
	{$L^\sudag = L$ means $\Bra{A}L\Ket{B} = \text{tr} A^\dag [H,B] =  \text{tr} [H,A]^\dag B = \Braket{L(A)}{B}$ without a sign. Taking the ordinary Hilbert-space adjoint of the \emph{result of the action} of $L$ on an operator $\bullet$ is an entirely different thing, $(L\bullet)^\dagger = [H,\bullet]^\dagger = -[H,\bullet]$
	with a minus sign.}
\begin{equation}
L^\sudag\bullet = L\bullet \lrsepa \cP L\cP\bullet = L\bullet\label{eq_local_liouvillian}\,
\end{equation}
to show that the Liouvillian satisfies a duality relation of simpler form than Eq.~\eqref{eq_duality_original} with $\dual{L}:=-L$,
\begin{align}
	(-iL)^\sudag - \mathcal{P} (-i \dual{L}) \mathcal{P} = 0
	.
\end{align}
Finally, taking the superadjoint of \eq{eq_born_markov_kernel} and using the properties \eq{eq_field_superoperator_properties} and \eq{eq_local_liouvillian},
we find the \emph{fermionic duality relation}
\begin{align}
(-iL+W) + \mathcal{P} ( i \dual{L} + \dual{W})^\sudag \mathcal{P} = -\Gamop 
\label{eq_duality_generators}
,
\end{align}
with the \emph{coupling superoperator}
\begin{align}
\Gamop(L) &
:=
\frac{i}{2\pi}\int_{-\infty}^\infty d E
\sumsub{\ell\ell'}{\eta q}
(G^{q}_{\eta \ell})^\sudag
\frac{\Gamma^{\eta}_{\ell\ell'}(E)}{\eta E - L + i0_+}G^q_{\eta \ell'}
\label{eq_coupling_superoperator}.
\end{align}
Equations \eq{eq_duality_generators} and \eq{eq_coupling_superoperator} constitute the main result of the paper. They represent the non-wideband fermionic duality relation between the \emph{generator} of the quantum master equation of interest,
\begin{align}
	\partial_t \rho(t) = (-iL+W) \rho(t)
	\label{eq_qme},
\end{align}
and the generator of a \emph{dual master equation}\footnote{Importantly, the offdiagonal elements must be treated consistently for both the original parameters ($L$, $W$) and the dual parameters ($\dual{L}$, $\dual{W}$), in the sense that secular coherences are systematically separated from nonsecular coherences, depending on the parameters $L$ and $-\dual{L}$ relative to $W$ and $\dual{W}$. The point is that nonsecular coherences contribute effectively \emph{beyond} the leading order approximation in the tunnel coupling~\cite{Leijnse2008Dec,Koller2010Dec}.
}
	\begin{align}
	\partial_t \dual{\rho}(t)
	&= (i\dual{L}+\dual{W}) \dual{\rho}(t).
	\label{eq_qme_dual}
	\end{align}
The kernel of Eq.~\eqref{eq_qme} is a function $W=F(L,\{\mu_\alpha \})$ [\Eq{eq_born_markov_kernel}], while the \emph{dual} kernel $\dual{W}$ of Eq.~\eqref{eq_qme_dual} is the same function of $\dual{L}=-L$ and
$\dual{\mu}_\alpha=-\mu_\alpha$ \emph{except for one difference}: just as for the single-mode example in \Eq{eq_inversion}, we do not invert the sign of the open-system energies (contained in $L$) at which the energy-dependence $\Gamma_{\alpha\nu;\ell\ell'}(E)$ of the coupling is 
evaluated in \eq{eq_born_markov_kernel}, see Appendix~\ref{sec_app_duality}. Note that the unexpected occurrence of $+\dual{L}$ in \Eq{eq_qme_dual} instead of $-\dual{L}$ is discussed further in Appendix~\ref{sec_unexpected_L}.
\par

We stress that the only \emph{practical} difference between \Eq{eq_duality_generators} and the wideband-limit duality \eq{eq_duality_original} is that the
scalar $\Gamma$ on the right-hand side has become the nontrivial superoperator $\Gamop$.
Remarkably, this coupling superoperator \eq{eq_coupling_superoperator} 
still does \emph{not} depend on the state of the reservoirs.
The function $\Gamop(L)$ is thus the same for all quantum master equations with \emph{different} electrochemical potentials $\mu_\alpha$ and temperatures $T_\alpha$,
but with the same \emph{lump sum} of energy-dependent coupling matrices
\begin{equation}
\Gamma_{\ell\ell'}^\eta(E)
:= \sum_{\alpha\nu}\Gamma^{\eta}_{\alpha\nu;\ell\ell'}(E)
.
\label{eq_lump_sum2}
\end{equation}
This highlights that the duality \eq{eq_duality_generators} is not just a trivial rewriting of the generator of the master equation in terms of two other superoperators; on the contrary, these other two superoperators are either directly related to $W$ by a well-defined parameter transformation ($\dual{W}$) or strongly restricted not to depend on the reservoir states ($\Gamop$).
In the wideband limit, $\Gamma_{\ell\ell'}^\eta(E)=\Gamma_{\ell\ell'}^\eta$, we recover\footnote
	{The anticommutation relations \eq{eq_anticomm} imply that the integrand in \Eq{eq_coupling_superoperator} is $\propto \delta(E-L) \delta_{\ell\ell'}$.}
$\Gamop = \Gamma \mathcal{I}$ in \eq{eq_duality_original},
where the lump sum \eq{eq_lump_sum2} is extended to
all orbitals $\ell$ in \Eq{eq_lump_sum}.
For our main result, Eq.~\eqref{eq_duality_generators}, valid for energy-dependent couplings, the operator structure \eq{eq_coupling_superoperator} of $\Gamop$ requires that we analyze the implications of the duality anew.

%%%%%%%%%%%%%%%%%%%%%%%%%%%%%%%%%%%%%%%%%%%%%%%%%%%%%%%%%%%%%%%
\begin{figure*}
	\includegraphics[width=0.8\linewidth]{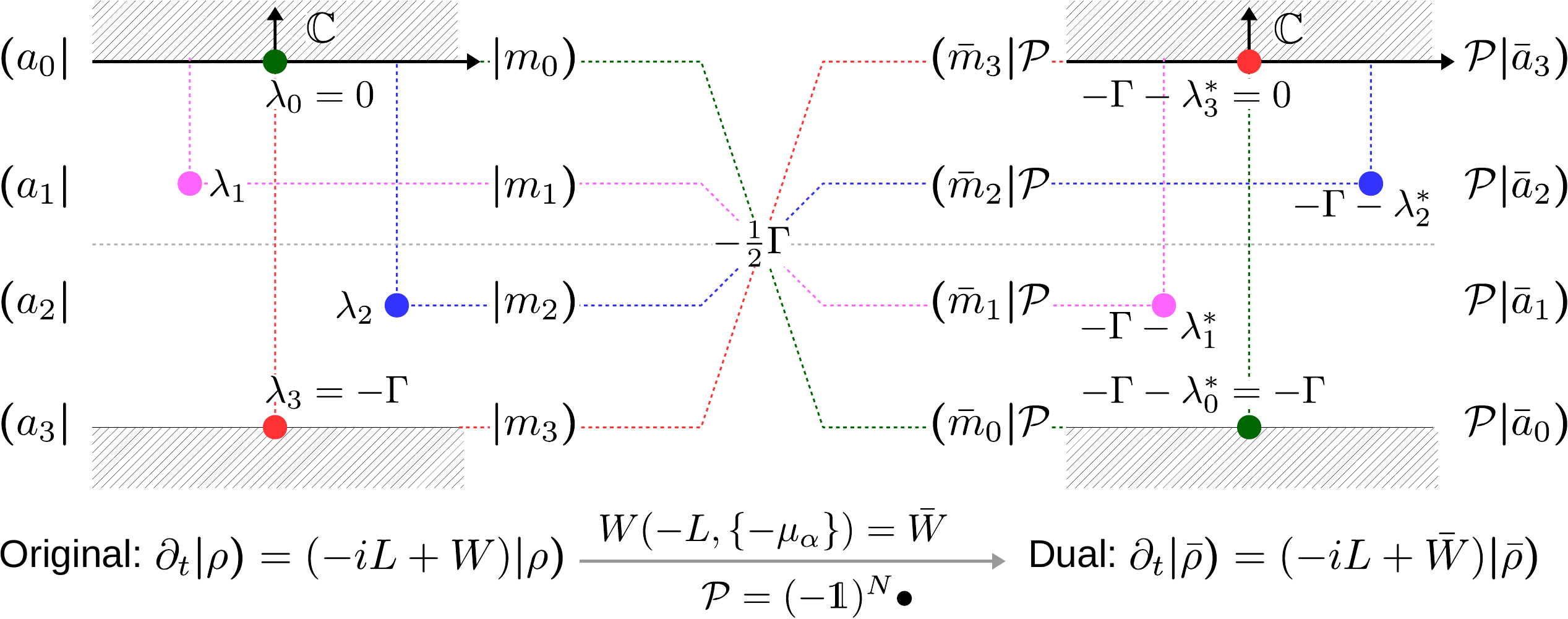}
	\caption{Fermionic duality for quantum-master equations \textit{in the wideband limit} \BrackEq{eq_duality_generators_wbl}.
		Schematically shown are the cross relations between amplitudes and modes,
		the left and right eigenvectors of the generator $-iL+W$ (left panel)
		and of its dual $i\dual{L}+\dual{W}$ (right panel) for \emph{different} eigenvalues.
		Each panel shows the eigenvalues (colored dots) in lower half of the complex plane.
		The eigenvalues are restricted to the strip $0 \geq \Re\sqbrack{\lambda_k} \geq -\Gamma$
		first, by the boundedness of the solutions $\rho$ and $\dual{\rho}$ (excluding the upper shaded area)
		and, second, by fermionic duality (excluding the lower shaded area bounded by the fermion-parity rate $-\Gamma$).
		The cross relation vertically mirrors the eigenvalues about the horizontal line $-\frac{1}{2}\Gamma$ (gray).
		We will show that even \textit{beyond the wideband limit} \BrackEq{eq_duality_generators} it is possible to obtain a fundamentally simpler matrix \eq{eq_wbl_basis} for the generator by applying the \emph{same} mapping to a subset of known eigenvectors [\Sec{sec_mode_mixing}].
		% to complete a basis [\Sec{sec_mode_mixing}].
		%In the \textit{rate-equation limit} one can simply restrict the crossrelations to the real eigenvalues of the rate matrix, $-iL+W \to W$ [\Sec{sec_duality_rate}].
		\label{fig_crossrelation}
		}	
\end{figure*}
%%%%%%%%%%%%%%%%%%%%%%%%%%%%%%%%%%%%%%%%%%%%%%%%%%%%%%%%%%%%%%%

%%%%%%%%%%%%%%%%%%%%%%%%%%%%%%%%%%%%%%%%%%%%%%%%%%%%%%%%%%%%%%%
\subsection{Mode-amplitude duality in the wideband limit}
\label{sec_duality_wbl}
%%%%%%%%%%%%%%%%%%%%%%%%%%%%%%%%%%%%%%%%%%%%%%%%%%%%%%%%%%%%%%%
To understand how the extended fermionic duality \eq{eq_duality_generators} can yield physical insights,
we first outline how this works in the simpler wideband limit.
We follow previous studies~\cite{Schulenborg2016Feb,Vanherck2017Mar} which have treated only weak-coupling rate equations for probabilities, but extend the procedure to the full density operator including coherences, i.e., offdiagonal elements with respect to the local energy eigenstates $\ket{i}$ of the open-system Hamiltonian $H$.
This brings in a novel aspect, since the time-evolution generator is the sum of the kernel $W$ and the system Liouvillian term $-iL$ .
The wideband duality relation \eq{eq_duality_original} for the generator $-iL + W$ then reads 
\begin{align}
(-iL+W) + \mathcal{P} ( i \dual{L} + \dual{W})^\sudag \mathcal{P} = -\Gamma \, \mathcal{I},
\label{eq_duality_generators_wbl}
\end{align}
as explained in Sec.~\ref{sec_duality_extended} without the wideband limit.
 Let us analyze the implications of this duality \eq{eq_duality_generators_wbl}.

%%%%%%%%%%%%%%%%%%%%%%%%%%%%%%%%%%%%%%%%%%%%%%%%%%%%%%%%%%%%%%%
\subsubsection{Cross relations}
%%%%%%%%%%%%%%%%%%%%%%%%%%%%%%%%%%%%%%%%%%%%%%%%%%%%%%%%%%%%%%%
The wideband-limit duality \eq{eq_duality_generators_wbl} between the two generators can be exploited analagously to our simple example [\Eq{eq_rho_expansion}]:
\Eq{eq_qme} is solved by diagonalizing~\footnote{We assume $-iL + W$ is diagonalizable, but note that the crossrelations can be extended to generalized eigenvectors in the nondiagonalizable case.} the \emph{generator},
\begin{align}
	-iL+W = \sum_{k=0}^{n} \lambda_k \Ket{m_k} \Bra{a_k}
	.\label{eq_generator_diagonalized}
\end{align}
One then inserts this into the exponential solution
\begin{gather}
\Ket{\rho(t)} = e^{(-iL+W)t} \Ket{\rho(0)}
= \sum_{k=0}^n \Ket{m_k} \, \Braket{a_k}{\rho(0)} \,  e^{\lambda_k t}
\notag
\\
= \Ket{m_0}\Braket{a_0}{\rho(0)} + ...
+ \Ket{m_n}\Braket{a_n}{\rho(0)} e^{-\Gamma t}
\label{eq_rho_sol},
\end{gather}
where we let $n+1$ denote the number of modes.
When one has computed the \emph{right} eigenvectors
as \emph{function} of $L$ and $W$,
one can find the remaining \emph{left} eigenvectors
as follows.
Assuming one knows mode $\Ket{m_k(-iL+W)}$, corresponding to an eigenvalue $\lambda_k(-iL+W)$,
this determines an amplitude
by multiplication by the parity operator $\mathcal{P}$,
substitution of parameters by their duals
and taking the superadjoint (see Appendix~\ref{sec_crossrel}):
\begin{subequations}
\begin{align}
	\Bra{a_{k'}(-iL+W)}
	 := \Big[ \, \Ket{\, \fpOp m_k(i\dual{L}+\dual{W}) \, } \, \Big]^\dag
	 .
	 \label{eq_crossrel_braa}
\end{align}
This is a relation between operators $a_{k'}(-iL+W)=(-\one)^N m_k(i\bar{L}+\dual{W})$ characterizing the bra and the ket, respectively.
Similarly, if instead the amplitude is known, a mode can be determined via
\begin{align}
\Ket{m_{k'}(-iL+W)}
:=
\Big[ \, \Bra{\, \fpOp a_k(i\dual{L}+\dual{W}) \, } \, \Big]^\dag
\label{eq_crossrel_ketm}.
\end{align}
In contrast to the originals (numbered $k$),
the resulting amplitude and mode belong to a \emph{different} eigenvalue (numbered $k'$),
\begin{align}
\lambda_{k'} (-iL+W)
=-[
\Gamma + \lambda_k^{*} (i\dual{L}+\dual{W})]
.
\label{eq_crossrel_eigenvalue}
\end{align}\label{eq_crossrel}\end{subequations}
More precisely, the imaginary parts of these different eigenvalues for dual parameters are equal,
\begin{align}
\Im\sqbrack{\lambda_{k'} (-iL+W)}
=\Im\sqbrack{\lambda_k (i\dual{L}+\dual{W})}
,
\label{eq_eigenvalue_imaginary_duality}
\end{align}
whereas the negatives of the real parts, the non-negative decay rates, are cross related
\begin{align}
- \Re\sqbrack{\lambda_{k'} (-iL+W)}
=\Gamma - [- \Re\sqbrack{\lambda_k(i\dual{L}+\dual{W})} ]
\geq 0
\label{eq_eigenvalue_real_duality}
.
\end{align}
This physically remarkable cross relation between the slowly decaying and quickly decaying modes is illustrated in \Fig{fig_crossrelation}.

We emphasize that the above described cross link presents a huge simplification even if the considered open system consists only of a few orbitals;  one can bypass essentially half of the eigenvalue problem by simple \emph{parameter substitutions} and the obtained expressions are much more compact than when algebraically solving the full eigenvalue problem in the conventional way.
Moreover, one is free to compute only the right eigenvectors or only the left ones or any computationally advantageous combination.

%%%%%%%%%%%%%%%%%%%%%%%%%%%%%%%%%%%%%%%%%%%%%%%%%%%%%%%%%%%%%%%
\subsubsection{Trace-preservation $\leftrightarrow$ fermion-parity mode}
%%%%%%%%%%%%%%%%%%%%%%%%%%%%%%%%%%%%%%%%%%%%%%%%%%%%%%%%%%%%%%%
The quantum master equation has one obvious 'universal' feature:
the amplitude covector for the \emph{zero decay rate} is the linear trace function, i.e., $\Bra{a_0}$ with $a_0=\one$.
This corresponds to a left eigenvector $\Bra{a_0}(-i L + W)=0$ expressing the trace preservation of the time-dependent state $\rho(t)$, independent of all parameters.
The fermionic duality implies that the \emph{fastest decaying mode} is similarly 'universal': it is the \emph{parity-mode} $\Ket{m_n}=\Ket{\fpOp}$, which is also independent of all parameters. In fact, this holds for \emph{any} fermionic quantum master equation derived for bilinear coupling in the wideband limit~\cite{Schulenborg2016Feb}.

The corresponding right eigenvector for the zero eigenvalue, the unique stationary state (or zero mode) obeys $(-i L + W)\Ket{m_0(-iL+W)}=0$.
Duality fixes the corresponding amplitude covector for the fastest decaying mode by a parameter substitution:
$\Bra{a_n}= \Bra{(-1)^N m_0(i\dual{L}+\dual{W})}$,
avoiding the calculation of this generally complicated expression.
The above two special cross relations are noted in our simple example [Table \ref{tab}], and we will see in the following that the involved four vectors continue to play a role beyond the wideband limit [\Eq{eq_suggestion}, \eq{eq_basis_quantum_dot}]. At this point, we can, however, already conclude that for energy-independent couplings, there are in fact $n-1$ additional nontrivial cross relations of the same type, as summarized in Eqs.~\eq{eq_crossrel}.

%%%%%%%%%%%%%%%%%%%%%%%%%%%%%%%%%%%%%%%%%%%%%%%%%%%%%%%%%%%%%%%
\subsection{Non-wideband coupling superoperator}
%%%%%%%%%%%%%%%%%%%%%%%%%%%%%%%%%%%%%%%%%%%%%%%%%%%%%%%%%%%%%%%

Having covered the wideband limit duality, let us now turn to the implications of the fermionic duality \eq{eq_duality_generators} for energy-dependent couplings.
We highlight upfront that while the replacement $\Gamma \to \Gamop$ on the right hand side of \Eq{eq_duality_generators}
prohibits a direct cross link between left and right eigenvectors, the duality nevertheless implies nontrivial relations for the real (or for the super-Hermitian) [\Eq{eq_duality_extended_real}] and for the imaginary (or for the anti super-Hermitian) part of $W$ [\Eq{eq_duality_extended_imag}], relying only on a few properties of the
coupling superoperator $\Gamop$. In this section, we present the required fundamental properties of $\Gamop$.

First, we note that $\Gamop$ as defined in \Eq{eq_coupling_superoperator} commutes with the parity superoperator $\cP$
by virtue of \eq{eq_field_superoperator_properties}
and \eq{eq_local_liouvillian}:
\begin{equation}
\cP\Gamop\cP
=
\Gamop
\label{eq_coupling_superoperator_properties1}.
\end{equation}
Second, taking the superadjoint is equivalent to taking the dual transform to inverted energy parameters,
\begin{equation}
[\Gamop(L)]^\sudag = \dual{\Gamop}(L)
\label{eq_coupling_superoperator_properties2}.
\end{equation}
As for $\dual{W}$ below \eq{eq_qme_dual}, $\dual{\Gamop}(L)$ equals $\Gamop(-L)$ except for the couplings $\Gamma_{\alpha\nu;\ell\ell'}(E)$ in \Eq{eq_coupling_superoperator}, which are still evaluated at the \emph{original} energies contained in $+L$, see Appendix~\ref{sec_app_duality}.
Equation \eq{eq_coupling_superoperator_properties2} follows most easily by taking the adjoint of the duality relation \eq{eq_duality_generators}, applying $\cP$ from the left and right, using $[\cP,\Gamop] = 0$ and $\cP^2 = \mathcal{I}$.

Third, the real part $\Re\sqbrack{\Gamop} := (\Gamop + \Gamop^\sudag)/2$ explicitly reads
\begin{gather}
\Re\sqbrack{\Gamop}
=
\frac{1}{2}
\sum_{ij}
\sumsub{\ell\ell'}{\eta q}
\Gamma^{\eta}_{\ell\ell'}(\eta E_{ij})
G^{q \sudag}_{\eta \ell}
\Ket{ij}\Bra{ij}G^q_{\eta \ell'}.
\label{eq_coupling_superoperator_real}
\end{gather}
This can be derived from \Eq{eq_coupling_superoperator_properties2} in two steps.
The first step is to insert the eigen decomposition of the Liouvillian $L=\sum_{ij} E_{ij} \Ket{ij} \Bra{ij}$ obtained from the many-body Hamiltonian $H=\sum_i E_i \ket{i}\bra{i}$ of the system. In this decomposition, we denote the matrix-elements and transition operators by
\begin{align}
	\Bra{ij} := \bra{i}\bullet \ket{j}
	,
	\qquad
	\Ket{ij} := \ket{i}\bra{j},
	\label{eq_energy_basis}
\end{align}
respectively, and the transition frequencies as $E_{ij} = E_{i} - E_{j}$.
In the second step, one then uses the assumed analyticity of $\Gamma_{\alpha\nu;\ell\ell'}(E)$, giving the result \Eq{eq_coupling_superoperator_real}.
Note that this result contains the wideband limit\footnote
	{Use completeness relation $\mathcal{I} = \sum_{ij} \Ket{ij} \Bra{ij}$
	and anticommutation relations \eq{eq_anticomm}.}, meaning
$\Re\sqbrack{\Gamop}\rightarrow \Gamma \mathcal{I}$
whereas $\Im\sqbrack{\Gamop} \rightarrow 0$. 

Furthermore, $\Re\sqbrack{\Gamop}$ is positive semidefinite,
\begin{align}
	\Re\sqbrack{\Gamop} \geq 0
	\label{eq_Gamop_pos}
	.
\end{align}
This follows from inserting \Eq{eq_coupling_rates} into \Eq{eq_coupling_superoperator_real}, giving a sum of positive semidefinite superoperators that altogether is positive semidefinite,
\begin{gather}
\Re\sqbrack{\Gamop}
=
2\pi
\sumsub{\eta q i j}{\kappa \alpha \nu}
\delta(\eta E_{ij} - \epsilon_{\kappa\alpha\nu})
\, 
\Ket{V^{\eta q i j}_{\kappa\alpha\nu}} \Bra{V^{\eta q i j}_{\kappa\alpha\nu}}
\geq 0
,
\label{eq_gamop_positive}
\end{gather}
where we have defined $\Ket{V^{\eta q i j}_{\kappa\alpha\nu}}
:=
\sum_{\ell}
\tau_{\kappa\alpha\nu;\ell}^{*}
\, 
G^{q \sudag}_{\eta \ell} \Ket{ij}
$.
%%%%%%%%%%%%%%%%%%%%%%%%%%%%%%%%%%%%%%%%%%%%%%%%%%%%%%%%%%%%%%%
\subsection{Duality bound on fermionic decay rates}
\label{sec_duality_general_implications_dissipativity}
%%%%%%%%%%%%%%%%%%%%%%%%%%%%%%%%%%%%%%%%%%%%%%%%%%%%%%%%%%%%%%%

The wideband-limit duality \eq{eq_duality_original} implies that the lump sum of couplings  $\Gamma$ is an upper bound on the relaxation rates \BrackEq{eq_eigenvalue_real_duality} which is even tight:
the exact decay rate of the fermion-parity mode $\Ket{m_n}=\Ket{\fpOp}$ is the largest and coincides with $\Gamma$.
For energy-dependent coupling, the superoperator $\Gamop$ plays an analogous but less direct role.

To show this, we use that the fermionic duality relation for the kernel $W$ \eq{eq_duality_generators} implies
a separate duality relation between the real parts of the kernel and its dual:
\begin{equation}
-\Re\sqbrack{W} - \cP \, \Re\sqbrack{\dual{W}} \, \cP
\label{eq_duality_extended_real}
=	
\Re\sqbrack{\Gamop}
,
\end{equation}
where we have used $\cP^\sudag = \cP$.
We also use that the real part $\Re\sqbrack{W}$ features in the necessary (but not sufficient) condition for the solution of the quantum master equation to be bounded~\footnote
	{Otherwise the eigenvalues of $\rho(t)$ diverge, violating positivity and trace-normalization, rendering $\rho(t)$ unphysical. This boundedness condition is called 'dissipativity'.},
i.e., 
$\Braket{\rho(t)}{\rho(t)}=\text{tr} \rho(t)^2 < \infty$.
To see this, one applies the master equation \eq{eq_qme} to arrive at
$\partial_t \Braket{\rho(t)}{\rho(t)} = 2 \Bra{\rho(t)} \Re\sqbrack{W} \Ket{\rho(t)} \leq 0$,
with the term $-iL$ dropping out.
Since this must hold for any initial condition $\rho(0)$,
$\Re\sqbrack{W}$ must have
only negative or zero eigenvalues, and thus
\begin{align}
	-\Re\sqbrack{W} \geq 0
	.\label{eq_duality_bound}
\end{align}
This can also be expressed as the well-known lower bound\footnote
	{In general the real (imaginary) part of an eigenvalue of a matrix is contained in the spectrum of its (anti)hermitian part~\cite{Garren1968}.}
on the negatives of the real parts of the eigenvalues $\lambda_k$ of the generator $-iL+W$ that are of interest, the decay rates:
\begin{align}
0 \leq
-\Re\sqbrack{\lambda_k(L,W)}
\leq
 \text{max} \big\{ \text{spec}(-\Re\sqbrack{W}) \big\},
\end{align}
where spec$(\cdot)$ denotes the spectrum of eigenvalues.
Importantly, since we assume \Eq{eq_duality_bound} to hold for \emph{all} values of the physical parameters,
the dual generator $i\dual{L} + \dual{W}$ also has only bounded solutions $\dual{\rho}(t)$, which means
\begin{align}
-\Re\sqbrack{\dual{W}} \geq 0.
\end{align}
Taking together the above relations now leads to a nontrivial \emph{upper bound} on the decay rates given by
\begin{align}
-\Re\sqbrack{\lambda_k(L,W)}
\leq  \text{max} \big\{ \text{spec}(\Re\sqbrack{\Gamop}) \big\}
\label{eq_decay_bound}
.
\end{align}
This follows from the real part of the duality \eq{eq_duality_extended_real},
together with
the properties \eq{eq_coupling_superoperator_properties1} and \eq{eq_Gamop_pos}
\begin{align}
\text{max} \big\{ \text{spec}(-\Re\sqbrack{W}) \big\}
&=
\text{max} \big\{ \text{spec}(-\mathcal{P} \Re\sqbrack{W} \mathcal{P} ) \big\}
\notag
\\
&=
\text{max} \big\{ \text{spec}( \Re\sqbrack{\Gamop+W} ) \big\}
\notag
\\
&
\leq
\text{max} \big\{ \text{spec}(\Re\sqbrack{\Gamop} ) \big\}
,
\end{align}
where we used that the maximal eigenvalue of a sum of hermitian matrices is bounded by the sum of their maximal eigenvalues. Importantly, the bound \eq{eq_decay_bound} is easier to compute analytically compared to the eigenvalues of $-iL + W$, since $\Re\sqbrack{\Gamop}$ has a much simpler structure  \BrackEq{eq_coupling_superoperator_real} than the kernel.

%%%%%%%%%%%%%%%%%%%%%%%%%%%%%%%%%%%%%%%%%%%%%%%%%%%%%%%%%%%%%%%
\subsection{Restrictions on renormalized transition frequencies} 
%%%%%%%%%%%%%%%%%%%%%%%%%%%%%%%%%%%%%%%%%%%%%%%%%%%%%%%%%%%%%%%

The dynamics of the coherences and their effect on physical quantities is crucially affected by $\Im\sqbrack{W} = (W - W^\sudag)/2i$.
The latter plays a central role for the so-called 'exchange-field'~\cite{Konig2003Apr,Braun2004Nov} and its generalizations~\cite{Misiorny2013Oct}
in quantum-dot spintronics and various other Lamb-shift effects mentioned in the introduction.
The terms $\Im\sqbrack{W}$ cause the Lamb-shift renormalization of the
transition frequencies of the system $(L)$ due to its coupling
to the reservoirs, and must be included consistently~\cite{Leijnse2008Dec,Koller2010Dec} even in the weak coupling limit~\cite{Konig2001Apr}.

The imaginary part of the duality \eq{eq_duality_generators},
\begin{equation}
(-L+\Im\sqbrack{W}) + \cP (L -\Im\sqbrack{\dual{W}}) \cP
=	
- \Im\sqbrack{\Gamop} 
\label{eq_duality_extended_imag}
,
\end{equation}
now implies that the renormalized oscillation frequencies obey nontrivial restrictions. 
To see this, we first note that the imaginary parts of the eigenvalues $\lambda_k$ of the generator $-iL+W$
are contained in the spectrum of the imaginary part of $-iL+W$:
\begin{align}
	\Im\sqbrack{\lambda_k(L,W)} \in \text{spec}(-L +\Im\sqbrack{W})
	.\label{eq_bounding_spectrum}
\end{align}
\color{black}
These $\Im\sqbrack{\lambda_k(L,W)}$ are the oscillation frequencies
in the solution $\Ket{\rho(t)}$, e.g. describing precession of the spin-accumulation vector in quantum-dot spin valves or Cooper pair oscillations on a quantum dot subject to a superconducting proximity effect.
Similarly, for the oscillation frequencies of the dual quantum master equation
\begin{align}
\Im\sqbrack{\lambda_k (L,\dual{W})} \in \text{spec}(-L + \Im\sqbrack{\dual{W}})
.\label{eq_bounding_spectrum_dual}
\end{align}
In the wideband limit we have $\Im\sqbrack{\Gamop} = 0$, and \Eq{eq_duality_extended_imag} implies
that the bounding spectra \eq{eq_bounding_spectrum} and \eq{eq_bounding_spectrum_dual} even coincide, in agreement with \Eq{eq_eigenvalue_imaginary_duality}.
For energy-dependent coupling, the duality establishes
that the bounding spectra of the frequency-renormalization terms are still strongly related to those of the dual quantum master equation,
albeit in a more complicated way,
\begin{align}
	\text{spec}( -L + \Im\sqbrack{W})
	=
\text{spec}( -L + \Im\sqbrack{\dual{W}} - \Im\sqbrack{\Gamop}).
\end{align}
This additionally involves the non-zero superoperator $\Im\sqbrack{\Gamop} = 0$, which is still simpler to compute than $\Im\sqbrack{W}$.

%%%%%%%%%%%%%%%%%%%%%%%%%%%%%%%%%%%%%%%%%%%%%%%%%%%%%%%%%%%%%%%
\subsection{Duality restrictions on the generator matrix}
\label{sec_mode_mixing}
%%%%%%%%%%%%%%%%%%%%%%%%%%%%%%%%%%%%%%%%%%%%%%%%%%%%%%%%%%%%%%%
In addition to these general restrictions imposed on the eigenvalue spectrum of the generators, the fermionic duality \eq{eq_duality_generators} 
can be exploited to substantially simplify matrix elements in practical computations.
Namely, \eq{eq_duality_generators} implies that for every right eigenvector $\Ket{m_k}$ of $-iL + W$ with eigenvalue
$\lambda_k= \lambda_k(L,W)$,
there is a left covector $\Bra{\mathcal{A}_{l(k)}}=\Bra{\fpOp \dual{m}_k}$
which is \emph{not} a left eigenvector,
but represents a simplification. More explicitly, the nontrivial left action of the kernel $W$ is replaced by a much simpler action of the coupling superoperator $\Gamop$,
\begin{subequations}
\label{eq_basis_vectors}
\begin{gather}
	\Bra{\mathcal{A}_{l(k)}}(-iL+W)
	= -\Bra{\mathcal{A}_{l(k)}}
	\big( \Gamop +\dual{\lambda}_k^* \, \mathcal{I} \big)
	.
	\label{eq_duality_mode_to_amplitude}
\end{gather}
Notably, thereby the entire nontrivial reservoir-dependence ($\mu_\alpha,T_\alpha$) of the left-action of $W$ is captured by this single known eigenvalue $\lambda_k(L,\dual{W})$.
Here, $l(k)$ denotes some way of numbering these vectors depending on $k$
with $\dual{\lambda}_k = \lambda_k(L,\dual{W})$.
Likewise, every known left eigenvector
$\Bra{a_k}(-iL + W) = \lambda_k \Bra{a_k}$,
defines a right vector
$\Ket{\mathcal{M}_{l(k)}} = \Ket{\fpOp \dual{a}_k}$,
which is not an eigenvector but fulfills
\begin{gather}
(-iL+W)\Ket{\mathcal{M}_{l(k)}} = 
-\big( \Gamop +\dual{\lambda}_k^* \, \mathcal{I} \big) \Ket{\mathcal{M}_{l(k)}}
\label{eq_duality_amplitude_to_mode}.
\end{gather}\label{eq_duality_amplitudes-modes}%
\end{subequations}

To clearly sketch the implications of these relations, we now
assume that $n+1$, the number of components of $\rho$, is even
and that one has computed only the first \emph{half} of the eigenvalues $\lambda_k$ [numbered by $k$, e.g., according to their magnitude]
and their corresponding amplitudes $\Bra{a_k}$ and modes $\Ket{m_k}$.
We can then ---without further calculation--- complete the set of left (right) vectors to a (not necessarily biorthogonal, see \Sec{sec_mode_mixing2}) basis by adding
the right vectors $\Ket{\mathcal{M}_{l(k)}}:=\mathcal{P} \Ket{\dual{a}_k}$ as well as 
the left vectors $\Bra{\mathcal{A}_{l(k)}}:= \Bra{\dual{m}_k} \mathcal{P}$, 
where $l(k)$ labels the second half of these basis vectors in terms of the first half.
In the wideband limit, these vectors would give the second half of the modes and amplitudes [\Sec{sec_duality_wbl}, \Fig{fig_crossrelation}],
but here they are \emph{not eigen}vectors of $-iL+W$ as indicated by the caligraphic labeling $\mathcal{A},\mathcal{M}$ instead of $a,m$.
Nevertheless, via Eqs.~\eqref{eq_basis_vectors}, the duality dictates that the matrix representation of the generator with respect to these left and right bases
has a much simpler form,
\begin{align}
	&-iL+W \cong\nonumber\\
	&\left[\begin{array}{c|c}
	\lambda_k\delta_{k k'}  &   \lambda_k \Braket{a_k}{\mathcal{M}_{l(k')}}\\
	\\\hline\\
	\Braket{\mathcal{A}_{l(k')}}{m_{k}} \lambda_{k}
	&  \Bra{\mathcal{A}_{l(k)}} \Gamop +  \delta_{k,k'} \dual{\lambda}_{k}  \mathcal{I} \Ket{\mathcal{M}_{l'(k')}} 
	\end{array}\right]\ .
	\label{eq_wbl_basis}
\end{align}
Here, we indicate  the type of matrix elements, which can occur in each of the four blocks of the matrix. 
By solving only \textit{half} of the eigenvalue problem, all nontrivial dependence on parameters has been incorporated entirely into the \emph{known} eigenvectors and eigenvalues and their wideband duals.
We stress that the goal of this procedure is not just to \emph{algebraically} simplify the matrix structure,
but especially to achieve \emph{analytically} more compact expressions by effectively setting $-iL+W \to \Gamop+\dual{\lambda}_k$.
In particular, the remaining nondiagonal structure of the matrix \eq{eq_wbl_basis} in the above introduced, duality-induced basis
is a consequence of the energy dependent coupling, i.e., $\Gamop \neq \Gamma \mathcal{I}$.
Equivalently, one can say the modes (amplitudes) computed in the wideband limit $\Gamop =\Gamma \mathcal{I}$
become mixed when the 'energy-dependent perturbation' $\Gamop -\Gamma \mathcal{I}$ is turned on.
This wideband mode mixing is studied in \Sec{sec_quantum_dot} for a simple example that can be described in the setting of rate equations, to which we turn next.

%% file: paper_duality_rate.tex
\section{Fermionic duality for rate equations}
\label{sec_rate_equation}

We now focus on the rate-equation limit assumed for the concrete example system in \Sec{sec_quantum_dot}
to illustrate the impact of energy-dependent couplings.

%%%%%%%%%%%%%%%%%%%%%%%%%%%%%%%%%%%%%%%%%%%%%%%%%%%%%%%%%%%%%
\subsection{Reduction to rate equation}
%%%%%%%%%%%%%%%%%%%%%%%%%%%%%%%%%%%%%%%%%%%%%%%%%%%%%%%%%%%%%

We first show how this rate-equation limit arises and display some of the general characteristics. In the quantum master equation, the probabilities
$P_i(t) :=\bra{i}\rho(t)\ket{i}=\Braket{ii}{\rho(t)}$
decouple from the coherences
$\bra{i}\rho(t)\ket{j}=\Braket{ij}{\rho(t)}$ ($i\neq j$)
when\footnote
	{This may be exact (due to selection rules/conservation laws) or approximate (after neglecting small nonsecular contributions belonging to the next-to-leading order in the coupling expansion~\cite{Leijnse2008Dec,Koller2010Dec}).}
the generator $-iL+W$ is block-diagonal in the energy eigenbasis $\{ \Ket{ij} \}$ [\Eq{eq_energy_basis}].
We can then restrict the analysis to the Liouville subspace spanned by energy-diagonal operators
$\ket{i}\bra{i}=\Ket{ii}$ in which the Liouvillian $L$ has only zero matrix elements.
In this subblock, the kernel $W$ has nonzero elements which we relabel as a \emph{rate matrix}:
for $i \neq j$,
\begin{align}
W_{i j}
&:= \Bra{ii} W \Ket{jj}
\equalbyeqn{eq_kernel_lead_sum} \sum_\alpha\Bra{ii}W_\alpha\Ket{jj}\notag\\
&= 
\sum_\alpha\Gamma_{\alpha,ij}^\eta(\eta E_{ij})
f^\eta_\alpha(\eta E_{ij})
\label{eq_Wij}
\end{align}\label{eq_Wij_form}
is the rate for transition $j \to i$. It includes the Fermi functions $f^\eta_\alpha(x)$ for reservoir $\alpha$ as introduced below \Eq{eq_born_markov_kernel}.
The \emph{coupling rates} are by \Eq{eq_coupling_rates} positive semidefinite:
\begin{subequations}
\begin{align}
\Gamma_{\alpha,ij}^\eta(\eta E_{ij} )
&=
\sum_{\nu}
\sum_{\ell\ell'} \Gamma^\eta_{\alpha\nu; \ell\ell'} (\eta E_{ij})
\, 
\bra{i} d_{\eta \ell} \ket{j}\bra{j}d_{\eta \ell'}^\dag \ket{i}
\\
&=
\sum_{\kappa\nu}
\Big|
\bra{i} \sum_{\ell} \tau_{\kappa\alpha \nu; \ell}^{{-\eta}} d_{\eta \ell} \ket{j}
\Big|^2
\geq 0.
\end{align}\label{eq_coupling-rates}\end{subequations}
The value of $\eta = N_i-N_j = \pm 1$ (denoting $N_i := \bra{i}N\ket{i}$) is imposed by the charge-selection rule on the matrix elements of $d_{\eta \ell}$,
reflecting the leading-order approximation $\mathcal{W} \approx \mathcal{W}_1$ in the bilinear coupling \eq{eq_hamiltonian_tun}.
Interchanging states $i\leftrightarrow j$ thus implies $\eta \to -\eta$, leaving
the energy argument $\eta E_{ij}=-\eta E_{ji}$ of the coupling rates invariant:
\begin{align}
\Gamma_{\alpha,ij}^\eta(\eta E_{ij} ) = \Gamma_{\alpha,ji}^{-\eta}(-\eta E_{ji} )
.
\label{eq_coupling_invariant}
\end{align}
Finally, we point out that the left zero eigenvector $\Bra{\one}=\text{tr} \sum_i \ket{i}\bra{i}=\sum_i \Bra{ii}$ of the kernel $W$
(preserving probability normalization) translates to the column-sum rule for the rate matrix
\begin{align}
	\sum_i W_{i j} = \sum_i \Bra{ii} W \Ket{jj}  = \Bra{\one} W \Ket{jj} =0,
\end{align}
thereby fixing the diagonal elements $W_{ii} = - \sum_{j} W_{ji}$.

%%%%%%%%%%%%%%%%%%%%%%%%%%%%%%%%%%%%%%%%%%%%%%%%%%%%%%%%%%%%%
\subsection{Duality for fermionic rate matrix}\label{sec_duality_rate}
%%%%%%%%%%%%%%%%%%%%%%%%%%%%%%%%%%%%%%%%%%%%%%%%%%%%%%%%%%%%%

As a next step, we show how the rate equation limit simplifies the fermionic duality.
The probabilities obey the rate equation $\partial_t P_i(t) = \sum_{j} W_{i j} P_j(t)$, and its solution requires the eigenvalues of the modes and amplitudes of $W$.
With the above notation, the fermionic duality applies to the rate matrix $W$ if one formally sets $L \to 0$.
For this, we need the parity right-action superoperator $\mathcal{P}\bullet = \fpOp\bullet$ \BrackEq{eq_parity},
which by parity superselection $[\fpOp,H] = 0$ is block-diagonal
in the energy eigenbasis $\{ \Ket{ij} \}$, just as $W$ itself:
\begin{gather}
\Bra{ii}\mathcal{P}\Ket{jj}
= \delta_{i j} \, (-1)^{N_i}.
\end{gather}
The required coupling superoperator $\Gamop$ \BrackEq{eq_coupling_superoperator}
even further simplifies to a fully diagonal matrix:
\begin{gather}
	\Bra{ii}\Gamop\Ket{jj} = 
	\delta_{ij} \,  \Gamma_i %\sum_{ll'n} 
	,
	\quad
	\Gamma_i : = 
	\sum_{\alpha j} \Gamma^\eta_{\alpha,ji}(\eta E_{ji}) 
	\label{eq_coupling_superoperator_diagonal},
\end{gather}%
where we sum the coupling rates for the transitions from state $i$ over all final states $j$.
Thus, in the rate-equation limit, the fermionic duality relation reduces to\footnote
	{The operations $\mathcal{P} \bullet \mathcal{P}$ and 'bar' operation (inversion of energy \emph{parameters}) applied to $W$ do not alter its block-diagonal structure.
	As a result, $\Gamop$ on the right hand side of the duality must have this structure also.}:
\begin{align}
	W_{ji} + (-1)^{N_i} \dual{W}_{ij} (-1)^{N_j} = - \Gamma_{i} \, \delta_{ij} \text{ for all $i,j$}.
	\label{eq_duality_rate}
\end{align}
All appearing quantities are available once the rate equation has been written down.
This generalizes the two simple observations made in \Sec{sec_idea}
for a single-mode fermion system in the following way:

(i) Due to the restriction to sequential tunneling for weak, bilinear couplings, the off-diagonal rate-matrix elements $i \neq j$
 all represent transitions with a parity-sign change ($|N_i-N_j|=1$). Thus, using \Eqs{eq_Wbar_def},\eq{eq_Wij_form} and $\eta':= N_j-N_i = \pm 1$, the energy-inverted dual rates read
\begin{align}
\dual{W}_{i j}
& :=
\sum_\alpha\Gamma_{\alpha,ij}^\eta(\eta E_{ij})
\, 
f^{-\eta}_\alpha(\eta E_{ij}),
\notag
\\
& =
\sum_\alpha\Gamma_{\alpha,ji}^{\eta'}(\eta' E_{ji})
\, 
f^{\eta'}_\alpha(\eta' E_{ji})
 = W_{ji}
.
\label{eq_duality_offdiagonal}
\end{align}
As in \Eq{eq_inversion}, the energy inversion $(\{E_{ij}\},\{\mua\}) \rightarrow (\{-E_{ij}\},\{-\mua\})$ in $\dual{W}_{ij}$ is defined not to affect the energies $\eta E_{ij}$ at which the couplings $\Gamma^{\eta}_{\alpha,ij}(E)$ are evaluated, but only the Fermi functions, leading to $f^{\eta}_{\alpha} \rightarrow f^{-\eta}_{\alpha}$ in the first line of \Eq{eq_duality_offdiagonal}.

(ii) The diagonal rate-matrix elements involve no parity sign,
and $\dual{W}_{ii}:=-\sum_j \dual{W}_{ji}$ (sum rule) is found to obey \Eq{eq_duality_rate} by using
$
W_{ij}+W_{ji}= \Gamma_{ij}(E_{ij})
$.

The above in fact constitutes a simple derivation of \Eq{eq_duality_rate} as a special case of the much more broadly applicable duality derived and discussed for quantum master equations in the previous section.
The important physical implication is that all quantities in the dual open system ---such as average particle number or fluctuations--- can be basically understood in the same way as in the wideband limit~\cite{Schulenborg2016Feb,Vanherck2017Mar,Schulenborg2017Dec}.
In particular, strong local Coulomb \emph{repulsion} in the model of interest translates to strong \emph{attraction} in its dual system which exhibits pronounced fermion-pairing effects.
These qualitatively different properties of the dual model enter via the statistical factors (Fermi functions)
in which the energy is inverted. This is the ultimate reason why duality remains useful beyond the wideband limit.
In \Sec{sec_quantum_dot}, we explore how the interesting interplay of the energy dependence of the couplings (at original energies) with the dual statistical factors (at inverted energies) determines the nontrivial parameter dependence of a problem of interest.

%%%%%%%%%%%%%%%%%%%%%%%%%%%%%%%%%%%%%%%%%%%%%%%%%%%%%%%%%%%%%
\subsection{Mode-amplitude relaton in the wideband limit}
%%%%%%%%%%%%%%%%%%%%%%%%%%%%%%%%%%%%%%%%%%%%%%%%%%%%%%%%%%%%%

For energy-independent couplings ($\Gamma_i \to \Gamma$), we recover the duality for rate equations of \Ref{Vanherck2017Mar,Schulenborg2017Dec}.
Hence, the amplitude covectors and mode vectors of the rate matrix $W$,
are cross related:
\begin{subequations}
\begin{align}
\Bra{a_{k'}(W)}
= \Big[ \Ket{\fpOp m_k(\dual{W})} \Big]^\dag
\label{eq_ak_duality_rate}
\\
\Ket{m_{k'}(W)}
=
\Big[ \Bra{\fpOp a_k(\dual{W})} \Big]^\dag
.
\end{align}\label{eq_crossrel_rateeq}%
These vectors belong to different \emph{real} eigenvalues
\begin{align}
- \lambda_{k'} (W)
=\Gamma - [- \lambda_k(\dual{W}) ]
\label{eq_eigenvalue_real_duality2}
,
\end{align}
\end{subequations}
which are either \emph{negative} ($k\neq 0$, decay rates) or zero ($k=0$, stationary state).
The operators involved here are all diagonal in the energy eigenbasis and include
the eigenvectors for the zero eigenvalue and $-\Gamma$, respectively:
\begin{align}
\Bra{a_0} &= \sum_i \bra{i} \bullet \ket{i} \lrsepa \Ket{m_0} = \sum_i P_i \ket{i}\bra{i}\label{eq_bounding_vectors}\\
\Bra{a_{n}} &= \sum_i (-1)^{N_i} \dual{P}_i \bra{i} \bullet \ket{i} \lrsepa \Ket{m_n} = \sum_i (-1)^{N_i}\ket{i}\bra{i}.\notag
\end{align}
The probabilities $P_i$ for occupying energy eigenstate $\ket{i}$ in the stationary state $\lim_{t\to \infty} \Ket{\rho(t)}=\Ket{m_0}$
are cross related by duality to the amplitude covector for the fastest decay, the parity amplitude $\Bra{a_{n}}$.
Remarkably, we will now find that the dual stationary probabilities $\dual{P}_i$ can be \emph{explicitly} expressed in the stationary probabilities $P_i$ (i.e. without parameter substitution) even for energy-dependent couplings, as long as detailed balance holds.

%%%%%%%%%%%%%%%%%%%%%%%%%%%%%%%%%%%%%%%%%%%%%%%%%%%%%%%%%%%%%
\subsection{Recurrence times from fermionic duality}
\label{sec_reversibility}
%%%%%%%%%%%%%%%%%%%%%%%%%%%%%%%%%%%%%%%%%%%%%%%%%%%%%%%%%%%%%

As mentioned in the introduction, fermionic duality is a novel concept.
In particular, since it is \emph{not} equivalent to classical detailed balance, it is interesting to combine the two in order to obtain even stronger restrictions on the underlying rate equation.

Detailed balance presupposes that the stationary state $\Ket{m_0}$ is unique with strictly positive probabilities
$P_{i} = \bra{i}m_0\ket{i} > 0$.
This is the case whenever the system is 
\emph{positively recurrent}, meaning that any two energy eigenstates $i$ and $j$ are connected via a sequence of
transitions $i'\leftarrow i''$ with strictly positive couplings $W_{i'i''} > 0$~\cite{Serfozo2009}.
In our case, this is satisfied if all involved states are connected by a nonzero coupling, $\Gamma_{i'i''}>0$ \BrackEq{eq_coupling-rates}, and all temperatures are nonzero \BrackEq{eq_Wij}.

The detailed balance relation
${{P}_i}/{{P}_j}
=
{{W}_{ij}}/{{W}_{ji}}
$
holds if and only if Kolmogorov's criterion~\cite{Kolmogoroff1936} is furthermore satisfied:
for \emph{any} sequence $i_1,i_2,\dotsc,i_N = i_1$ of states forming a graph-theoretical \emph{closed loop}~\cite{Schnakenberg1976Oct},
the product of transition rates must be equal in both directions:
$W_{i_1i_2}\cdot W_{i_2i_3}\dotsc W_{i_{N-1}i_1}=W_{i_1i_{N-1}}\cdot W_{i_{N-1}i_{N-2}}\dotsc W_{i_{2}i_1}$.

Importantly, when constructing the dual rate matrix $\dual{W}$, one does not break the above two conditions if they hold for $W$. Hence, the dual system then also has a unique stationary state, and satisfies a detailed balance with the \emph{inverse} ratio:
\begin{equation}
\frac{\dual{P}_i}{\dual{P}_j}
=
\frac{\dual{W}_{ij}}{\dual{W}_{ji}}
=
\frac{W_{ji}}{W_{ij}}
=
\frac{P_j}{P_i}
\quad
\text{for any $i \neq j$}
\label{eq_detailed_balance}
.
\end{equation}
In the crucial second step, we have used the fermionic duality \eq{eq_duality_offdiagonal}, $\dual{W}_{ij} = W_{ji}$.  \Eq{eq_detailed_balance} implies that the product of each stationary probability for the system and for its dual
is given by some function $C(W)$, which is independent of the considered energy eigenstate $\ket{i}$:
\begin{equation}
P_{i}\cdot \dual{P}_{i} = C(W)
\label{eq_dual_state_inversion}.
\end{equation}
Although $C(W)$ depends nontrivially on the rate matrix of the original problem,
it can be expressed in the stationary occupations or its duals, using probability conservation $\sum_i P_i= \sum_i \dual{P}_i=1$:
\begin{equation}
C(W) =
\Big[ \sum_j (P_j)^{-1} \Big]^{-1}
=
\Big[ \sum_j (\dual{P}_j)^{-1} \Big]^{-1}
.
\label{eq_dual_state_overlap} 
\end{equation}
Thus, the simple observation \eq{eq_pbar} made for a single fermion mode holds generally:
\begin{equation}
\dual{P}_i
=
\frac{ (P_i)^{-1} }
{ \sum_j (P_j)^{-1}}
\label{eq_dual_state_probabilities}.
\end{equation}
The product \eqref{eq_dual_state_inversion} has an upper bound\footnote
	{The bound is found from the arithmetic-harmonic mean inequality $d \,  (\sum_{i=1}^d (P_i)^{-1} )^{-1} \leq \frac{1}{d} \sum_{i=1}^d P_i$, here written for probabilities $P_i>0$. It follows from the Cauchy-Schwarz inequality $x\cdot y \leq |x| |y|$ applied to $d$-dimensional vectors
	$x=(\sqrt{P_1},\ldots,\sqrt{P_d})$ and $y=(1/\sqrt{P_1},\ldots,1/\sqrt{P_d})$.}
 given by the number of states $d$,
\begin{equation}
C(W)\leq \frac{1}{d^2}
\label{eq_CW_bound},
\end{equation}
which is independent of all other parameters.  The bound is only attained\footnote{The inequality turns into an equality if  $x= r y$. For $P_i>0$ this implies that $P_i=r > 0$ for each $i$. With $\sum_iP_1=1$ it follows $P_i=1/d$.} if the original and the dual system have equal, uniform probabilities $1/d$ at stationarity. This happens, for example, in the infinite temperature limit,
\begin{align}
P_i \dual{P}_i \leq \frac{1}{d^2} = \lim_{T_\alpha \to \infty} \dual{P}_i P_i  
\label{eq:product},
\end{align}
but also under specific nonequilibrium conditions [\Sec{sec_transient}].

Summing over $i$, we find that the overlap of the stationary state and its dual is bounded by the infinite temperature value
\begin{align}
\Braket{\dual{m}_0}{m_0}
\leq \frac{1}{d} = \lim_{T_\alpha \to \infty} \Braket{\dual{m}_0}{m_0}.
\end{align}
This explicit bound on the overlap makes quantitatively clear that the system and its dual are physically very different. For systems with many states ($d\gg 1$), these supervectors are nearly orthogonal independently of the choice of parameters.

\Eq{eq_dual_state_probabilities} allows for an intriguing interpretation of the probabilities of the dual stationary state.
The inverse probabilities $(P_i)^{-1}$ have a direct physical meaning: they quantify the recurrence of the Markov stochastic process on the energy-eigenstates of the system induced by tunneling transitions.
By Kac's lemma~\cite{Kac1947Oct}, 
$(P_i)^{-1}$ equals the long-time limit of the mean recurrence time of the state $i$ relative to the average time that the original system spends in that state.
Therefore, \Eq{eq_dual_state_probabilities} expresses that whenever detailed balance holds, the dual stationary probability $\dual{P}_i$ quantifies the \emph{relative stationary rareness} of energy eigenstate $\ket{i}$ in the original stationary fermionic system. In other words, it yields the recurrence time of state $j$ compared to the sum over all recurrence times.

Specifically for the wideband limit, inserting \Eq{eq_dual_state_probabilities} into the evolution formula
\eq{eq_rho_sol} furthermore shows how under detailed balance, the stationary occupations $P_i$ not only determine the slowest decay (stationary mode), but also the fastest decay (parity amplitude) of the system, given any initial probabilities $P_j(0)$:
\begin{gather}
	\Ket{\rho(t)} 
	= \Ket{m_0}\Braket{a_0}{\rho(0)} + ...
	+ \Ket{m_n}\Braket{a_n}{\rho(0)} e^{-\Gamma t}
	\label{eq_unstable}\\
	=
	\sum_i P_i \Ket{i}
	+
	...
	+
	\sum_i (-1)^{N_i} \Ket{i} \, \sum_j \frac{(-1)^{N_j} P_j(0)}{P_j\sum_l P_l^{-1}} e^{-\Gamma t}\ .\notag
\end{gather}
Finally, \Eq{eq_dual_state_probabilities} means that the stationary state of the inverted model $\Ket{\dual{m}_0}$ is the \emph{least recurring} state in the stationary limit under detailed balance and it is therefore of direct physical interest. This is consistent with previous findings that the dual stationary state is the \emph{maximally unstable state} relative to the stationary state of the original system, see Ref.~\cite{Schulenborg2016Feb,Vanherck2017Mar}. 
This can easily be understood remembering that the energy inversion of the duality also reverses the sign of the interaction [\Eq{eq_duality_offdiagonal} ff.] from, say repulsive to attractive. One therefore expects electron-pairing at low bias in the dual model. 
As a consequence, only the dual probabilities $\dual{P}_i$ which differ by an even number of electrons $N_i$ are sizable, and we may approximate $\Braket{a_n}{\rho(0)} \propto \Braket{\dual{m}_0}{\rho(0)}$ in \Eq{eq_unstable}, see also \Eq{eq_bounding_vectors}. This means that in the wideband limit, an \emph{initial} state which has a large overlap with the \emph{dual} of the \emph{final} stationary state will have the fastest rate of decay.

%%%%%%%%%%%%%%%%%%%%%%%%%%%%%%%%%%%%%%%%%%%%%%%%%%%%%%%%%%%%%
\subsection{Duality restrictions on the rate matrix}
\label{sec_mode_mixing2}
%%%%%%%%%%%%%%%%%%%%%%%%%%%%%%%%%%%%%%%%%%%%%%%%%%%%%%%%%%%%%

In the simpler wideband limit, the duality cross relations \eq{eq_crossrel_rateeq} naturally suggest to represent the kernel $W$ in a basis of right eigenvectors including
the stationary state $\Ket{m_0}$ and the parity mode $\Ket{m_n}=\Ket{\fpOp}$,
and a basis of left eigenvectors including the trace $\Bra{a_0}=\Bra{\one}$ and the parity amplitude covector $\Bra{a_n}=\Bra{\fpOp \dual{m}_0}$, where the latter is known if $m_0$ has been computed.
In particular, the orthogonality
\begin{equation}
\Braket{\fpOp \dual{m}_0}{m_0}=0
\label{eq_zero}
\end{equation}
always holds. For the more complicated situation with energy-dependent couplings, we have shown in \Sec{sec_mode_mixing} that 
$W\Ket{m_0}=0$ no longer guarantees that
$\Bra{\fpOp \dual{m}_0}$ is a left eigenvector of $W$, and it thus might seem at first sight that \Eq{eq_zero} fails to hold.
However, the restriction \eq{eq_dual_state_inversion} emerging from the \emph{combination} of fermionic duality and detailed balance \emph{still} guarantees that \Eq{eq_zero} holds. 
In fact, such orthogonality holds for any
product of an energy-diagonal operator $A$ with the stationary state $m_0$,
\begin{equation}
\Braket{\fpOp \dual{m}_0 }{A m_0} = 0
\label{eq_orthogonalities_detailed_balance}
,
\end{equation}
provided only that $A$ is orthogonal to the fermion-parity operator:
$\Braket{\dual{m}_0 \fpOp}{A m_0}=\sum_i (-1)^{N_i} \dual{P}_i \bra{i}A\ket{i} P_i= C(W) \cdot  \Braket{\fpOp}{A}=0$
by \Eq{eq_dual_state_inversion}.

The validity of \Eq{eq_orthogonalities_detailed_balance} has already been observed in \Ref{Schulenborg2017Dec}
---ignorant of its origin---
and extensively used
to simplify coefficients of linear response to applied voltage or thermal bias in the wideband limit.
Here we have derived a much stronger result: it is valid at \emph{any bias} and for energy-dependent couplings, as long as the detailed balance relation \eq{eq_detailed_balance} holds.
The main point is that the above result simplifies\footnote
	{The properties \eq{eq_parity_result} and \eq{eq_inverted_state} imply nontrivial relations for $\Gamop$:
	$
	\Bra{\one}\Gamop\Ket{\fpOp} = 0
	$
	and
	$
	\Bra{\fpOp \dual{m}_0} \Gamop\Ket{m_0} = 0
	$
}
the general way of exposing the duality restrictions on matrix elements discussed in \Sec{sec_mode_mixing}. In \Sec{sec_quantum_dot_master_equation}, we shall exploit this to study the wideband-mode mixing induced by the energy dependent coupling.
Therefore, instead of solving for half of the exact eigenvectors of $W$,
we proceed with the minimal effort that is sufficient for our application [\Eq{eq_basis_quantum_dot}].
The only explicit calculation we do  is to obtain \emph{one eigen}vector, the stationary state $\Ket{m_0}=\lim_{t\to \infty}\Ket{\rho(t)}$.
This defines a left vector $\Bra{\mathcal{A}_n}=\Bra{\fpOp \dual{m}_0}$ with a simple action of $W$, which is  independent of $\{\mu_\alpha,T_\alpha \}$ [\Eq{eq_duality_mode_to_amplitude}]:
\begin{align}
\Bra{\fpOp \dual{m}_0}W = -\Bra{\fpOp \dual{m}_0 } \Gamop
\label{eq_inverted_state}
.
\end{align}
We already know the corresponding left zero mode, the \emph{eigen}vector given by the trace operation $\Bra{a_0}=\Bra{\one}$,
from which we can construct a right vector $\Ket{\mathcal{M}_n} = \Ket{(-\one)^N}$ which is independent of any details.
The action of $W$ on this right vector yields  [\Eq{eq_duality_amplitude_to_mode}]:
\begin{align}
W \Ket{\fpOp} = -\Gamop\Ket{\fpOp}
\label{eq_parity_result}
.
\end{align}
Thus, a fundamental structure of any fermionic rate matrix satisfying detailed balance is already exposed
\begin{align}
\left[\begin{array}{c|ccc|c}
0 &        \ldots & 0 &\ldots                                           & 0 \\
\hline
\vdots &   	 & &          & \vdots\\   
0 &  	& \Bra{..} W \Ket{..} & & \Bra{..} \Gamop \Ket{(-\one)^N}\\  
\vdots & & & & \vdots\\\hline
0 &   \cdots     & \Bra{ (-\one)^N\bar{m}_0 } \Gamop \Ket{..} &\cdots & \Bra{ (-\one)^N \bar{m}_0 } \Gamop \Ket{(-\one)^N}								
\end{array}\right]
\label{eq_suggestion}
\end{align}
 with respect to any right basis
$\{ \Ket{m_0},\ldots,\Ket{(-\one)^N} \}$
and any left basis
$\{ \Bra{a_0},\ldots,\Bra{(-\one)^N\bar{m}_0} \}$.
Crucially, these two bases can be made biorthogonal due to detailed balance [\Eq{eq_zero}].

%% file: paper_quantum_dot.tex
\section{Interacting single-level quantum dot}
\label{sec_quantum_dot}
In this final section
we illustrate how the fermionic duality can be applied to problems with energy-dependent coupling
and 
the insights it provides into both time-dependent as well as stationary charge- and energy transport.
We focus on the simplest nontrivial example of a weakly tunnel-coupled single-level quantum dot with strong local Coulomb repulsion
described by rate equations [\Sec{sec_rate_equation}].
Thereby, we go beyond   the extensively studied~\cite{Splettstoesser2010Apr,Contreras-Pulido2012Feb,Saptsov2012Dec,Juergens2013Jun,Gergs2015May,Sanchez2016Dec,Kennes2013Jun} wideband limit
and focus on the mixing of wideband-limit relaxation modes governing the transient decay~\cite{Splettstoesser2010Apr,Contreras-Pulido2012Feb,Schulenborg2016Feb} after a switch \BrackSec{sec_transient}
and on stationary thermoelectric transport~\cite{Schulenborg2017Dec} in the linear regime \BrackSec{sec_stationary}.

%%%%%%%%%%%%%%%%%%%%%%%%%%%%%%%%%%%%%%%%%%%%%%%%%%%%%%%%%%%%%
\subsection{Model and energy dependence of couplings}
%%%%%%%%%%%%%%%%%%%%%%%%%%%%%%%%%%%%%%%%%%%%%%%%%%%%%%%%%%%%%

%%%%%%%%%%%%%%%%%%%%%%%%%%%%%%%%%%%%%%%%%%%%%%%%%%%%%%%%%%%%%
\begin{figure}
	\includegraphics[width=\linewidth]{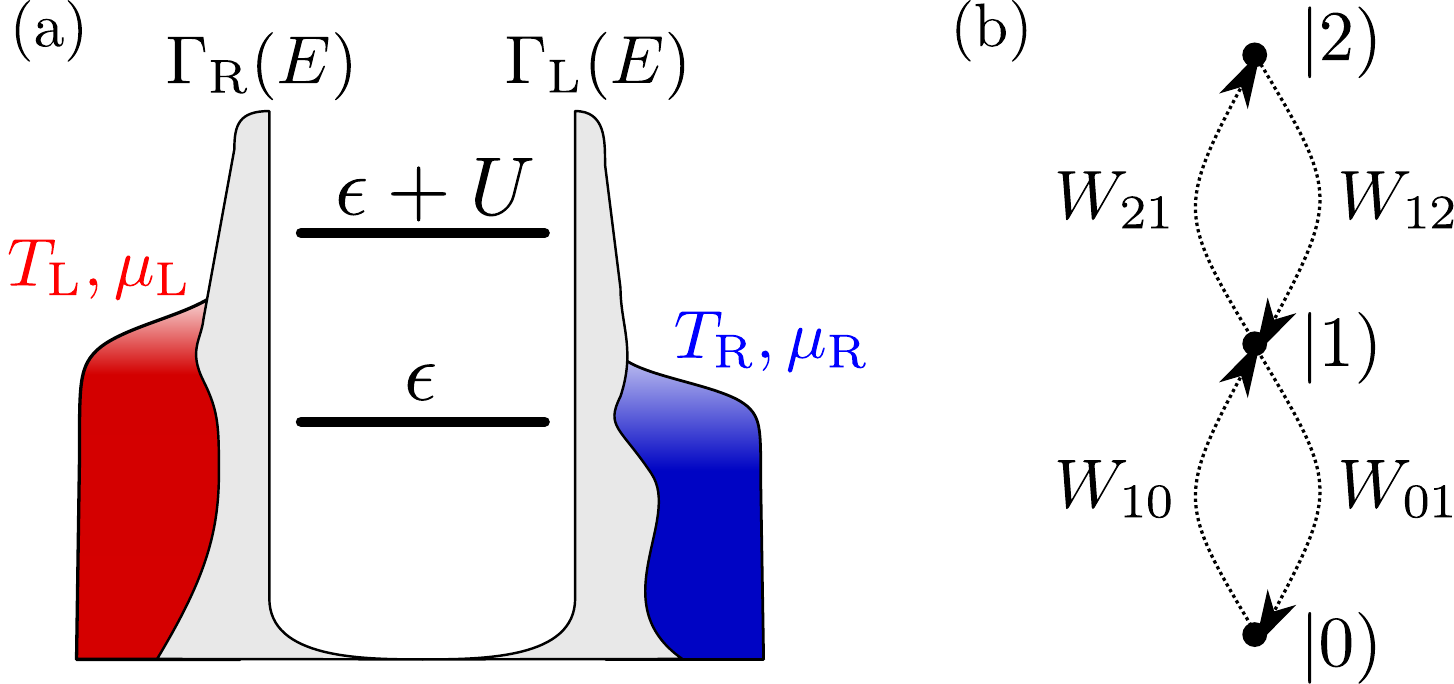}
	\caption{
		(a) Spin-degenerate single-level quantum dot with level-position $\epsilon$, local Coulomb interaction $U$, and energy-dependent tunnel barriers $\GamL(E) \neq \GamR(E)$ to a left and right reservoir at different electrochemical potentials $\muL \neq \muR$ and temperatures $\TL \neq \TR$.
		(b) In the weak-coupling, sequential-tunneling limit, the rate matrix of the \emph{spin-degenerate} system can be mapped to a simple tree graph, which by Kolmogorov's criterion~\cite{Kolmogoroff1936}
		guarantees detailed balance.
	}
	\label{fig_quantum_dot}
\end{figure}
%%%%%%%%%%%%%%%%%%%%%%%%%%%%%%%%%%%%%%%%%%%%%%%%%%%%%%%%%%%%%

\begin{figure}
	\centering
	\includegraphics[width=0.7\linewidth]{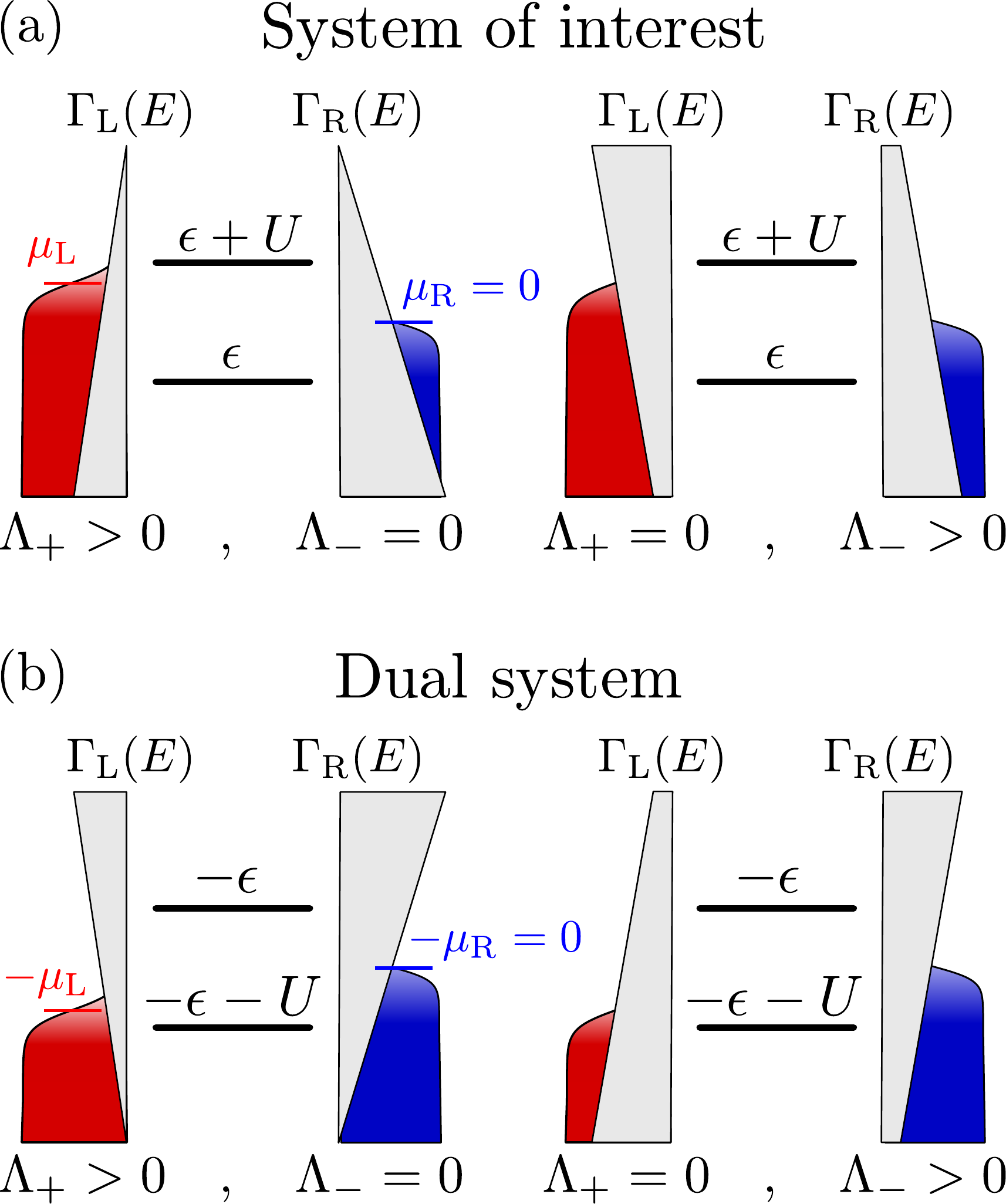}
	\caption{
		Schematic tunnel-barrier profiles for the extreme cases of the asymmetries $\Lambda_\pm$.
		(a) Original system of interest: the tunnel coupling is evaluated at the many-body transition energy $E=\epsilon$ or $E=\epsilon+U$ for the transition $\Ket{0} \to \Ket{1}$ and $\Ket{1} \to \Ket{2}$, respectively.
		(b) Dual system with inverted system energies, inverted electrochemical potentials, and with the tunnel coupling $\Gamma_{\alpha}(E)$ evaluated at the original energies $E = \epsilon, \epsilon + U$. The latter implies that the energy profile of the barrier is also inverted, see Appendix~\ref{sec_app_duality}.
	}
	\label{fig_important_quantities1}
\end{figure}

The quantum dot of interest is sketched in \Fig{fig_quantum_dot}(a)
and is described by the general model
introduced in \Sec{sec_model}. With the specific Hamiltonian
\begin{equation}
H = \epsilon N + U N_\uparrow N_\downarrow,\label{eq_hamiltonian_quantum_dot}
\end{equation}
it encompasses a spin-degenerate orbital with single-particle states labeled by
$\ell = \sigma = \uparrow,\downarrow$, where $\epsilon$ is the tunable level position and $U > 0$ the on-site Coulomb repulsion strength. The operators $N_\sigma = d_\sigma^\dagger d_\sigma$ are the spin-resolved occupation number operators, where $N = N_\uparrow + N_\downarrow$.
The energies are $E_0 = 0$, $E_1 = \epsilon$, and $E_2 = 2\epsilon + U$ 
with $0,1,2$ electrons\footnote{We do not express the state in terms of \emph{pure} states, therefore requiring care with the normalization factors.}. The solution of the rate equation can be expressed\footnote
	{Charge- and spin conservation decouple probabilities and coherences in this model.} as $\Ket{\rho(t)} = \sum_{i=0,1,2} P_i(t) \Ket{i}$
with time-dependent probabilities $P_i(t)$ and the corresponding spin-symmetric many-body energy \emph{density operators}
\begin{equation}
 \Ket{0} = \ket{0}\bra{0} \lrsepa \Ket{1} = \frac{1}{2}\sum_{\sigma=\uparrow,\downarrow}\ket{\sigma}\bra{\sigma} \lrsepa \Ket{2} = \ket{2}\bra{2}.\label{eq_eigenstates_quantum_dot}
\end{equation}
When employed as supervectors, the latter decompose the identity superoperator with a spin-degeneracy factor
\begin{equation}
 \mathcal{I} = \Ket{0}\Bra{0} + 2 \, \Ket{1}\Bra{1} + \Ket{2}\Bra{2}
 .
 \label{eq_liouville_identity}
\end{equation}
We fix $\muR \equiv \mu$ and $\TR \equiv T$ as references and apply a bias voltage $V < 0$ as well as temperature bias $\Delta T > 0$ to the left:
$\muL = \mu - V$ and $\TL = T + \Delta T$. 
The tunnel couplings \eq{eq_coupling_rates} are spin-conserving\footnote
	{Since we assume the coupling Hamiltonian to be spin-conserving, the off-diagonal terms in \Eq{eq_coupling_rates} vanish, $\Gamma_{\alpha\nu;\uparrow\downarrow} = \Gamma_{\alpha\nu;\downarrow\uparrow} = 0$.}
and depend on energy via the coupling functions
\begin{align}
	\Gamma_\alpha(E) :=\sum_{\nu,\ell = \uparrow,\downarrow}\Gamma_{\alpha\nu;\ell\ell}(E).
\end{align}
The values for these rates and their reservoir sums at the two possible many-body transition energies $E_{10}=\epsilon$ and $E_{21}=\epsilon + U$ ---the Coulomb resonances--- are in the following denoted by the shorthands
\begin{subequations}
\begin{alignat}{2}
\Gamepsa &:= \Gamma_\alpha(\epsilon)
,
\qquad
&
\Gameps &:= \sum_\alpha\Gamepsa
,
\\
\GamUa &:= \Gamma_\alpha(\epsilon + U)
,
\qquad
&
\GamU &:= \sum_\alpha\GamUa
.
\label{eq_coupling_rates_quantum_dot}
\end{alignat}%
\end{subequations}
%and are required to be small, $\Gamepsa,\GamUa \ll T_\alpha$.
The deviations from the wideband limit are quantified by
\begin{equation}
\Delta\Gama := \GamUa - \Gamepsa \lrsepa
\Delta\Gamma := \GamU - \Gameps = \sum_\alpha\Delta\Gamma_\alpha
\label{eq_coupling_rates_quantum_dot2}.
\end{equation}%
Figure \fig{fig_important_quantities1} schematically indicates this energy dependence of the couplings
by the corresponding barrier profiles. It is characterized by two important coupling asymmetry parameters:
\begin{align}
\Lambda_+ := \frac{\GamepsL\GamUL - \GamepsR\GamUR}{\Gameps\GamU}
,
\quad
\Lambda_- := \frac{\GamepsL\GamUR - \GamepsR\GamUL}{\Gameps\GamU}.
\label{eq_coupling_asymmetries}
\end{align}
The left panels in (a) and (b) show that the factor $\Lambda_+$ quantifies the left-right coupling asymmetry
which for $\Lambda_- = 0$ (implied by the wideband limit)
simplifies to the usual asymmetry factor, $\Lambda_+=\frac{\GamL - \GamR}{\GamL + \GamR}$.
The right panels illustrate the cross asymmetry $\Lambda_-$ which emerges \emph{only} for energy-dependent couplings. 
For example, if $\Lambda_+ = 0$ then the couplings have a perfect cross asymmetry between the two Coulomb resonances, quantified by $\Lambda_- = (\GamepsL - \GamepsR)/\Gameps = -(\GamUL - \GamUR)/\GamU$.

%%%%%%%%%%%%%%%%%%%%%%%%%%%%%%%%%%%%%%%%%%%%%%%%%%%%%%%%%%%%%
%%%%%%%%%%%%%%%%%%%%%%%%%%%%%%%%%%%%%%%%%%%%%%%%%%%%%%%%%%%%%

%%%%%%%%%%%%%%%%%%%%%%%%%%%%%%%%%%%%%%%%%%%%%%%%%%%%%%%%%%%%%
\subsection{Constructing the rate matrix by duality}
\label{sec_quantum_dot_master_equation}
%%%%%%%%%%%%%%%%%%%%%%%%%%%%%%%%%%%%%%%%%%%%%%%%%%%%%%%%%%%%%

The system can be described by a rate equation $\partial_t \Ket{\rho(t)}=W \Ket{\rho(t)}$
for the probabilities
$P_i(t)$; everything discussed in \Sec{sec_born_markov} applies.
In particular, detailed balance holds for \emph{any} value of the applied biases $V$ and $\Delta T$, since the many-body states are connected in the simple tree graph shown in \Fig{fig_quantum_dot}(b).\par

Instead of approaching the solution of this problem in the standard way, namely by constructing $W$ using Fermi's Golden rule,
then computing its eigenvalues and eigenvectors, etc.
we here follow a \emph{constructive} approach analogously to \Ref{Schulenborg2017Dec}, based entirely on the duality relation.
Given the considered approximations,
the evolution kernel of this model is fixed by the restrictions of the duality \eq{eq_duality_rate}.
It is these restrictions that yield physically insightful expressions for the required matrix elements of $W$,
\emph{not} the parameter dependences one inherits from Fermi's Golden rule rates.
Notably, this holds for any energy-dependent weak coupling to any number of reservoirs that are arbitrarily biased.

The starting point is the coupling superoperator $\Gamop = \sum_\alpha\Gamopa$,
whose contributions from the two junctions can be written down immediately using \Eq{eq_coupling_superoperator_diagonal}:
\begin{align}
 \Gamopa &= \Gameps\Ket{0}\Bra{0} + (\Gamepsa + \GamUa)\Ket{1}\Bra{1} + \GamUa\Ket{2}\Bra{2}\notag\\
 &= \sqbrack{\frac{2 - N}{2}\Gamepsa + \frac{N}{2}\GamUa}\cdot\mathcal{I}.
 \label{eq_coupling_superoperator_quantum_dot}
\end{align}
The second form of $\Gamopa$ follows from the identity \eq{eq_liouville_identity} and the left action of the particle number operator $N$ on the energy eigenstates \eq{eq_eigenstates_quantum_dot}.\par
Next, we follow \Sec{sec_mode_mixing2} and base our construction only on the guaranteed existence of \emph{one} pair of exact eigenvectors of $W$,
the stationary state $\Ket{m_0}=\Ket{z}$, denoted from hereon as $z$ero mode $z$,
and the trace operation $\Bra{a_0}=\Bra{z'}=\Bra{\one}$ as the corresponding left zero eigenvector.
Applying the non-wideband limit duality mapping, two further basis vectors then immediately suggest themselves: the $p$arity vector $\Ket{\mathcal{M}_2} = \Ket{p} = \Ket{\fpOp}$ and the dual covector $\Bra{\mathcal{A}_2} = \Bra{p'} = \Bra{\zin\fpOp}$ containing the stationary state of the \emph{dual} system $\Ket{\zin} = \Ket{\dual{z}}$. Importantly, while neither $\Ket{p}$ nor $\Bra{p'}$ is an eigenvector of $W$, both are nevertheless biorthogonal to $\Bra{z'} = \Bra{\one}$ and $\Ket{z}$ by virtue of detailed balance dictating \Eq{eq_orthogonalities_detailed_balance}. 
With this insight, the remaining basis vectors are fixed up to normalization by biorthogonality, yielding the $c$harge deviation $\Bra{\mathcal{A}_1} = \Bra{c'} = \Bra{N} - n_z\Bra{\one}$ from the stationary average $\nz$ and corresponding right vector $\Ket{\mathcal{M}_1} = \Ket{c} = \frac{\fpOp}{2}\sqbrack{\Ket{N} - \nzi\Ket{\one}}$. In summary, the left and right bases states are given by  
\begin{align}
 \Bra{\one} &\lrsepa \Ket{z}
 \label{eq_basis_quantum_dot}
 \\
 \Bra{p'} = \Bra{\fpOp \zin} &\lrsepa \Ket{p} = \Ket{\fpOp}\notag\\
 \Bra{c'} = \Bra{N} - n_z\Bra{\one} &\lrsepa \Ket{c} = \frac{\fpOp}{2}\sqbrack{\Ket{N} - \nzi\Ket{\one}}
 \notag.
\end{align}
Note that everything can in the end be traced back to the stationary state $\Ket{z}$, since the dual state $\Ket{\zin}$ and its average particle number $\nzi = \Braket{N}{\zin}$ required in \Eq{eq_basis_quantum_dot} are ultimately determined by $\Ket{z}$ together with the appropriate parameter transform.

In the wideband limit $\Gamepsa = \GamUa$, the vectors \eq{eq_basis_quantum_dot} are by construction all the known eigenvectors of $W$, see \Ref{Vanherck2017Mar}. For energy-dependent couplings, they are still convenient basis vectors because they enable us to construct the kernel $W$ [\Eq{eq_suggestion}]
entirely from the matrix elements of the simple coupling superoperator $\Gamop = \sum_\alpha\Gamopa$ \BrackEq{eq_coupling_superoperator_quantum_dot}, as shown in \Sec{sec_mode_mixing2}.
This is precisely what leads to the following, very compact form of the kernel:
\begin{subequations}
\begin{align}
W = &
-\gamceff\Ket{c}\Bra{c'} - \gampeff\Ket{p}\Bra{p'}
\\
& - \Delta\Gamma\sqbrack{\Ket{c}\Bra{p'} + \frac{\delnsqzi}{4}\Ket{p}\Bra{c'}}
.
\end{align}\label{eq_kernel_quantum_dot}\end{subequations}
This form contains the effective rates
\begin{subequations}
\label{eq_gammaeff}
\begin{align}
\gampeff = 
\frac{2-\nzi}{2}\Gameps + \frac{\nzi}{2}\GamU
&= \Gameps +  \frac{\nzi}{2}\Delta\Gamma
\label{eq_gammapeff}
\\
\gamceff  = \zeta
\sqbrack{\frac{\nzi}{2}\Gameps + \frac{2-\nzi}{2}\GamU}
&= \zeta \sqbrack{ \Gameps + \frac{2-\nzi}{2}\Delta\Gamma },
\label{eq_gammaceff}
\end{align}
with the effective spin-degeneracy factor\footnote
	{For singular parameter combinations at which $\nzi = n_z$, $\zeta$ needs to be calcualted from \Eq{eq_charge_rate_in_reservoir}, as the latter is an analytic continuation of $(\nzi - 1)/(\nzi - n_z)$ at these singular points.}
\begin{align}
\zeta & =
\frac{\nzi - 1}{\nzi - n_z}\lrsepa 0 < \zeta < 1,
\label{eq_zeta}
\end{align}\label{eq_effective_rates}%
\end{subequations}
accounting for the fraction of tunnel processes available for the transition [\Sec{sec_zeta}]. Furthermore, the stationary (dual) particle number fluctuations
$\delnsqzi(\epsilon,U,V)=\delnsqz(-\epsilon,-U,-V)$ enter \Eq{eq_kernel_quantum_dot} and are given by
\begin{subequations}
\begin{align}
\delnsqzi & 	=
\Braket{N^2}{\zin} - \Braket{N}{\zin}^2 =
\nzi\cdot(2 - \nzi)\cdot\zeta
\label{eq_dual_fluctuations}\\
\delnsqz &= \Braket{N^2}{z} - \Braket{N}{z}^2 = n_z\cdot (2 - n_z)\cdot(1 - \zeta).
\label{eq_fluctuations}
\end{align}%
\end{subequations}

More explicitly, the stationary state of the original and the dual model entering the kernel and determining its properties are contained in the expressions
\begin{align}
\Bra{p'} = \Bra{\zin\fpOp} &= \frac{2 - \nzi}{2}\Bra{0} + \frac{\nzi}{2}\Bra{2}
\label{eq_parity_amplitude_quantum_dot}
\\
&\phantom{=}-\frac{1-\zeta}{2\zeta}\delnsqzi \Big[  \Bra{0} + 2\Bra{1} + \Bra{2} \Big]
\notag
\end{align}
deriving from~\footnote
	{Note that $\zeta \rightarrow 1 - \zeta$ under the dual transform by \Eq{eq_zeta}.}%
\begin{align}
\Ket{z} &= \frac{2 - n_z}{2}\Ket{0} + \frac{n_z}{2}\Ket{2}
\notag\\
&\phantom{=}-\frac{\zeta\cdot \delnsqz}{2(1-\zeta)} \Big[  \Ket{0} - 2\Ket{1} + \Ket{2} \Big]
\label{eq_stationary_state_quantum_dot}.
\end{align}
Equation \eq{eq_kernel_quantum_dot} together with Eqs. \eq{eq_gammaeff}-\eq{eq_stationary_state_quantum_dot} form the central result of our application of the fermionic duality.
It leads to a remarkable conclusion:
apart from the four coupling constants $\Gamepsa, \GamUa$,
the physically-motivated stationary particle numbers $n_z, \nzi$ are the \emph{only two} variables for the complex dependence of $W$
on $\epsilon-\mu$, $U$, $T$, $\Delta T$ and $V$.
Although in full accord with a Golden-Rule calculation of $W$,
this systematic, \emph{unambiguous} reduction of all parameter dependencies seems virtually impossible to achieve in such a brute-force approach.
Note that computer \emph{algebra} is also of little help, since the duality makes use of parameter substitutions of the \emph{functions} involved.
Instead, we highlight that we obtain Eqs. \eq{eq_gammaeff}-\eq{eq_stationary_state_quantum_dot} \emph{without at all} knowing how $\Ket{z}$ depends on the system parameters; in analogy to \Ref{Schulenborg2017Dec}, we only use the duality \eq{eq_duality_rate}, positive recurrence \BrackSec{sec_reversibility}, the restriction to sequential tunneling causing the pair transitions $\Bra{0}W\Ket{2} = \Bra{2}W\Ket{0} = 0$ to disappear, and finally, detailed balance guaranteeing the orthogonality \eq{eq_orthogonalities_detailed_balance}. 

%%%%%%%%%%%%%%%%%%%%%%%%%%%%%%%%%%%%%%%%%%%%%%%%%%%%%%%%%%%%%
\subsection{Average charge occupations}\label{sec_occ}
%%%%%%%%%%%%%%%%%%%%%%%%%%%%%%%%%%%%%%%%%%%%%%%%%%%%%%%%%%%%%

The result \eq{eq_kernel_quantum_dot} of the fermionic duality implies that, in order to understand experimentally relevant parameter dependencies,
we only need to know the behavior of the stationary average charge of the system $n_z$, 
and, in particular, of the dual system $\nzi$. The latter implies that
that much of the analysis of the repulsive quantum dot model of interest
relies on the well-known~\cite{Anderson1975Apr} and experimentally demonstrated~\cite{Hamo2016Jul,Prawiroatmodjo2017Aug,Placke2018} behavior of quantum dots with effectively attractive interaction. In fact, we see below that it even dominates much of the behavior of the original system.
This includes the case of finite thermal gradients $\Delta T$ and voltages for which we will assume $\mu_L \geq \muR$, i.e., $V\leq0$.\par

Remembering now that we have not yet \emph{explicitly} determined $\Ket{z}$ as a function of the system parameters\footnote
	{At nonzero reservoir bias, the stationary state $\Ket{z}$ differs from the already known wideband-limit expression by its separate dependence on the \emph{four} energy-dependent couplings ($\Gamepsa$, $\GamUa$) instead of just two ($\Gamma_\alpha$).}, the advantage is that we can already argue \emph{qualitatively}, realizing that the occupations $n_z,\nzi$ can be understood from basic considerations of many-body energies relative to the electrochemical potentials.
Namely, for the system of interest, the repulsive interaction $U$ causes the average charge $n_z $ to be quantized deep inside subsequent Coulomb blockade regimes,
\begin{subequations}
\begin{eqnarray}
n_z=0 & \hspace{1cm} & \muL,\muR \ll \epsilon
,
\\
n_z=1 & \hspace{1cm} & \epsilon \ll \muL,\muR \ll \epsilon + U
,
\\
n_z=2 & \hspace{1cm} &\epsilon + U \ll \muL,\muR
,
\end{eqnarray}%
\end{subequations}
respectively.
In constrast, when only one many-body energy $\epsilon$ ($\epsilon+U$) lies well within the bias window $[\muL,\muR]$,
the occupation takes on a constant intermediate value $n_z \in [0,1]$ ($n_z \in [1,2]$).
When both $\epsilon$ and $\epsilon+U$ lie well within $[\muL,\muR]$,
we have the intermediate average value $n_z =1$ in the wideband limit (by electron-hole symmetry).
One expects that introducing energy-dependent coupling into this consideration
only affects these nonequilibrium intermediate values,
unless one of the many-body levels becomes effectively decoupled from the reservoirs.

The behavior of $\nzi$ is, instead, governed by the \emph{attractive} interaction $-U<0$ in the dual system, quantizing $\nzi$ to $0$ and $2$ in two subsequent \emph{inverted} Coulomb blockade regimes:
\begin{subequations}
	\begin{eqnarray}
\nzi=0& \hspace{1cm} &	-(\epsilon+U/2) \ll -\muL, -\muR
	\\
\nzi=2& \hspace{1cm} &	 -\muL,-\muR \ll -(\epsilon+U/2)
	,
	\end{eqnarray}%
\end{subequations}
respectively, expressed in terms of the inverted many-body energies and electrochemical potentials,
see \Fig{fig_important_quantities1}(b).
For a bias below the inverted Coulomb gap, $|V| \lesssim |-U|=U$, this leads to a direct, thermally-broadened transition of $\nzi$ between 0 and 2 when $\epsilon$ passes through
\begin{equation}
\epsilon + U/2 =
\tfrac{1}{2}(\muL + \muR) = \mu - \tfrac{1}{2}V =:\widetilde{\mu} .\label{eq_particle_hole_symmetry_point}
\end{equation}
Transitions involving the one-electron state are allowed only via thermal excitations, due to  the inverted order of the energies $-\epsilon-U < -\epsilon$ for the $1\to 2$ and $0\to 1$ transitions, respectively.

Clearly, at higher bias, $|V| \gtrsim U$, these transitions become enhanced for\footnote{In the original variables, this nonequilibrium regime occurs at intermediate values $\muR < \epsilon < \epsilon +U < \muL$.} $-\muL < -\epsilon -U < -\epsilon < - \muR$.
Here, $\nzi$ takes on intermediate values, depending on details of the rates,
because the electron-pairing is overcome by nonequilibrium processes.
Physically, we expect that energy-dependent couplings will not qualitatively counteract these pairing effects for $|V| < U$, unless we effectively decouple one of the many-body levels.

%%%%%%%%%%%%%%%%%%%%%%%%%%%%%%%%%%%%%%%%%%%%%%%%%%%%%%%%%%%%%
\begin{figure*}
	\centering
	\includegraphics[width=\linewidth]{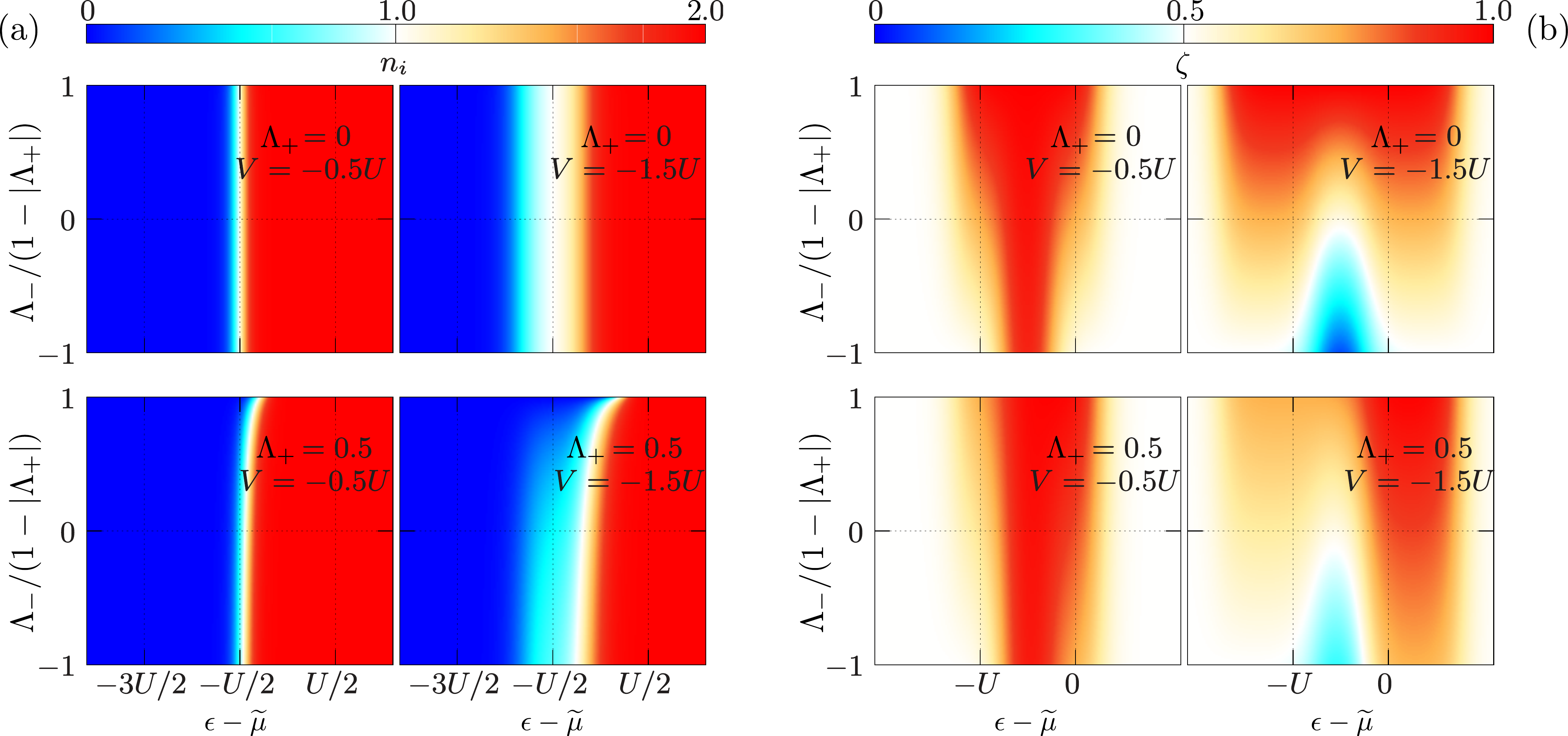}
	\caption{
		(a) Average dual occupation $\nzi$ \BrackEq{eq_charge_number_in_reservoir} for $\TL = \TR = 0.1\, U$
		as a function of the level position $\epsilon$ relative to $\widetilde{\mu}$ and the cross asymmetry $\Lambda_-$ of the energy dependence of the coupling \BrackEq{eq_coupling_asymmetries}.
		%\\
		Left panels: moderate bias $|V| < U$.
		The sharp step between $\nzi = 0$ and $2$ around $\epsilon - \widetilde{\mu} = -U/2$ is hardly affected by $\Lambda_-$,
		even for significant left-right coupling asymmetry $\Lambda_+ \neq 0$ \BrackEq{eq_coupling_asymmetries}
		due to the pairing effect in the dual system.
		Right panels: high bias $|V| > U$. An intermediate nonequilibrium regime $\muR < \epsilon < \epsilon + U < \muL$ appears in which the pairing effect is overcome.
		Without left-right coupling asymmetry ($\Lambda_+=0$), the dual stationary probabilies are uniform so that \emph{on average} $\nzi = 1$, even for large cross antisymmetry $|\Lambda_-|$.
		Pronounced deviations occur only when one many-body level is completely eliminated ($\Lambda_- \approx \Lambda_+$ for $\muL \geq \muR$).
		(b) Corresponding plots of the factor $\zeta$ in \Eq{eq_effective_rates} which accounts for the relative number of available tunneling processes [\Sec{sec_zeta}].
		Upper panels: Without left-right asymmetry $\Lambda_+ = 0$, \emph{positive} cross asymmetry $\Lambda_{-}$
		extends the $\zeta=1$ regime.
		Here an electron likely tunnels into an empty dot from the left reservoir ($\GamepsL > \GamepsR$)
		with bias $\muL > \epsilon$,
		but then remains there for a long time even though $\epsilon > \muR$,
		thereby leading to a preferred single occupation $\Ket{z} = \Ket{1}$.
		Likewise, $\GamUL < \GamUR$ implies the same preference when $\epsilon + U$ lies in the bias window:
		one of two electrons on the dot tunnels out faster to the right
		than a second electron tunnels back into a single-occupied dot from the left.
		%\\
		By complementary arguments, \emph{negative} cross asymmetries $\Lambda_-$ favor $\Ket{z} = \Ket{0}$ or $\Ket{2}$, and thereby extend the $\zeta=1/2$ regime.
		Lower panels: Positive left-right asymmetry $\Lambda_+ $ diminishes the effects of negative~$\Lambda_- $.
	}
	\label{fig_important_quantities2}
\end{figure*}
%%%%%%%%%%%%%%%%%%%%%%%%%%%%%%%%%%%%%%%%%%%%%%%%%%%%%%%%%%%%%

To study the behavior of $n_z$ and $\nzi$ \emph{quantitatively}, we combine the fermionic duality with the decomposition approach of \Sec{sec_linear_response}, that is, without \emph{ever} explicitly calculating the \emph{full} state $\Ket{z}$.
Instead, we use $W = \sum_\alpha W_\alpha$ \BrackEq{eq_kernel_lead_sum}
and note that \emph{each} reservoir-resolved kernel $W_\alpha$ \emph{separately} obeys the duality \eq{eq_duality_rate}
with $\Gameps,\GamU \rightarrow \Gamepsa,\GamUa$.
This enables us to apply the same constructive procedure that led to \Eq{eq_kernel_quantum_dot}
for each $W_\alpha$ separately.
The decisive difference is that the reservoir-resolved stationary state $\Ket{z_\alpha}$ describes an equilibrium with a single reservoir $\alpha$, and is therefore simply given by \Eq{eq_boltzmann_factor_lead}. This decomposes each average charge\footnote
	{
	 Our approach allows to express the kernel $W_\alpha$ only by $\Gamepsa,\GamUa$ and the reservoir-resolved \emph{equilibrium} occupations $\nza = \Braket{N}{z_\alpha}$ and $\nzia = \Braket{N}{\zia}$.
	One then identifies $(c'|W = \sum_\alpha (c'_\alpha|W_\alpha$ and rewrites the obtained expressions in terms of $n_i$, $\zeta$ and $n_z$ for the total as well as for the reservoir-resolved system. Acting from the right with the linearly independent vectors $\frac{1}{2}[|N) - |\one)]$ and
$\frac{1}{4}|\one)$ leads to two independent equations determining the quantities $n_z$ and $n_i$ (or $\zeta$) as given in \Eq{eq_charge_number_in_reservoir}.
	}
\begin{subequations}
\begin{align}
\nz - 1 &=  \sum_\alpha \frac{\zeta_\alpha}{\zeta} 
\sqbrack{\frac{\nza}{2} \frac{\Gamepsa}{\Gameps} - \frac{2 - \nza}{2} \frac{\GamUa}{\GamU}}
\label{eq_charge_number_in_reservoir}
\\
\nzi - 1 &=  \sum_\alpha \frac{1-\zeta_\alpha}{1-\zeta}
\sqbrack{\frac{\nzia}{2} \frac{\Gamepsa}{\Gameps} - \frac{2 - \nzia}{2} \frac{\GamUa}{\GamU}}
,\label{eq_dual_charge_number_in_reservoir}
\end{align}\label{eq_average_charges}%
\end{subequations}
at \emph{arbitrarily biased reservoirs} into the values obtained when the system is in equilibrium with reservoir $\alpha$ separately,
\begin{subequations}
\begin{align}
\nza &=\frac{2f^+_\alpha(\epsilon)}{f^+_\alpha(\epsilon) + f^-_\alpha(\epsilon + U)}
,
\\
\nzia &= \frac{2f^-_\alpha(\epsilon)}{f^-_\alpha(\epsilon) + f^+_\alpha(\epsilon + U)}
\label{eq_statistical}
\\
	& \approx 2\left.f^{-}_\alpha (\epsilon + \tfrac{1}{2} U)\right|_{T_\alpha\rightarrow T_\alpha/2} \qquad \text{for $U \gg T_\alpha$}
	,\label{eq_statistical_approx}
\end{align}
\end{subequations}
weighted by equilibrium values\footnote
	{One verifies from \Eqs{eq_charge_number_in_reservoir}-\eq{eq_app_charge_rate_number_reservoir} that for $\Gamepsa = \GamUa$, we retain the reservoir sums derived in \Ref{Schulenborg2017Dec} for the wideband limit.}
 of the factors \eq{eq_zeta},
\begin{subequations}
\begin{align}
\zeta_\alpha &=
\frac{\nzia - 1}{\nzia - \nza} = \frac{1}{2}\nbrack{f^+_\alpha(\epsilon) + f^-_\alpha(\epsilon + U)}
\label{eq_app_charge_rate_number_reservoir}
,
\\
\zeta & := \sum_\alpha
\zeta_\alpha\sqbrack{\frac{\nza}{2}\frac{\Gamepsa}{\Gameps} + \frac{2 - \nza}{2}\frac{\GamUa}{\GamU}}
\label{eq_charge_rate_in_reservoir}
.
\end{align}
\end{subequations}
In \Fig{fig_important_quantities2}(a), we plot the dual charge $\nzi$ as a function of the level position $\epsilon- \widetilde{\mu}$
as we vary\footnote
	{The coupling asymmetries \eq{eq_coupling_asymmetries} are taken as independent parameters. Implicitly, they depend on $\epsilon$ as discussed in \Sec{sec_stationary}. Note also that in general in the non-linear regime, the specific form of the coupling asymmetry as a function of energy depends on the bias and reservoir temperatures, see Refs.~\cite{Christen1996Sep,Benenti2017Jun}.},
the energy dependence of the coupling through the asymmetry parameters $\Lambda_\pm$ defined in \Eq{eq_coupling_asymmetries} and illustrated in figure~\ref{fig_important_quantities1}.
Overall, \Fig{fig_important_quantities2}(a) confirms the above qualitative analysis, stating that $\nzi$ is dominated by statistical effects [\Eq{eq_statistical_approx}] due to the strong negative interaction $-U$.
The energy dependence of the couplings mainly determines the relative reservoir weights in the intermediate nonequilibrium regime
$\muR < \epsilon < \epsilon + U <  \muL$ at large biases $|V| > U$.
Thus, in the following two applications, we can indeed make use of the above described, \emph{intuitive} understanding of the variables $n_z$ and $\nzi$, just as in~\Ref{Vanherck2017Mar} for the wideband limit .

%%%%%%%%%%%%%%%%%%%%%%%%%%%%%%%%%%%%%%%%%%%%%%%%%%%%%%%%%%%%%
\subsection{Transient decay after a switch}\label{sec_transient}
%%%%%%%%%%%%%%%%%%%%%%%%%%%%%%%%%%%%%%%%%%%%%%%%%%%%%%%%%%%%%

\begin{figure}
	\centering
	\includegraphics[width=\linewidth]{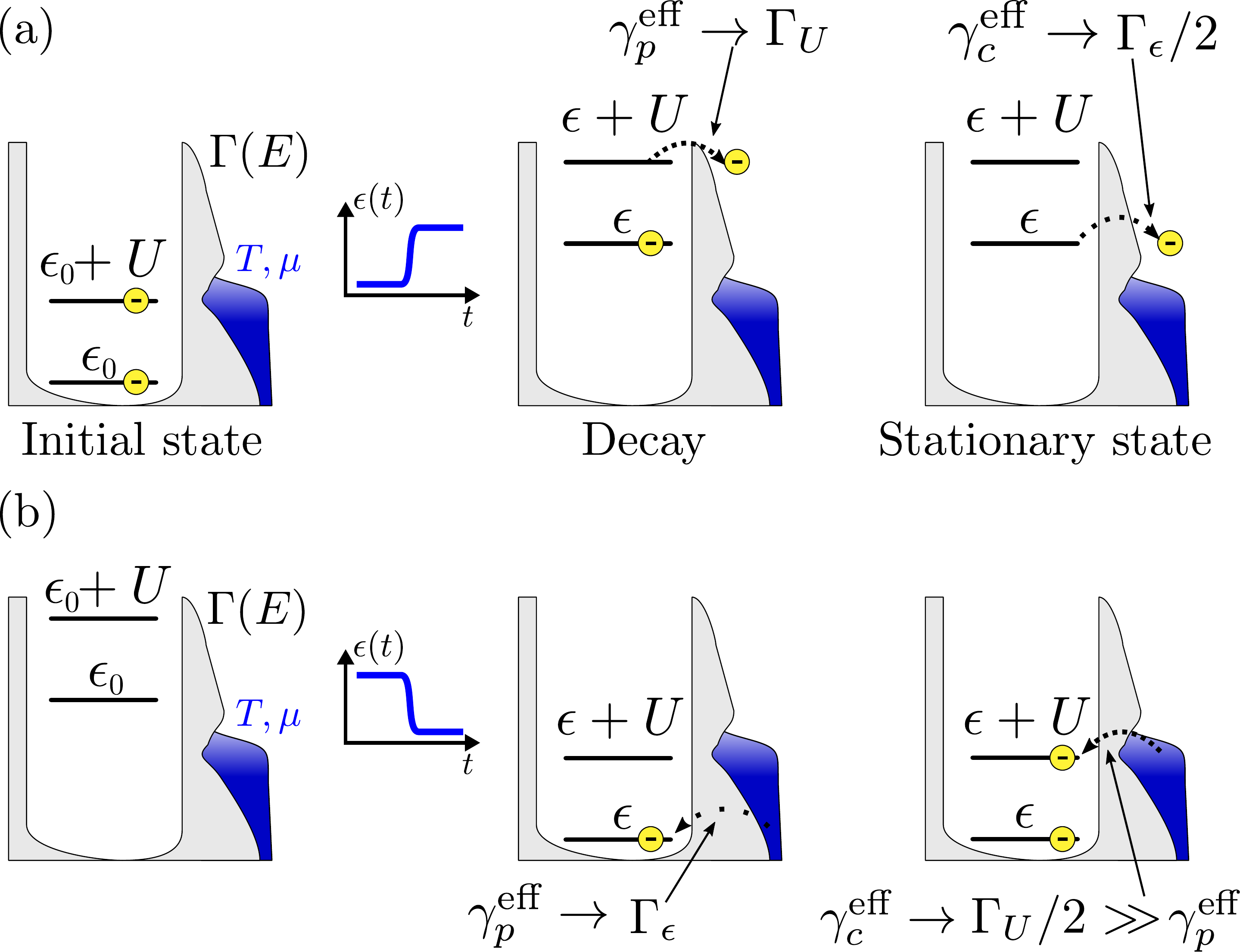}
	\caption{
		Sequences of tunneling processes between the quantum dot and the reservoir after a sudden upward (a) or downward (b) shift of $\epsilon_0 \to \epsilon$ by a gate-voltage switch.
		The rate expressions \eq{eq_effective_rates} then give correspondingly
		$\gampeff \sim \GamU$ (first electron out) and $\gamceff \sim \Gameps$ (second electron out)
		for $\nzi \approx 2$ [(a)],
		and
		$\gampeff \sim \Gameps$ (first electron in) and $\gamceff \sim \GamU$ (second electron in)
		for $\nzi \approx 0$ [(b)].
		As before, the barrier thickness represents its transparency $\Gamma(E)$,
		the factor $\zeta \approx 1/2$  in the final transitions ---accounted for in $\gamceff$---
		stems from the fact that the tunneling electron may have only one out of two possible spin polarizations.}
	\label{fig_temporal}
\end{figure}

\subsubsection{Mixing due to energy-dependent coupling}

Our first application of the central result \eq{eq_kernel_quantum_dot}
concerns the transient time evolution of the quantum dot.
This presents the simplest non-trivial extension of our example in \Sec{sec_idea} to the case of two modes with strong local Coulomb interaction and two reservoirs.
The transient decay is caused by a switch between two values of the level position, $\epsilon_0 \to \epsilon$,
as sketched in a simplified way with only one reservoir in \Fig{fig_temporal}. 

\emph{Decay rates.}
We first consider the relaxation rates of the 
exponential decay of the state $\Ket{\rho(t)}$ towards the stationary state $\Ket{z}$.
Diagonalizing \eq{eq_kernel_quantum_dot}, we obtain the general expressions for the two involved decay rates for any system parameter values and for an arbitrary number of coupled reservoirs:
\begin{equation}
\gamma_\pm =
\frac{1}{2}\sqbrack{\gamceff + \gampeff \pm \sqrt{\nbrack{\gampeff - \gamceff}^2 + \Delta\Gamma^2\delnsqzi}}.
\label{eq_relaxation_rates_quantum_dot}
\end{equation}
Due to the energy-dependent coupling, $\Delta\Gamma = \GamU - \Gameps$, the modes are mixed and the amplitudes are mixed,
causing their eigenvalues to repel each other.
Remarkably, all nontrivial parameter\footnote
	{The parameters of $W$ define the final state of the switch experiment.}
dependence of the strength of this mode mixing is controlled by just one quantity, the dual charge fluctuations $\delnsqzi$ [\Eq{eq_dual_fluctuations}]. This finding is a direct result of the duality-based representation of the kernel \eqref{eq_kernel_quantum_dot}. How this mixing behaves is straightforwardly understood based on the attractive physics of the dual model [\Sec{sec_occ}].
For a moderate voltage bias $|V| < U$ between the two reservoirs, the sharp step $0 \leftrightarrow 2$ of $\nzi(\epsilon)$ for $|V| < U$  causes the dual fluctuation $\delnsqzi \sim \nzi(2 - \nzi)$ \BrackEq{eq_dual_fluctuations} to vanish except for a peak around the \emph{electron-pair resonance} at $\epsilon + U/2 = \widetilde{\mu}$ in the dual system.
Apart from this special level position ---completely unexpected from the point of view of the original repulsive model---
there is negligible mixing between the effective decay rates in \Eq{eq_relaxation_rates_quantum_dot}:
\begin{align}
	\gamma_{\pm} \approx \max/\min  \{ \gamceff, \gampeff \} + \mathcal{O}(\Delta\Gamma^2 \delnsqzi)
	.
\end{align}
To understand the physics underlying the effective rates $\gamceff,\gampeff$, let us first look at the mixing of the corresponding amplitudes.

\emph{Parity amplitude.}
In the wideband limit, the effective 'parity' rate \eq{eq_gammapeff}
reduces to $\gampeff \to \gamp = \Gamma$. This is indeed the decay rate
for the mode given by the fermion-parity operator,
and has been noted to depend \emph{only} on the lump sum of couplings~\cite{Splettstoesser2010Apr,Contreras-Pulido2012Feb,Saptsov2012Dec,Saptsov2014Jul,Schulenborg2016Feb}.
Beyond the wideband limit, we here find it to still be independent of the \emph{statistical factors}, and hence temperatures and chemical potentials for most open-system parameters,
except at the pair resonance \eq{eq_particle_hole_symmetry_point} where the constant value changes from $\Gameps$ and $\GamU$
due to the step-like behavior of $\nzi$ in \Eq{eq_effective_rates}.
Away from this step, the only parameter dependence comes from the lump sums of ---now energy-resolved--- couplings $\Gameps$ or $\GamU$ themselves.

In full agreement with $\gampeff$ being a decay rate to a good approximation, the corresponding left vector $\Bra{p'} = \Bra{\zin\fpOp}$ is also still an approximate \emph{eigen}vector of the kernel in this regime,
up to corrections $\mathcal{O}(\Delta\Gamma^2 \delnsqzi)$. Importantly, this fact can directly be read from the duality-based construction of the kernel \eq{eq_kernel_quantum_dot}, where the prefactor of the coupling of the parity mode to the charge mode is suppressed by exactly this small factor. 
Thus, much of what we know about the \emph{parity rate} remarkably remains true as long as $V \leq U$,
due to \emph{strong pairing effects in the dual model}.

\emph{Charge amplitude.}
In contrast to the parity, the understanding of the 'charge' mode needs to be revised.
In the wideband limit~\cite{Splettstoesser2010Apr,Contreras-Pulido2012Feb,Vanherck2017Mar}
the charge deviation $\Bra{c'} = \Bra{N} - n_z\Bra{\one}$
is an exact left eigenvector of $W$,
and as a result, the time-dependent particle number $n(t) = \Braket{N}{\rho(t)}$ shows \emph{single} exponential decay.
For this reason, $\gamceff = \gamma_c=\zeta \, \Gamma$ has been referred to as 'charge decay rate'.
We now find that for energy-dependent couplings, the covector $\Bra{c'}$ mixes with $\Bra{p'}$
to an extent that grows \emph{linearly} in $\Delta\Gamma/(\Gameps + \GamU)$, and that \emph{is not} suppressed by the factor $\delnsqzi$, see again the revealing form of the kernel \eqref{eq_kernel_quantum_dot}.
In contrast to the parity amplitude $\Bra{\zin\fpOp}$,
the mixing is thus sizable for most system-parameter values. This rules out any interpretation of $\gamceff$ as an effective 'charge' rate, despite our tentative labeling by the index $c$.

Finally considering the nonequilibrium regime $\muR < \epsilon < \epsilon + U < \muL$ at large $|V| > U$, even the parity rate can no longer be understood as in the wideband limit.
In particular, for small cross asymmetry $|\Lambda_-| \ll 1$, the stationary probabilities of the original and dual model are both uniform and thus coincide, $\Ket{z} \approx \Ket{\zin}$
($P_0=P_2=1/4$ but $P_1=P_\uparrow+P_\downarrow=1/2$ due to spin).
This implies large nonequilibrium dual charge fluctuations $\delnsqzi \sim 1$ around the average $\nzi = 1$,
causing the exact decay rates $\gamma_\pm$ [\Eq{eq_relaxation_rates_quantum_dot}] to strongly differ\footnote
	{Consequently, there is no ``most unstable state'' anymore as for the medium bias regime. Therefore the processes clearly related to either $\gampeff$ or $\gamceff$ for $|V| < U$, mix in a nontrivial way.}
from the effective rates $\gamceff$ and $\gampeff$.
We note that in this regime, the overlap $\Braket{\zin}{z}$ of the dual stationary state with the stationary state is maximal:
the general upper bound \eq{eq:product} set by duality and detailed balance is saturated\footnote{Note that in order to calculate this bound one has to consider the probabilities of the energy eigenmodes, $P_0,P_\uparrow,P_\downarrow,P_2$. },
an indication that $\Ket{\zin}$ can no longer define a maximally unstable initial state.

\subsubsection{Maximally unstable state and steps toward stationarity}
\label{sec_unstable}
From the previous subsection, it is clear that
the physics underlying the decay rates in the wideband limit according to \Refs{Splettstoesser2010Apr,Contreras-Pulido2012Feb} need to be reconsidered, in particular for the effective rate $\gamceff$.
We focus on the regime $|V| < U$ where one can associate $\gampeff,\gamceff$ with the specific transitions during the transient decay towards the stationary state $\lim_{t \to \infty} \Ket{\rho(t)}=\Ket{z}$.

The key idea is that any nonstationary initial state $\Ket{\rho(0)}$ is \emph{unstable} with respect to the tunneling of electrons.
Since the quantum dot can host at most two electrons, the stationary state $\Ket{z}$
can be reached from any initial state $\Ket{\rho(0)}$ by the tunneling of either one or two electrons in sequence.
As noted in general, the dual stationary state $\Ket{\zin}$ corresponds to the \emph{most unstable} initial state \emph{relative}
to $\Ket{z}$  [\Eq{eq_unstable} ff.].
In particular for $U\gg T$, if $\rho(0)=\zin$ is initially prepared, then on average \emph{more than one} electron sequentially tunnel in / out to reach the stationary state $\Ket{z}$.

This determines how strongly the rate $\gampeff$ influences the transient evolution:
the amplitude function $\Braket{p'}{\rho(0)}$ of $\exp(- \gampeff t)$ compares an \emph{arbitrary} initial state with the most unstable one since $\Bra{p'} \approx \Bra{\zin}$ by \Eq{eq_parity_amplitude_quantum_dot}.
Duality makes clear why this is a good approximation for $|V| < U$:
due to the attractive interaction governing the dual stationary state,
$\Bra{\zin\fpOp} \approx \Bra{\zin}$ projects mostly on either the empty or doubly occupied state [\Eq{eq_parity_amplitude_quantum_dot}].
The decay rate $\gampeff$ thus only appears with significant amplitude when $\Ket{\rho(0)} \approx \Ket{\zin} = \Ket{0}$ or $\Ket{2}$, respectively.
In agreement with this, corrections in \Eq{eq_parity_amplitude_quantum_dot} are important only when the charge can fluctuate in the dual system [$\sim\delnsqzi$].

In \Fig{fig_temporal}, we illustrate the decay to a stationary state for which \emph{two} electrons are required:
in this case the rate $\gampeff$ ($\gamceff$) is always tied to the first (final) electron tunneling
due to their opposite ordering of the weights $\nzi/2$ and $(2 - \nzi)/2$ \BrackEq{eq_effective_rates}.
Thus, instead of being associated with the decay of a particular dot observable,
the effective rates $\gamceff,\gampeff$ rather represent the \emph{average effect}
of a \emph{temporal} sequence of tunneling events on $\rho(t)$.

A possible source of confusion in \Fig{fig_temporal}(b)
is that the contribution to $n(t) = \Braket{N}{\rho(t)}$
of the final transition $\sim \exp(-\gamceff t)$ 
decays much \emph{faster} than of the first one $\sim \exp(-\gampeff t)$.
This does not mean that causality is broken.
The faster decaying contribution rather reflects that transport is correlated:
the decay $n(t) \rightarrow 2[1 - \expfn{-\gampeff t}]$
hinges on average on the first tunneling event, 
the second electron following almost immediately on the scale given by the rate $\gampeff$ of the first (bunching).

For $|V| > U$ in the nonequilibrium regime $\muR < \epsilon < \epsilon + U < \muL$, 
there is no simple relation between the decay rates and individual transitions towards stationarity.
Also in this regime, this becomes clear from duality.
Namely, tunneling processes in the attractive dual model
are able to undo the pairing effects by inducing large charge fluctuations $\delnsqzi \sim 1$, and this
correlates with the mixing of the decay rates $\gampeff$ and $\gamceff$ due to energy-dependent coupling [\Eq{eq_relaxation_rates_quantum_dot} ff.].

\subsubsection{Number of available decay processes}
\label{sec_zeta}

As long as the above, temporal distinction can be made,
the factor $\zeta$ [\Eq{eq_zeta}] accounts for the number of available processes in the \emph{second} transition [rate $\gamceff$, \Eq{eq_gammaceff}] towards the stationary state $\Ket{z}$. To demonstrate this, \Fig{fig_important_quantities2}(b) shows $\zeta$ as we vary
the level position $\epsilon$ and the energy dependence of the coupling.
For moderate bias $|V| \leq U$ (left panels) there are three regimes visible wich differ by the final state $\Ket{z}$ of the transient decay.

$\zeta \approx 1/2$ (white):
Reaching the final state $\Ket{z} \approx \Ket{2}$ or $\Ket{0}$
always requires a final transition from $\Ket{1}$ with rate $\gamceff \sim \GamU$ ($\Gameps$).
Since this final transition starts in one definite spin state out of two possible spin states, 
only a fraction $\zeta \approx 1/2$ of the available processes contribute by spin-conservation relative to a process starting from a zero-spin state.

$\zeta \approx 1$ (red):
The final state $\Ket{z} \approx \Ket{1}$ is reached \emph{on average} after at most one tunneling transition.
In this case, the $\gampeff$-decay has negligible amplitude since there is only one final transition,
leaving only the decay with rate $\gamceff \approx \Gameps$ [$\GamU$] when starting from $\Ket{\rho(0)} \approx \Ket{0}$ or $\Ket{2}$.
Here, $\zeta \approx 1$ reflects that tunneling of either spin polarization contributes.

$1/2 < \zeta < 1$ (orange): In this regime, $\Ket{z}$ is a mixture of $\Ket{1}$ and either $\Ket{0}$ or $\Ket{2}$
due to thermal broadening and nonequilibrium, resulting in intermediate values between the two previous cases.

Finally, we turn to large biases $|V| \geq U$ [right panels in  \Fig{fig_important_quantities2}(b)]
and $\muR < \epsilon < \epsilon + U < \muL$, where the clear temporal association between decay rates and tunneling processes breaks down. Here, a fourth regime appears:

$\zeta \approx 0$ (blue).
As $\Lambda_- \to -1$, the vanishing of $\zeta$ causes the rate $\gamceff$ to vanish,
i.e., an approximate second zero eigenvalue emerges. This reflects that the singly occupied state $\Ket{1}$ becomes a quasi-stationary state\footnote{Referred to as ``dark state'' in quantum optics, or ``probability sink'' in classical statistics.} with vanishing charge fluctuation. The reason is that in the limit $\GamUL,\GamepsR \rightarrow 0$, a single particle that is on the dot can neither escape to the right nor be accompanied by a second particle from the left.

%% file: paper_quantum_dot2.tex
\subsection{Stationary charge and energy transport in linear response}
\label{sec_stationary}

Our second application of the central result \eq{eq_kernel_quantum_dot}
concerns stationary thermoelectric transport in linear response.
In the wideband limit, the fermionic duality has also been used to analyze this problem~\cite{Schulenborg2017Dec}, simplifying the analysis of the linear and nonlinear response of the particle current $I_N$ and heat current $J$ through the quantum dot to the applied voltage $-V = \muL - \muR$ and temperature gradient $\Delta T = \TL - \TR$.
Using rate equations, these currents have been obtained by separately considering the corresponding flow from the reservoir $\alpha = \text{L,R}$ into the dot\footnote
	{See, e.g., the appendix to \Ref{SchulenborgLic} for a derivation of these expressions in the rate-equation limit.}:
\begin{subequations}
\begin{gather}
\INa = \Bra{N}W_\alpha\Ket{z}
,
\quad
\IEa = \Bra{H}W_\alpha\Ket{z},
\label{eq_stationary_particle_current}
\\
J^\alpha = \IEa - \mua\INa.
\end{gather}
\end{subequations}

Here, we extend this analysis to energy-dependent coupling, focussing entirely on the \emph{linear} Seebeck coefficient $S = \lim_{V,\Delta T \to 0} \left.V/\Delta T\right|_{I_N = 0}$ ---the voltage $V$ required to compensate a particle current
induced by a temperature gradient $\Delta T$--- and the \emph{linear} thermal conductance $K = \lim_{V,\Delta T \to 0} \left. J/\Delta T\right|_{I_N = 0}$ in absence of a net charge current.

%%%%%%%%%%%%%%%%%%%%%%%%%%%%%%%%%%%%%%%%%%%
\subsubsection{Coefficients in the wideband limit}
%%%%%%%%%%%%%%%%%%%%%%%%%%%%%%%%%%%%%%%%%%%

As a reference, we start from the duality-based expressions of these quantities in the wideband limit, as obtained by \Ref{Schulenborg2017Dec}:
\begin{align}
S_\text{WBL}T
&= \epsilon - \mu + \frac{2 - \nzieq}{2}U
\\
K_\text{WBL}
&= \frac{\GamL\GamR}{\Gamma}
\, \frac{U^2}{4T^2}
\, \delnsqzeq\delnsqzieq.
\label{eq_stationary_thermoelectric_wbl}
\end{align}
Regarding the Seebeck coefficient $S$, this shows that the only contribution from the strong Coulomb interaction
is governed by the equilibrium average $\nzieq$ of the \emph{dual} stationary particle number.
Based on duality, it is immediately \emph{quantitatively} clear that this leads to the experimentally observed~\cite{Staring1993Apr,Dzurak1993Sep,Dzurak1997Apr} sawtooth-like step as function of the level-position, right around $\epsilon - \mu = -U/2$ where $\nzi$ jumps sharply between $0$ and $2$ [\Fig{fig_important_quantities2}(a)]. In \Ref{Schulenborg2017Dec}, we compare this interpretation with the usual explanation of this effect~\cite{Beenakker1991Jul,Beenakker1992Oct}.

The behavior of the linear \emph{thermal} conductance $K$ gives, according to \Eq{eq_stationary_thermoelectric_wbl}, an even more interesting twist to the theoretical explanations from earlier studies~\cite{Zianni2007Jan,Erdman2017Jun}.
Whereas the fluctuation-dissipation theorem dictates that the electrical conductance $G \sim \delnsqzeq$ is proportional to the equilibrium charge fluctuations $\delnsqzeq$, the heat conductance $K$ \BrackEq{eq_stationary_thermoelectric_wbl} is additionally proportional to the \emph{dual} fluctuations $\delnsqzieq$.
In fact, the latter fluctuations dominate the former since the sharply peaked behavior of $\delnsqzieq$ around $\epsilon - \mu = -U/2$ deduced from \Eq{eq_dual_fluctuations} suppresses any feature of the regular Coulomb resonances in the fluctuations $\delnsqzeq$, see \Ref{Schulenborg2017Dec}.

One might suspect that the above surprising qualitative and quantitative insights critically depend on the assumed wideband limit.
Using our linear response formula \eq{eq_state_linearization} together with $2\times2$ kernel matrix \eq{eq_kernel_quantum_dot}, we now show that this is not the case. 
The key is that precisely by \emph{avoiding} brute-force derivatives of nonequilibrium quantities, we can again express all parameter-dependences of the transport quantities other than the couplings compactly in terms of the equilibrium charge $\nzeq$ and its dual $\nzieq$, as well as their fluctuations. As these quantities do not depend on $\Gamepsa,\GamUa$ at equilibrium, this enables us in the following to analyze the effect of the energy-dependent couplings separately from the environment statistics entering $\nzeq,\nzieq$. 

%%%%%%%%%%%%%%%%%%%%%%%%%%%%%%%%%%%%%%%%
\subsubsection{Seebeck coefficient beyond wideband}
%%%%%%%%%%%%%%%%%%%%%%%%%%%%%%%%%%%%%%%%
As in the wideband limit, the Seebeck coefficient for energy-dependent couplings can be expressed as a characteristic energy $\Eeq = ST$ consisting of the excess energy of the quantum-dot level $\epsilon-\mu$ plus an interaction-contribution with nontrivial $\epsilon$ dependence:
\begin{align}
	S T
	&=
	\epsilon - \mu + \frac{\frac{2 - \nzieq}{2} \frac{\GamUL\GamUR}{\GamU}}
	{\frac{\nzieq}{2} \frac{\GamepsL\GamepsR}{\Gameps}+\frac{2 - \nzieq}{2}\frac{\GamUL\GamUR}{\GamU}}
	\, U
	\label{eq_stationary_seebeck}\\
	&=
	\epsilon - \mu + \frac{(1+\Lambda)(2 - \nzieq)}{(1-\Lambda)\nzieq + (1+\Lambda)(2 - \nzieq)} \, U
	\notag.
\end{align}
The energy dependence of the couplings is entirely captured by the single coupling-asymmetry factor
\begin{align}
 \Lambda &= \frac{\tau_\epsilon - \tau_U}{\tau_\epsilon + \tau_U},
 \label{eq_coupling_asymmetry}
\end{align}
determined by the net time scales $\tau_{\epsilon} =  \frac{1}{\GamepsL} + \frac{1}{\GamepsR}$ and $\tau_{U} =  \frac{1}{\GamUL} + \frac{1}{\GamUR}$ for tunneling at energy $\epsilon$ and $\epsilon + U$, respectively.
These times are dominated by the slowest process, and, when multiplied by the temperature $T$,
can be understood as quantifying the highest effective tunneling resistance. In particular, these times enter \Eq{eq_coupling_asymmetry} such that a left-right asymmetry in the junction alone does not lead to a nonzero $\Lambda$.
Thus, in the linear regime, only an asymmetry in the \emph{energy dependence} of the couplings matters. This is in contrast to the nonlinear regime analyzed above \BrackEq{eq_average_charges}, where the (energy-dependent) couplings to different resevoirs contribute with different weights.

More explicitly, while we recover the wideband limit result \eq{eq_stationary_thermoelectric_wbl} for $\Lambda = 0$,
a strong energy-asymmetry $\Lambda\rightarrow\pm 1$ effectively suppresses one of the two Coulomb resonances,
making the dependence of $S$ on $\epsilon$ purely linear.
The value of $ST$ then coincides with the single \emph{many-body} energy that effectively remains, meaning $ S T \approx \epsilon$ for $\Lambda\rightarrow -1$ or $ S T \approx \epsilon + U$ for $\Lambda\rightarrow +1$.

\begin{figure}
	\centering
	\includegraphics[width=\linewidth]{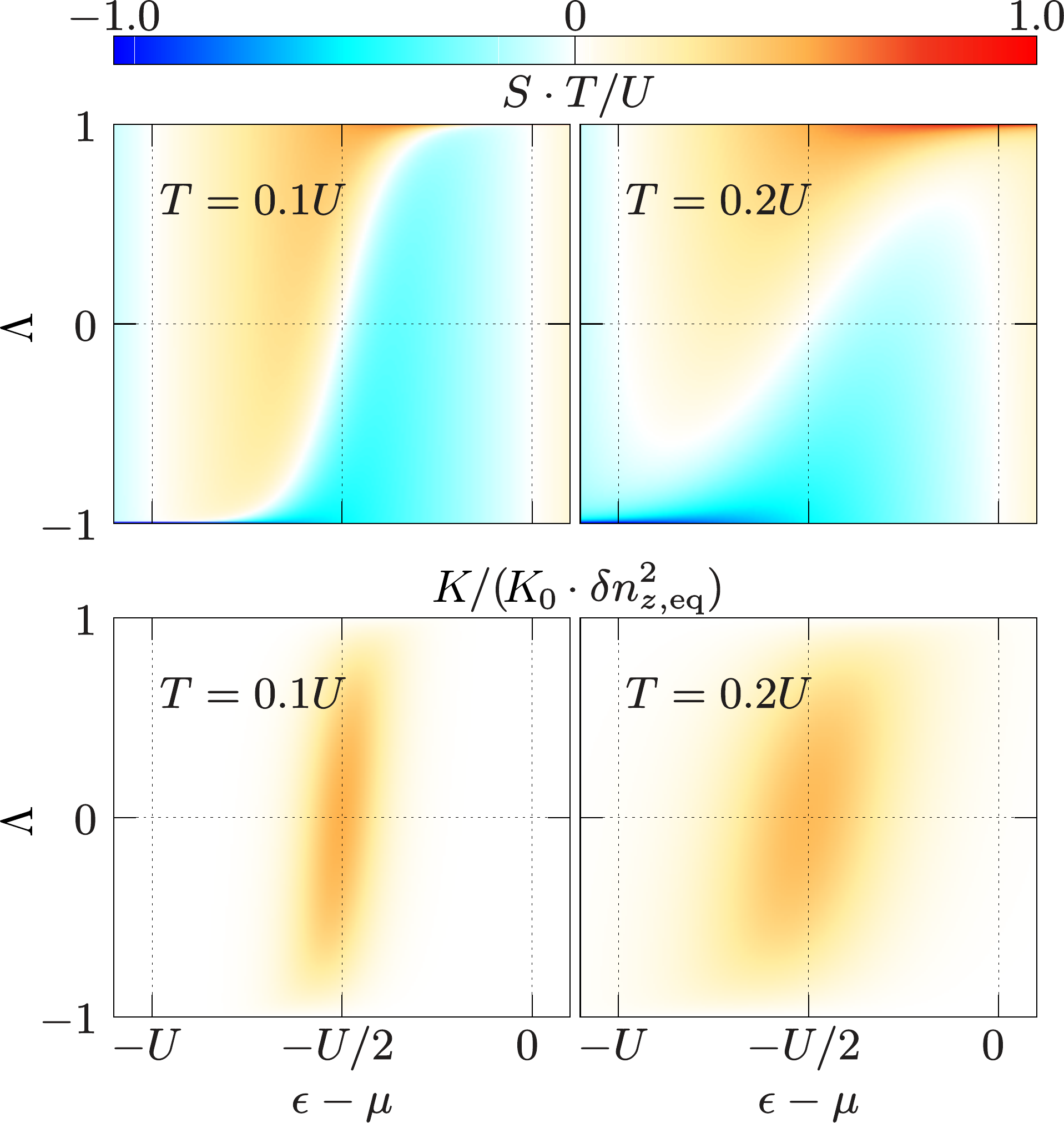}
	\caption{
		Linear Seebeck coefficient $S$ \BrackEq{eq_stationary_seebeck} and thermal conductance $K$ \BrackEq{eq_stationary_fourier} as a function of the level-position $\epsilon$ and the coupling asymmetry $\Lambda$ \BrackEq{eq_coupling_asymmetry}.
		The physical scale of the heat conductance $K_0$ is defined in \Eq{eq_stationary_fourier}. We have furthermore divided $K$ by the fluctuations $\delnsqzeq$, since these are strongly suppressed between the Coulomb resonances, and otherwise add very little structure to the resonant behavior of $K(\epsilon)$.
	}
	\label{fig_stationary_linear_response}
\end{figure}
At intermediate asymmetries $0 < |\Lambda| < 1$, the sharp step in $\nzieq(\epsilon)$ enters the behavior of the Seebeck coefficient, as plotted in \Fig{fig_stationary_linear_response} with $\epsilon-\mu$ and $\Lambda$ as independent parameters.
For low $T/U = 0.1$ (left panels), the width and shape of this sawtooth step are hardly affected.
The only effect of finite energy-asymmetry ---even up to sizable $|\Lambda|$--- is that the central root $S = 0$ shifts away from its  position in the wideband limit, $\epsilon \rightarrow \mu-U/2$, by an amount $\Delta\epsilon_S$.
Analytically, we obtain this shift by first solving $S = 0$ for $\Lambda(\epsilon)$ to find $\Lambda \sim (\epsilon - \mu + \frac{U}{2}) + \mathcal{O}[(\epsilon - \mu + \frac{U}{2})^3]$, and by then inverting the result. This shows that the electron-hole symmetry breaking is approximately linear in $\Lambda$:
\begin{align}
	\frac{\Delta\epsilon_S}{U} \approx \frac{1}{2}\sqbrack{ \frac{U}{4T}\sqbrack{1 + \tanh\nbrack{\frac{U}{4T}}}-1}^{-1} \times \Lambda.\label{eq_asym_shift}
\end{align}
The estimate holds up to $\Lambda \sim 0.9$ in the plot for $T/U = 0.1$ (white line feature), which corresponds to an order of magnitude in the coupling ratio $\Gamma_U / \Gamma_\epsilon$.

We conclude,
that the energy dependence of the coupling does not prevent the pairing physics of the dual attactive model
from dominating the observable properties of the original model of interest, which here leads to the sharp step in $S(\epsilon)$.
This is also consistent with our observations in \Fig{fig_important_quantities2} for $T = U/10$.

The right panels in \Fig{fig_stationary_linear_response} show the effect of an increased temperature  $T/U = 0.2$.
As long as the saw-tooth step is present, it remains relatively sharp with the effective $T/2$-broadening~\cite{Schulenborg2017Dec} of the dual charge $\nzi$ \BrackEq{eq_statistical_approx}.
However, the step disappears already for smaller asymmetries $|\Lambda| \gtrsim 0.5$, i.e.,
coupling ratios $\GamU/\Gameps \gtrsim 3$ or $\GamU/\Gameps \lesssim 1/3$. This agrees with \Eq{eq_asym_shift} indicating that the factor at which the shift $\Delta\epsilon_S$ linearly increases with $\Lambda$ is a function of $U/4T$; it means that a temperature increase $T = U/10 \rightarrow 2T = U/5$ almost corresponds  to an order of magnitude in this ratio. In fact, for large enough temperatures, the central root even merges with one of the $T$-broadened roots close to the Coulomb resonances $\epsilon - \mu = 0,-U$ at a threshold asymmetry $0 < |\Lambda_\text{th}| \leq 1$. 
We can estimate this threshold by the solution of $\Delta\epsilon_S/U \approx \pm(1/2 - T/U)$, with \Eq{eq_asym_shift} yielding $|\Lambda_\text{th}| \approx 0.8$ for $T = 0.2U$, which is in reasonable agreement with \Fig{fig_stationary_linear_response}.
Physically speaking, the above observed $T$-dependence reflects that rising temperatures tend to average the contributions from the two Coulomb resonances $\epsilon - \mu = 0, -U$ over a wider range of level-positions. This enhances the effect of energy-asymmetries $\Lambda\neq 0$ favoring processes at either resonance, and hence of the effective single-resonance behavior of $S$ in the limit $|\Lambda|\rightarrow 1$.\par

Next, we note that while figure~\fig{fig_stationary_linear_response} treats the asymmetry $\Lambda$ as an independent parameter ---and thereby gives a useful overview of all the possible effects of energy-dependent coupling--- $\Lambda$ actually depends on $\epsilon$
through the \emph{energy profiles} of the tunnel barriers $\Gamma_{\epsilon\alpha}$ and $\Gamma_{U\alpha}$ \BrackEq{eq_coupling_rates_quantum_dot}.
We see, for instance, that if both junctions $\alpha$ are coupled with an exponential energy profile 
$\Gamma_\alpha(E) \propto e^{(E-E_0)/D}$, as, e.g., assumed in \Refs{Matveev1996Oct,Kaestner2015Sep}, 
this corresponds to an energy-asymmetry factor
$\Lambda = \tanh(U/2D)$ which is independent of $\epsilon$ and profile parameter $E_0$.
The independent variation of $\Lambda$ that we have assumed in the discussion so far
then practically corresponds to varying $D$.
Note that in the case of a (near) exponential energy profile and a known $U$, an estimate of the tunnel barrier energy-profile parameter $D$ is possible from a measurement of $S$. Namely, the Seebeck coefficient in the middle between the two Coulomb resonances $\epsilon - \mu = 0,-U$, directly gives the value of the energy-asymmetry factor at that energy:
\begin{align}
\Lambda|_{\epsilon=\mu-U/2} = 2 \, \frac{S T}{U} \Big|_{\epsilon=\mu-U/2}.
\end{align}
We stress that this simple result is not due to particle-hole symmetry, which is broken for $\Lambda \neq 0$. It arises due to the simplifications incurred by the fermionic duality.

If the energy profile is instead assumed to be linear~\cite{Sothmann2012May,Hartmann2015Apr},
$\Gamma_\alpha(E) \propto (E-E_0)/D$ for both $\alpha$,
then
$\Lambda(\epsilon) = [2 (\epsilon-E_0)/U +1]^{-1}$.
Experimentally varying $\epsilon$ in this subexponential case, we traverse a \emph{curve} $(\epsilon,\Lambda(\epsilon))$ in \Fig{fig_stationary_linear_response}
starting at $\Lambda(\epsilon=0) \neq 0$
and bending towards $\Lambda \to 0$ for $\pm(\epsilon -E_0) \gg U$.
For the opposite case of superexponential energy dependence, the asymptotes are reversed.

%%%%%%%%%%%%%%%%%%%%%%%%%%%%%%%%%%%%%%%%
\subsubsection{Thermal conductance beyond wideband limit}
%%%%%%%%%%%%%%%%%%%%%%%%%%%%%%%%%%%%%%%%

The result for the thermal conductance beyond the wideband limit reads
\begin{align}
	K 
	&=
	\frac{
	\frac{\GamepsL\GamepsR}{\Gameps} \cdot \frac{\GamUL\GamUR}{\GamU}}
	{\frac{\nzieq}{2} \frac{\GamepsL\GamepsR}{\Gameps}+\frac{2 - \nzieq}{2}\frac{\GamUL\GamUR}{\GamU}}
	\frac{U^2}{4T^2}
	\, 
	\delnsqzeq\delnsqzieq\nonumber\\
	&=K_0
	\frac{(1-\Lambda)(1+\Lambda)}{(1 - \Lambda)\nzieq + (1 + \Lambda)(2 - \nzieq)}
	\delnsqzeq\delnsqzieq
	\label{eq_stationary_fourier}
	. 
\end{align}
It is plotted in the lower panels in \Fig{fig_stationary_linear_response}
as function of $\epsilon$ and $\Lambda$ for the same parameter sets as above. Note that in the figure, we have already scaled $K$ by the fluctuations $\delnsqzeq$, realizing from the above discussion that $\delnsqzeq$ mostly leads to an overall suppression of $K$ while not adding any features to $K(\epsilon)$ as long as $U \gg T$.

The main observation is that the peak in $K$ obtained in the wideband limit persists
with some modifications. Regarding the effect of the energy-dependent couplings,
we note that it enters the heat conductance in two ways.
First, it sets the overall scale of the peak
\begin{align}
K_0 = \nbrack{\frac{1}{\tau_\epsilon} + \frac{1}{\tau_U}}
\, \frac{U^2}{4T^2}
.
\end{align}
The addition of inverse tunneling times $\tau_{\epsilon}$ and $\tau_{U}$, dominated by the fastest process,
supports the picture of two parallel transport channels that is, e.g., illustrated in \Ref{Schulenborg2017Dec}; it highlights that the thermal conductance in a weak-coupling situation is entirely due to Coulomb interaction and can be understood as a two-particle resonance in the inverted model.  We hence  again observe the above mentioned ``twist'' in the explanation for the parameter-dependence of $K$.

The second way in which energy-dependent couplings affect the thermal conductance is through the prefactor in \Eq{eq_stationary_fourier}; it both shifts and suppresses the peak generated by the dual fluctuation $\delnsqzieq$
as seen in \Fig{fig_stationary_linear_response}.
To understand the suppression, we note that the dual fluctuation has its maximum as function of $\epsilon$
when $\nzieq \approx1$.
Thus, for moderate peak shifts, the denominator in \Eq{eq_stationary_fourier} remains constant
and leaves the numerator $1-\Lambda^2$ to suppress the peak amplitude.

What remains to be analyzed is thus how much the peak is shifted for growing asymmetries $|\Lambda| > 0$ and different temperatures. Similar to the shift of the central root of the Seebeck coefficient $S$, \Fig{fig_stationary_linear_response} suggests an approximately linear relation between $\Lambda$ and the peak-position shift $\Delta\epsilon_K$ of $K/\delnsqzeq$ away from the wideband position $\epsilon - \mu = -U/2$. Indeed, by solving $\partial_\epsilon(K/\delnsqzeq) = 0$ for $\Lambda$ and proceeding as for the Seebeck coefficient, we obtain the simple relation
\begin{equation}
 \frac{\Delta\epsilon_K}{U} \approx \frac{1}{2}\cdot \frac{2T}{U}\cdot \sqbrack{1 - f\nbrack{\frac{U}{2T}}}\times\Lambda,\label{eq_peak_shift}
\end{equation}
with $f(x) = \sqbrack{\expfn{x} + 1 }^{-1}$. The shift is a function of $U/2T$,  eventually leading to  increase $|\Delta\epsilon_K|$ for a fixed $\Lambda$ at higher temperatures,  in agreement with \Fig{fig_stationary_linear_response}. However, the  shift is not as pronounced as the shift of the Seebeck coefficient $S$, the latter being instead a function of $U/4T$. For instance, \Eq{eq_peak_shift} gives $\Delta\epsilon_K/U \approx 0.1\Lambda$ for $T = 0.1U$ and $\Delta\epsilon_K/U \approx 0.2\Lambda$ for $T = 0.2U$.
Physically, this again reflects that $K$ cannot be tied to a single Coulomb resonance. Therefore, the effect of  a finite energy-asymmetry $\Lambda\neq 0$ favoring either of the two Coulomb resonances, $\epsilon - \mu = 0$ or $-U$ is less pronounced than for the Seebeck coefficient.

In conclusion, despite the challenge of dealing with many parameters ($\epsilon$, $U$, $T$, $V$, $\Delta T$, $\Gamepsa$, $\GamUa$),
the fermionic duality enabled a transparent derivation and complete physical analysis of the linear thermoelectric effects in a strongly interacting quantum dot for any energy profile of the coupling.

%% file: paper_summary.tex
\section{Summary and outlook}

In this paper we have extended the recently discovered fermionic duality relation in two ways,
by allowing for strong energy dependence of the system-reservoir coupling matrix
and by accounting for the full density matrix of probabilities and coherences.
Most importantly, our results confirm that both the techniques and insights
of the earlier developed fermionic duality~\cite{Schulenborg2016Feb,Vanherck2017Mar,Schulenborg2017Dec}
extend \emph{far beyond} the wideband limit.
This promotes it further to a general tool for significantly simplifying quantum transport calculations for strongly interacting systems.

In particular, we have obtained insight into
the spectrum of eigenvalues and eigenvectors of general fermionic quantum master equations.
We therefore exploit a mapping between two systems
related by our duality, which have energies with opposite signs, in particular for the strong interactions,
allowing pairing effects in one system to explain physical parameter dependencies in the other.
We also established an explicit formula relating the stationary states of a fermionic rate equation for 
such two related models  when additionally detailed balance holds (Kolmogorov criterion), even at large voltage and/or thermal bias.

We applied our approach to a strongly interacting quantum dot and clearly identified the effects specific to the breakdown of the wideband-limit approximation.
Importantly, we achieved this directly on the level of the time-evolution \emph{kernel}
of rate equations, \emph{before} computing final expression for the decay rates, modes, amplitudes and transport quantities of interest.

For the transient decay induced by an instant switch of the energy level of a quantum dot coupled to reservoirs,
we showed that the mixing of wideband-limit decay rates is highly confined:
it appears only at parameters where the \emph{attractive dual} quantum dot shows a resonance due to electron pairing.
The corresponding mixing of eigenvectors can nevertheless be strong.
For example, the so-called 'charge' mode becomes a completely ill defined concept beyond the wideband limit due to strong interaction effects.
Remarkably, despite the strong energy-dependence of the coupling,
one of the exact time-evolution amplitudes remains very close to the 'parity'-amplitude in the wideband limit,
which could be rationalized by fermionic duality.

For linear thermoelectric transport through the quantum dot
we derived simple analytic formulas for the Seebeck and Fourier-heat coefficients that clearly distinguish the quantum-statistical effects from those due to the energy-dependence of the couplings.
This allowed us to quantify and physically understand how and why the breaking of electron-hole symmetry  affects these observables in different ways, when going far beyond the wideband limit.

%% file: paper_ack.tex
\acknowledgments
We acknowledge discussions with
R. Saptsov, N. Dittmann, M. Kataoka, J. Fletcher,
and financial support of the Deutsche Forschungsgemeinschaft (RTG 1995) (M. W., J. Sp.),
the Swedish Research Council (VR)
and the Knut and Alice Wallenberg foundation (J.Sc., J. Sp.).

%% file: paper_app_linearization.tex
\section{State linearization beyond the wideband limit}
\label{sec_app_state_linearization}
We here prove the state linearization formula \eq{eq_state_linearization} from the main text:
\begin{equation}
\evalAtEqui{\partial_x\Ket{m_0}} = \evalAtEqui{\frac{1}{W}}\sum_\alpha\evalAtEqui{W_\alpha}\evalAtEqui{\partial_x\Ket{m_{0,\alpha}}}.\label{eq_notes_14}
\end{equation}
The starting point is the lead sum \eq{eq_kernel_lead_sum}, 
\begin{equation}
 W = \sum_\alpha W_\alpha,\label{eq_notes_1}
\end{equation}
with reservoir index $\alpha$. Furthermore, we assume a non-degenerate eigenvalue $0$ for $W$ and at least one eigenvalue $0$ for each $W_\alpha$. This implies that there is only one stationary state $\Ket{m_0}$ and at least one reservoir-resolved stationary state $\Ket{m_{0,\alpha}}$ fulfilling
\begin{equation}
W\Ket{m_0} = 0 \lrsepa W_\alpha\Ket{m_{0,\alpha}} = 0\label{eq_notes_2}.
\end{equation}
Finally, we assume that each reservoir $\alpha$ has at least one stationary state $\Ket{m_{0,\alpha}}$ that equals $\Ket{m_0}$ \emph{at equilibrium}:
\begin{equation}
 \evalAtEqui{\Ket{m_0}} = \evalAtEqui{\Ket{m_{0,\alpha}}} \equiv \Ket{m_{0,\text{eq}}} \quad\forall\alpha.\label{eq_notes_4}
\end{equation}
The non-degenerate zero eigenvalue for $W$ implies that a non-Hermitian, \emph{reflexive generalized inverse} $W^{-1}$ exists~\cite{Ben-Israel2003,Hunter2014Apr} such that
\begin{equation}
 \frac{1}{W}W = W\frac{1}{W} = \mathcal{I} - \Ket{m_0}\Bra{\one}.\label{eq_notes_5}
\end{equation}
The right hand side projects onto the space orthogonal to the left and right zero-spaces. If $W$ is diagonalizable, this generalized inverse written in our notation \BrackEq{eq_generator_diagonalized} reads
\begin{equation}
 \frac{1}{W} = \sum_{k>0} \frac{1}{\lambda_k}\Ket{m_k}\Bra{a_k}.\label{eq_notes_7}
\end{equation}
Linearizing the first equation in \eq{eq_notes_2},
\begin{equation}
 0 = \partial_x\evalAtEqui{\sqbrack{W\Ket{m_0}}} \equalbyeqns{eq_notes_2}{eq_notes_4} \evalAtEqui{\sqbrack{\partial_x W}}\Ket{m_{0,\text{eq}}} + \evalAtEqui{W}\evalAtEqui{\partial_x\Ket{m_0}},\label{eq_notes_8}
\end{equation}
we insert \Eq{eq_notes_1} to obtain
\begin{equation}
 \evalAtEqui{W}\cdot\evalAtEqui{\partial_x\Ket{m_0}} \equalbyeqns{eq_notes_8}{eq_notes_1} -\sum_\alpha\evalAtEqui{\sqbrack{\partial_x W_\alpha}}\Ket{m_{0,\text{eq}}}.\label{eq_notes_9}
\end{equation}
Applying the generalized inverse from the left, we use
\begin{equation}
 \Bra{\one}\evalAtEqui{\partial_x\Ket{m_0}} = \evalAtEqui{\partial_x\underbrace{\Braket{\one}{m_0}}_{=1}} = 0,\label{eq_notes_10}
\end{equation}
to advance to
\begin{align}
\evalAtEqui{\sqbrack{\frac{1}{W}W}}\evalAtEqui{\partial_x\Ket{m_0}} &\equalbyeqn{eq_notes_5}\big[\mathcal{I} - \Ket{m_{0,\text{eq}}}\Bra{\one}\big]\evalAtEqui{\partial_x\Ket{m_0}}\notag\\
&\equalbyeqn{eq_notes_10} \evalAtEqui{\partial_x\Ket{m_0}}.\label{eq_notes_11}
\end{align}
Thus
\begin{equation}
 \evalAtEqui{\partial_x\Ket{m_0}} \equalbyeqns{eq_notes_9}{eq_notes_11} -\evalAtEqui{\frac{1}{W}}\sum_\alpha\evalAtEqui{\sqbrack{\partial_x W_\alpha}}\Ket{m_{0,\text{eq}}}.\label{eq_notes_12}
\end{equation}
Repeating the above procedure with the second equation in \eq{eq_notes_2}, we find
\begin{align}
 \evalAtEqui{\sqbrack{\partial_x W_\alpha}}\Ket{m_{0,\text{eq}}} &\equalbyeqn{eq_notes_4}\evalAtEqui{\partial_x\sqbrack{W_\alpha\Ket{m_{0,\alpha}}}}\notag\\
 &\phantom{\equalbyeqn{eq_notes_4}}- \evalAtEqui{W_\alpha}\evalAtEqui{\partial_x\Ket{m_{0,\alpha}}}\notag\\
 &\equalbyeqn{eq_notes_2} -\evalAtEqui{W_\alpha}\evalAtEqui{\partial_x\Ket{m_{0,\alpha}}}.\label{eq_notes_13}
\end{align}
Inserting \eq{eq_notes_13} into \eq{eq_notes_12} yields the desired result \eq{eq_notes_14}.

%% file: paper_app_duality.tex
\section{Dual Kernel}
\label{sec_app_duality}

We now illustrate how the \emph{dual} kernel $\dual{W}$ and its Hermitian conjugate is properly constructed for the fermionic duality \eq{eq_duality_generators} with explicitly energy-dependent couplings $\Gamma_{\alpha\nu;\ell\ell'}(E)$ \BrackEq{eq_coupling_rates}.\par

We start from expression \eq{eq_born_markov_kernel} for the kernel $W$ in terms of the fermionic superoperators defined in \Eq{eq_field_superoperator}.
First, we take the superhermitian adjoint of this expression and apply the superoperators $\cP$, where we use the relations \eq{eq_field_superoperator_properties} for the superoperators $G^q_{\eta \ell}$ and $L^\sudag = L, \cP L\cP = L$ according to
\Eq{eq_local_liouvillian}. Subsequent energy-inversion then leads to
\begin{align}
&\cP W^\sudag(-L,\{-\mua\})\cP\notag\\
&=
\frac{-1}{2\pi i}\int_{-\infty}^\infty d E
\sumsub{\ell\ell'}{\eta q}
\sum_{\alpha\nu}
\Gamma^{-\eta}_{\alpha\nu;\ell\ell'}(E)\notag\\
&\phantom{=}\times \sqbrack{f^{-\eta}_\alpha(-E) - qf^{{\eta}}_\alpha(-E)}G^{q}_{-\eta \ell'}\frac{1}{\eta E + L - i0_+}G^+_{{\eta}\ell}\notag\\
&= \frac{1}{2\pi i}\int_{-\infty}^\infty d E
\sumsub{\ell\ell'}{\eta q}
\sum_{\alpha\nu}
\Gamma^{\eta}_{\alpha\nu;\ell\ell'}(-E)\notag\\
&\phantom{=}\times \sqbrack{f^{-\eta}_\alpha(E) - qf^{{\eta}}_\alpha(E)}G^{q}_{-\eta \ell}\frac{1}{\eta E - L + i0_+}G^+_{{\eta}\ell'}
\label{eq_app_born_markov_dual_kernel_step},
\end{align}
where $\left.f^{\eta}_\alpha(E)\right|_{\mua \rightarrow -\mua} = f^{-\eta}_\alpha(-E)$ and $\sqbrack{\Gamma^{\eta}_{\alpha\nu;\ell\ell'}(E)}^* = \Gamma^{-\eta}_{\alpha\nu;\ell\ell'}(E)$ \BrackEq{eq_coupling_conjugate} have been used in the first step, and the second step follows by relabeling the summation indices $\ell \leftrightarrow \ell'$, then using $\Gamma^{-\eta}_{\alpha\nu;\ell'\ell}(E) = \Gamma^{\eta}_{\alpha\nu;\ell\ell'}(E)$, and finally by transforming the integration variable $E \rightarrow -E$.\par

The crucial point which still prevents us from adding $W$ and the expression in \Eq{eq_app_born_markov_dual_kernel_step} to prove the duality \eq{eq_duality_generators} is the minus sign in the energy-dependent couplings $\Gamma^{\eta}_{\alpha\nu;\ell\ell'}(-E)$. One solves this by \emph{defining} the dual kernel from the start with an inverted barrier energy profile $\dual{\Gamma}^{\eta}_{\alpha\nu;\ell\ell'}(E) = \Gamma^{\eta}_{\alpha\nu;\ell\ell'}(-E)$.
Namely, with $W = F(L,\{\mua\},\{\Gamma^{\eta}_{\alpha\nu;\ell\ell'}(E)\})$, we write
\begin{align}
\dual{W} &= F(\dual{L},\{\dual{\mu}_\alpha\},\{\dual{\Gamma}^{\eta}_{\alpha\nu;\ell\ell'}(E)\})\notag\\
&= F(-L,\{-\mua\},\{\Gamma^{\eta}_{\alpha\nu;\ell\ell'}(-E)\})\notag\\
&=
\frac{1}{2\pi i}\int_{-\infty}^\infty d E
\sumsub{\ell\ell'}{\eta q}
\sum_{\alpha\nu}
\dual{\Gamma}^{\eta}_{\alpha\nu;\ell\ell'}(E)\notag\\
&\phantom{=}\times \sqbrack{f^{-\eta}_\alpha(-E) - qf^{{\eta}}_\alpha(-E)}G^+_{{-\eta}\ell}\frac{1}{\eta E - \dual{L} + i0_+}G^q_{\eta \ell'}\notag\\
&=
\frac{-1}{2\pi i}\int_{-\infty}^\infty d E
\sumsub{\ell\ell'}{\eta q}
\sum_{\alpha\nu}
\Gamma^{\eta}_{\alpha\nu;\ell\ell'}(E)\notag\\
&\phantom{=}\times \sqbrack{f^{-\eta}_\alpha(E) - qf^{{\eta}}_\alpha(E)}G^+_{{-\eta}\ell}\frac{1}{\eta E - L - i0_+}G^q_{\eta \ell'}.
\label{eq_app_born_markov_kernel_dual}
\end{align}
Calculating $\cP \dual{W}^\sudag\cP$ analogously to \Eq{eq_app_born_markov_dual_kernel_step}, we indeed find the desired duality $W + \cP \dual{W}^\sudag\cP = -\Gamop$ with the coupling superoperator $\Gamop$ given by \Eq{eq_coupling_superoperator}. In other words, for general energy-dependent couplings, the dual model is not only defined in terms of a mere inversion of local energies $L \rightarrow -L$ and chemical potentials $\mua \rightarrow -\mua$, but also in terms of energy-inverted barriers $\Gamma^{\eta}_{\alpha\nu;\ell\ell'}(E) \rightarrow \Gamma^{\eta}_{\alpha\nu;\ell\ell'}(-E)$, as illustrated in \Fig{fig_important_quantities1}: when evaluating these energy-inverted barrier profiles at inverted local energies, the two energy sign changes compensate each other. 

Interestingly, recontemplating the definition of the couplings $\Gamma_{\alpha\nu;\ell\ell'}(E)$ given in \Eq{eq_coupling_rates}, the inverted dual barrier profile $\dual{\Gamma}^{\eta}_{\alpha\nu;\ell\ell'}(E) = \Gamma^{\eta}_{\alpha\nu;\ell\ell'}(-E)$ can be realized on a microscopic level by defining the dual model also in terms of inverted \emph{reservoir} energies $-\epsilon_{\kappa\alpha\nu}$, prior to taking the continuum limit.
This is achieved by $\Hlead \rightarrow \dual{H}_{\text{res}} = -\Hlead$, both in the reservoir density of states entering $\Gamma_{\alpha\nu;\ell\ell'}(E)$ \emph{and} in the initial reservoir state $\rho^\lead_0$ \BrackEq{eq_lead_state}. The energy inversion in $\rho^\lead_0$ is then necessary to still obtain the same energy signs in the Fermi functions $f^{\eta}_\alpha(E)$ in \Eq{eq_app_born_markov_kernel_dual}.

%% file: paper_app_unexpected_L.tex
\section{Generator of the dual master equation}
\label{sec_unexpected_L}
The dual master equation \eq{eq_qme_dual} in \Sec{sec_duality_extended} reads
	\begin{align}
	\partial_t \dual{\rho}(t)
	&= (i\dual{L}+\dual{W})) \dual{\rho}(t)
	=
	(-iL+\dual{W}) \dual{\rho}(t)\ .
	\label{eq_qme_dual_L}
	\end{align}
Note the different ---unexpected--- sign of the Liouvillian in this dual quantum master equation: the dual equation does not follow by inverting all energies $L \to \dual{L}$,
but \emph{only} those in the kernel $W \to \dual{W}$.
This means that in the solution $\dual{\rho}(t)$ of \Eq{eq_qme_dual_L}, the coherences are 'precessing' with the same orientation (dictated by $i\dual{L}=-iL$) as in the solution $\rho(t)$,
even though in the computation of the kernel $\dual{W}$, the coherences are precessing in the opposite direction (by setting $L \to \dual{L}=-L$ in the microscopic expression \eq{eq_born_markov_kernel}).
While formally perfectly valid, this requires attention
when considering $\dual{\rho}$ as the state of a physical system.

%% file: paper_app_crossrelation.tex
\section{Cross relations between eigenvectors in the wideband limit}
\label{sec_crossrel}
The wideband limit duality \eq{eq_duality_generators}
\begin{align}
(-iL+W) + \mathcal{P} ( i \bar{L} + \dual{W})^\sudag \mathcal{P} = -\Gamma
\label{eq_app_duality_generators2}
\end{align}
establishes a relation between the generator of the quantum master equation of interest,
\begin{align}
	\partial_t \rho(t) = (-iL+W) \rho(t)
	\label{eq_app_qme},
\end{align}
and a dual master equation
\begin{align}
\partial_t \dual{\rho}(t) = (i\dual{L}+\dual{W}) \dual{\rho}(t).
	\label{eq_app_qme_dual}
\end{align}
Using this, we can bypass the algebraic solution of the left eigenvalue problem
by simple parameter substitutions:

(i) Find a \emph{right} eigenvector $\Ket{m_k(L,W)}$ of the generator of interest $-iL+W$:
\begin{align}
( -i L + W) \Ket{m_k(L,W)}
=
\lambda_k(L,W) \, 
\Ket{m_k(L,W)}.
\end{align}

(ii) From this construct a \emph{right} eigenvector $\Ket{m_k(-\dual{L},\dual{W})}=\Ket{m_k(L,\dual{W})}$ of the
dual generator $i \dual{L} + \dual{W}$ by simple parameter substitution $W\to \dual{W}$:
\begin{align}
( i \dual{L} + \dual{W}) \Ket{m_k(-\dual{L},\dual{W})}
=
\lambda_k(-\dual{L},\dual{W}) \, 
\Ket{m_k(-\dual{L},\dual{W})}.
\label{eq_step2}
\end{align}

(iii) Right multiply the duality relation \eq{eq_app_duality_generators2} by $\mathcal{P}$ using $\mathcal{P}^2= \mathcal{I}$,
\begin{align}
(-iL+W) \mathcal{P} =  - \mathcal{P} \Big[ \Gamma + ( i \dual{L} + \dual{W})^\sudag \Big] \ .
\end{align}
Now apply this to $\Ket{m_k(-\dual{L},\dual{W})}$, 
\begin{align}
&\Bra{m_k(-\dual{L},\dual{W})}\mathcal{P}(-iL+W)\notag\\
=&  - \Bra{m_k(-\dual{L},\dual{W})} \, \mathcal{P} \Big[ \Gamma + \lambda_k^{*}  (-\dual{L},\dual{W}) \, \mathcal{I} \Big].
\end{align}

We have thus found one of the \emph{left} eigenvectors of the generator of interest of $-iL+W$,
\begin{align}
\Bra{a_{k'}(L,W)} =
\Bra{m_k(-\dual{L},\dual{W})}
\mathcal{P}
=
\Bra{\fpOp m_k(-\dual{L},\dual{W})},
\end{align}
which in general belongs to a \emph{different} eigenvalue, numbered by $k' \neq k$,
\begin{align}
	\lambda_{k'} (L,W)
	=-\Big[
	\Gamma + \lambda_k^{*} (-\dual{L},\dual{W})
	\Big]
	.
\end{align}

(iv)
Following the corresponding steps, one can equally derive a right eigenvector from a left one
with the same relation between their eigenvalues:
\begin{align}
\Ket{m_{k'}(L,W)} =
\mathcal{P} \,
\Ket{a_k(-\dual{L},\dual{W})}
=
\Ket{\fpOp a_k(-\dual{L},\dual{W})}.
\end{align}
We note that the above formulation is tailored for the weak-coupling limit and therefore slightly differs from that in \cite{Schulenborg2016Feb}.